\documentclass[11pt]{article}

\usepackage[utf8]{inputenc}
\usepackage[T1]{fontenc}
\usepackage[english]{babel}
\usepackage[margin=1in]{geometry}
\usepackage{mathptmx}
\usepackage{microtype}
\usepackage{graphicx}
\graphicspath{{figures/}}
\usepackage{amsmath}
\usepackage{amssymb}
\usepackage{amsthm}
\usepackage{booktabs}
\usepackage{tabularx}
\usepackage{array}
\usepackage{rotating}
\usepackage{float}
\usepackage{placeins}
\usepackage{caption}
\usepackage{enumitem}
\setlist[itemize]{itemsep=2pt,topsep=3pt}
\setlist[enumerate]{itemsep=2pt,topsep=3pt}
\usepackage{tikz}
\usetikzlibrary{positioning, arrows.meta, shapes.geometric, calc}
\usepackage{xcolor}
\usepackage{tcolorbox}
\usepackage[round]{natbib}
\usepackage{hyperref}
\hypersetup{colorlinks=true,linkcolor=black,citecolor=black,urlcolor=blue}

\theoremstyle{definition}
\newtheorem{theorem}{Theorem}
\theoremstyle{plain}
\newtheorem{corollary}{Corollary}
\newtheorem{remark}{Remark}

\title{Per-Prediction Uncertainty Is Not a Regime Detector:\\
       A Detectability Bound and Two-Level Deployment\\
       for Cross-Sectional Equity Rankers}
\author{Ursina Sanderink\\
        \small Quant Signal Lab}
\date{July 2026}

\begin{document}
\maketitle

\begin{abstract}
\noindent
Quantitative finance is increasingly exploring AI-native and agent-mediated
allocation: firms built around automated research and deployment rather
than legacy workflows with models retrofitted on top. Whether such stacks
operate as one ranker or many, a prior reliability question remains: when
should an allocator trust a cross-sectional model after a \emph{silent}
regime shift that volatility controls fail to register? Aleatoric scaling
addresses market noise; it does not detect epistemic invalidity when
ranking structure breaks without a volatility footprint. Building on
\citet{sanderink2026alpha} (\emph{When Alpha Breaks}), a prior walk-forward
DEUP deployment study on a LightGBM US AI-exposure panel that established
per-stock $\hat{e}(x)$, rank-geometry coupling, and a two-level gate at
single-strategy scale, this paper asks a \emph{structural} follow-on:
can per-stock epistemic scores be pooled into a portfolio-wide health
signal, or must strategy-level abstention condition on realised outcomes?
Three research questions organise the evaluation.
\textbf{RQ1 (Separation):} Does per-stock $\hat{e}(x)$ capture
model-specific failure risk distinct from volatility (and can score-only
aggregates detect strategy-level failure)?
\textbf{RQ2 (Decision value):} Does an architecture built on $\hat{e}$
beat volatility-only and ungated baselines on risk-adjusted return and
drawdown under frozen development rules?
\textbf{RQ3 (Robustness):} Do the RQ1--RQ2 answers hold across crisis and
calm regimes, a second base ranker, falsification tests, and a
universe-readiness gate?

\textbf{New contributions (N1--N4).}
We prove a detectability bound (Theorem~\ref{thm:detectability}) in
Le~Cam total-variation form \citep{gama2014,hinder2023}: for fixed
per-prediction error predictors $g(x)$, score-only aggregates
$\{g(x_i)\}$ cannot beat chance under strict concept shift
($\delta_{N}=0$). The bound's covariate-versus-concept dichotomy is
confirmed cross-domain (N2): CIFAR-10-C aggregation succeeds at
$N=10{,}000$ (AUROC $0.955$) but is exactly at chance under relabelled
strict concept shift, while equity cross-sectional aggregates of the same
rule remain near chance on the long ALL panel (bad-day AUROC $\approx 0.55$)
despite valid per-stock $\hat{e}(x)$ (N1). The constructive escape is
$Y$-conditioning: partial-horizon RankIC monitoring with a sequential
CUSUM ($k=5$; AUROC $0.796$), together with a matured-RankIC regime-trust
index whose economic readout is the deployment ablation and multi-crisis
panel, not headline AUROC alone. A falsification battery (HAC/PSR/DSR,
permutation nulls, Fama--French attribution) and a universe-readiness
diagnostic with an explicit \emph{stop} verdict (N3--N4) bound downstream
allocator claims.

\textbf{Consolidated architecture (C1--C2; re-tested).}
Rank-geometry coupling ($\rho(\hat{e},|\text{score}|)=0.616$) makes
inverse-uncertainty sizing counterproductive; a discrete $\hat{e}$-cap and
a strategy gate on realised efficacy are reframed as forced architectural
consequences of the bound rather than optional overlays (RQ2). All evidence
is historical walk-forward simulation on one thematic universe
($N \lesssim 100$) through February~2025; no live performance validation
is reported. The base ranker's FINAL median RankIC failed its
pre-registered threshold. The validated object is the reliability
architecture and its impossibility boundary, not transportable trading
alpha.
\end{abstract}


\medskip
\noindent\textbf{Keywords:} epistemic uncertainty; aleatoric uncertainty;
regime detection; selective prediction; cross-sectional equity ranking;
distribution shift; regime-trust monitoring.

\section{Introduction}
\label{sec:intro}

Cross-sectional rankers deployed in systematic equity strategies face a
reliability problem that conventional risk frameworks were not designed to
solve: when the market regime shifts (through a sector rotation, a thematic
rally, or a structural change in the cross-sectional relationships that
govern stock returns), a model that has performed reliably can become
unreliable in ways that volatility-based controls fail to register. The
standard response, scaling exposure to realised or anticipated volatility,
conflates two structurally distinct sources of uncertainty: aleatoric
uncertainty, the irreducible noise of the data-generating process, and
epistemic uncertainty, the model's own reducible ignorance about novel
conditions \citep{kendall2017,hullermeier2021}. The two carry opposite
operational implications: elevated aleatoric uncertainty motivates exposure
scaling proportional to expected variance, whereas elevated epistemic
uncertainty signals that the ranker is operating outside the support of its
training distribution and motivates abstention. A single conflated estimate
cannot prescribe both correctly.

\paragraph{Motivation and scope.}
Multi-model and agent-mediated allocation stacks face the same reliability
question studied here at single-strategy scale: when to trust an underlying
ranker after a silent regime shift. Standard volatility scaling addresses
aleatoric noise; it does not detect epistemic invalidity when cross-sectional
structure breaks without a volatility footprint. This paper develops that
distinction for one walk-forward GBT ranker and states an impossibility
boundary for score-only health aggregation; it is not a performance study of
autonomous fund architecture.

The setting where this matters most sharply is the gradient-boosted-tree
(GBT) cross-sectional ranker, a leading architecture in empirical asset
pricing \citep{gu2020,leippold2022}. Widespread deployment elevates a risk
that volatility controls were not designed to handle: silent erosion of
model validity, in which the learned ranking structure ceases to
characterise the market while observed volatility remains benign
\citep{allen2025,cui2025}. In the AI-exposed US equity universe studied
here, the base ranker's mean calendar-year RankIC at the 90-day
horizon fell to approximately $-0.07$ in 2021 (the meme-stock rotation) and
to approximately zero in 2024 ($-0.006$; the AI-thematic rotation) while its
development-sample 20-day median RankIC had been a stable $\approx 0.09$
through 2023; ranking efficacy collapsed without a volatility footprint.
Two systematic reviews confirm that applied finance rarely operationalises
the epistemic--aleatoric distinction, and more rarely still tests its
economic consequences \citep{blasco2024,eggen2025}.

This paper develops and stress-tests an uncertainty-aware deployment
framework for a single cross-sectional long--short equity ranker. Relative
to the author's prior deployment study \citep{sanderink2026alpha}, the new
material is theoretical and boundary-testing; the DEUP adaptation,
rank-geometry coupling, and two-level gate are consolidated as background.

\paragraph{Research questions.}
Three questions structure the evaluation; Theorem~\ref{thm:detectability}
supplies the formal spine for the negative part of RQ1.
\begin{description}[
  leftmargin=!,
  labelwidth=!,
  labelsep=0.5em,
  itemsep=0.35em,
  parsep=0pt,
  topsep=0.35em,
  font=\normalfont
]
  \item[\textbf{RQ1 (Separation).}] Does per-stock epistemic
    uncertainty $\hat{e}(x)$ capture model-specific ranking failure
    distinct from volatility---and can cross-sectional aggregates of
    $\{g(x_i)\}$ serve as a strategy-level regime detector?
  \item[\textbf{RQ2 (Decision value).}] Does a two-level deployment
    rule (strategy gate on realised efficacy plus position-level
    $\hat{e}$-cap) improve risk-adjusted outcomes relative to
    volatility-only and ungated baselines under frozen development rules?
  \item[\textbf{RQ3 (Robustness).}] Do the RQ1--RQ2 answers survive
    multi-crisis windows, a second base ranker (RA2), a five-test
    falsification battery, and a universe-readiness diagnostic with an
    explicit \emph{stop} verdict?
\end{description}

\paragraph{Contributions.}
Table~\ref{tab:contribution-map} maps six contributions to the research
questions; items marked \emph{consolidated} were first reported in
\citet{sanderink2026alpha} and are re-evaluated here under the stronger
tests the bound demands.

\begin{table}[!htbp]
  \centering
  \caption{Contribution map. \emph{Consolidated} items extend the prior
    deployment study; \emph{new} items are developed in this paper.}
  \label{tab:contribution-map}
  \small
  \setlength{\tabcolsep}{4pt}
  \begin{tabular}{@{}clp{5.6cm}l@{}}
    \toprule
    &
    \textbf{Contribution} &
    \textbf{Summary} &
    \textbf{Primary RQ} \\
    \midrule
    \multicolumn{4}{@{}l}{\textit{Consolidated from \citet{sanderink2026alpha}}} \\
    \addlinespace[2pt]
    C1 & Rank-geometry coupling &
      $\rho(\hat{e},|\text{score}|)=0.616$; discrete tail-cap not
      inverse sizing &
      RQ1--2 \\
    C2 & Two-level architecture &
      Strategy gate on $H(t)$ plus position-level $\hat{e}$-cap &
      RQ1--2 \\
    \addlinespace[4pt]
    \multicolumn{4}{@{}l}{\textit{New in this paper}} \\
    \addlinespace[2pt]
    N1 & Detectability bound &
      Theorem~\ref{thm:detectability} + corollaries; partial-horizon
      $Y$-escape &
      RQ1 \\
    N2 & CIFAR-10-C boundary &
      $2\times 2$ covariate- vs.\ concept-shift test of aggregation &
      RQ1 \\
    N3 & Falsification battery &
      HAC, PSR/DSR, permutation, Fama--French attribution &
      RQ3 \\
    N4 & Universe-readiness gate &
      Tier ladder with explicit \emph{stop} on allocator claims &
      RQ3 \\
    \bottomrule
  \end{tabular}
\end{table}

\textbf{New in this paper.}
\begin{enumerate}[label=(\roman*),leftmargin=*,itemsep=3pt]
  \item \textbf{A detectability bound (N1).} Theorem~\ref{thm:detectability}
  formalizes the real-versus-virtual drift limit familiar from
  streaming ML \citep{gama2014,hinder2020,hinder2023} in Le~Cam
  form for fixed error predictors $g(x)$ on low-$N$ equity
  cross-sections. \emph{Strict concept shift} means $\delta_{N}=0$
  (joint input-law invariance), not marginal-only invariance ($\eta=0$),
  which can leave dependence-structure movement and residual
  second-moment detectability (Appendix~\ref{app:detectability},
  scope item~1). The novelty is not this $\delta_{N}=0$ impossibility
  itself but four extensions:
  Corollary~\ref{cor:mean-suppressed} (double suppression of the
  mean under correlated low~$N$), a fold-conditional reading for
  walk-forward-refitted~$g$ (Appendix~\ref{app:detectability}),
  the completed $2\times 2$ empirical boundary (N2;
  Section~\ref{sec:cifar_external}), and the constructive
  partial-horizon $Y$-conditioning escape
  (Section~\ref{sec:partial-horizon}).
  \item \textbf{A cross-domain boundary (N2).} The identical aggregation rule
  that is near-chance in the equity cross-section succeeds at
  $N = 10{,}000$ near-i.i.d.\ items on CIFAR-10-C (Spearman
  $\rho = 0.926$, AUROC $0.955$) and is at chance (AUROC exactly $0.500$)
  under relabelled-CIFAR strict concept shift, matching the bound's
  prediction (Section~\ref{sec:cifar_external}).
  \item \textbf{Falsification and universe-readiness (N3--N4).} A frozen-rule
  statistical battery (HAC Sharpe, PSR/DSR, permutation nulls,
  Fama--French attribution) and a breadth diagnostic with explicit
  \emph{stop} verdict bound downstream allocator claims
  (Sections~\ref{sec:res-falsification}
  and~\ref{sec:res-readiness}).
\end{enumerate}

\textbf{Consolidated from \citet{sanderink2026alpha}.}
\begin{enumerate}[label=(\roman*),leftmargin=*,itemsep=3pt,start=4]
  \item \textbf{DEUP for ranking (C1 groundwork).} We adapt Direct Epistemic
  Uncertainty Prediction \citep{lahlou2023} to cross-sectional ranking by
  substituting rank displacement for regression loss; the signal survives
  residualisation, strengthens from development to holdout on $\hat{e}$
  (Table~\ref{tab:rho-canonical}: $0.142 \to 0.192$), and is strictly
  quintile-monotone (Sections~\ref{sec:method-deup}
  and~\ref{sec:res-deup}).
  \item \textbf{Two-level deployment (C1--C2).} The
  rank-geometry coupling $\rho(\hat{e}, |\text{score}|) = 0.616$ and the
  strategy-level gate paired with a position-level $\hat{e}$-cap are
  evaluated against an eight-window multi-crisis panel
  (Sections~\ref{sec:method-twolevel}, \ref{sec:res-gate},
  and~\ref{sec:res-crisis}).
\end{enumerate}

\begin{tcolorbox}[colback=blue!4,colframe=blue!40!black,
  title=\textbf{Practitioner summary: what works}]
\begin{itemize}[leftmargin=1.2em]
  \item Gate the \emph{strategy}, not the positions: a health index built
  from matured realised RankIC discriminates good from bad trading days at
  AUROC $0.721$ / $0.750$ (ALL/FINAL; block-bootstrap 95\% CIs
  $[0.671, 0.775]$ / $[0.624, 0.845]$; $N_{\mathrm{eff}}\approx 111/14$)
  against $0.449$ / $0.504$ for a VIX-percentile rule
  (Table~\ref{tab:regime_auroc})---a persistence measurement, not a
  standalone economic proof; validate via the ablation and crisis panel.
  That discrimination is not Sharpe-free on the frozen policy grid
  (Table~\ref{tab:ablation}): deployed Variant~6 attains
  $1.877$/$0.925$ (ALL/FINAL) versus $2.730$/$1.650$ for ungated raw,
  with crisis MaxDD improving only from $-7.4\%$ to $-6.7\%$.
  \item Binarise the gate: trade when $\Gamma(t) \geq 0.20$, hold zero gross
  otherwise ($47.2\%$ abstention at $80.0\%$ bad-day precision); continuous
  throttling destroys recovery convexity.
  \item Cap, do not size: reduce weight $\times 0.70$ only on positions with
  $\hat{e}$ above the 85th cross-sectional percentile
  (Table~\ref{tab:monitoring}).
  \item Platt-calibrate the health index into $P(\text{good day})$ before
  feeding a sizing rule (reliability error $0.073 \to 0.002$).
  \item Normalise conformal scores by $\hat{e}$: conditional-coverage spread
  falls from $20.2\%$ to $0.81\%$ (oracle), $4.44\%$ deployable
  (Section~\ref{sec:res-conformal}).
  \item Only outcome-conditioned ($Y$-measurable) channels detect
  concept-shift breaks on taxonomy-defined events: a partial-horizon RankIC
  channel fires a median $25$ trading days earlier on those events than the
  slow efficacy EWMA (Section~\ref{sec:partial-horizon}; labeling excludes the
  partial-horizon signal).
\end{itemize}
\end{tcolorbox}

\begin{tcolorbox}[colback=red!4,colframe=red!50!black,
  title=\textbf{What does not work: seven pre-registered or robust nulls}]
\begin{enumerate}[leftmargin=1.4em]
  \item \emph{Continuous inverse-$\hat{e}$ sizing}: the rank-geometry
  coupling $\rho(\hat{e},|\text{score}|) = 0.616$ mechanically de-levers the
  score tails that generate the long--short spread.
  \item \emph{Cross-sectional aggregates of $g(x)$ as regime detectors}:
  bad-day AUROC $\approx 0.55$ on ALL, near chance; a short FINAL lift under
  concurrent $\delta_N > 0$ is not a deployable substitute for $H(t)$
  (Table~\ref{tab:ood_finance}; Theorem~\ref{thm:detectability}).
  \item \emph{VIX-percentile gating}: false alarms on three profitable
  high-VIX windows in the multi-crisis panel (2019, H1~2023, and H2~2023
  rate-hiking; Table~\ref{tab:multi_crisis_windows}).
  \item \emph{HMM regime features in $g(x)$}: $\Delta\rho = -0.00009$
  (augment) and $-0.00063$ (replace), both below the pre-registered
  $|\Delta\rho| > 0.005$ materiality threshold.
  \item \emph{Continuous $\Gamma(t)$ exposure throttling}: FINAL-holdout
  Sharpe $-0.34$ versus $+0.925$ for the binary-gate variant.
  \item \emph{Conformal-width economic overlays}: null across all eight
  pre-registered levers (Section~\ref{sec:res-conformal}); the strongest
  candidate delivers out-of-sample
  $\Delta\text{Sharpe} = +0.013$ (95\% CI $[-0.19, +0.23]$).
  \item \emph{Literature UA-Sort and Trail-IC sizing}: match Gate~$+$~Raw on
  FINAL Sharpe ($-0.322$); kill criterion K4 triggered
  (Table~\ref{tab:comparators}; Appendix~\ref{app:experiment-registry}).
\end{enumerate}
\end{tcolorbox}

\paragraph{Relation to prior work.}
\citet{sanderink2026alpha} reported the consolidated architecture (C1--C2
in Table~\ref{tab:contribution-map}); this paper adds N1--N4 and re-tests
the full stack under Theorem~\ref{thm:detectability}. For positioning
relative to streaming drift limits
\citep{gama2014,hinder2023}, see Section~\ref{subsec:concept-drift} and
Appendix~\ref{app:detectability}.

\paragraph{Scope.}
The empirical perimeter is deliberately narrow: a single curated universe of
at most 100 AI-exposed US equities (February 2016 -- February 2025), one
base-ranker family, and a primary 20-day horizon. All reported results are
historical walk-forward simulation on matured labels through that freeze;
\textbf{no live trading P\&L or paper-trading alpha validation} is reported.
A preliminary post-freeze monitoring audit trail exists in the private
research archive; Section~\ref{sec:res-monitoring-corr} summarises whether
\emph{monitoring alerts} fired as designed (illustrative; not statistical
evidence for gate AUROC). Portfolio metrics in that audit are too sparse
and execution-noisy for inferential use. Two headline caveats are stated here rather than buried:
the base ranker's FINAL-holdout median RankIC of $+0.017$
\textbf{failed} its pre-registered $\geq 0.04$ acceptance threshold
(Table~\ref{tab:signal_quality_531}), and the FINAL partition spans only 14
months. The framework, not the trading alpha, is the validated object.

\paragraph{Roadmap.}
Section~\ref{sec:related} positions the work; Section~\ref{sec:method}
develops the method, the theorem, and the deployment architecture;
Section~\ref{sec:data} documents data and protocol; Section~\ref{sec:results}
reports results. \textbf{RQ mapping:}
Sections~\ref{sec:res-deup} and~\ref{sec:agg_gap} answer RQ1 (stock-level
separation and the aggregation ceiling);
Sections~\ref{sec:res-gate} (ablation) answer RQ2;
Sections~\ref{sec:res-crisis},~\ref{sec:res-falsification},
and~\ref{sec:res-readiness} answer RQ3;
Section~\ref{sec:cifar_external} tests the bound cross-domain;
Section~\ref{sec:partial-horizon} reports the constructive $Y$-channel
escape (RQ1).
Section~\ref{sec:nulls-limits} consolidates nulls and limitations before
Section~\ref{sec:conclusion} states scoped verdicts on RQ1--RQ3.
Appendix~\ref{app:experiment-registry} maps each falsification test to its
verdict and location in the text.

\section{Related Work}
\label{sec:related}

\subsection{Uncertainty Quantification and Epistemic--Aleatoric Decomposition}

\citet{hullermeier2021} provide the canonical conceptual map: aleatoric
uncertainty is irreducible noise in the data-generating process, epistemic
uncertainty is reducible model ignorance, and terminological slippage across
subfields assigns incompatible mathematical objects to the same labels. The
Bayesian lineage, evidence frameworks \citep{mackay1992}, full MCMC
posteriors \citep{neal1996}, MC~Dropout \citep{gal2016}, deep ensembles
\citep{lakshminarayanan2017,fort2019,gustafsson2020}, heteroscedastic-head
decompositions \citep{kendall2017}, and latent-variable BNN decompositions
\citep{depeweg2018}, presupposes differentiable, dropout- or
ensemble-capable architectures and is therefore inapplicable to the
gradient-boosted trees that dominate tabular asset pricing
\citep{gu2020}; \citet{lefolgoc2021} additionally contest MC~Dropout's
Bayesian interpretation. \citet{kotelevskii2025a} give the most general
modern formalisation, decomposing pointwise risk into a Bayes (aleatoric)
term and a Bregman-divergence excess-risk (epistemic) term under any strictly
proper scoring rule; DEUP \citep{lahlou2023} is the finite-sample plug-in
estimator of that excess risk: a secondary error predictor $g(x)$ trained on
held-out residuals, minus an aleatoric floor $a(x)$. DEUP is model-agnostic
and requires no posterior inference, which is what makes it deployable on a
GBT ranker. Two caution strands temper the programme:
\citet{mucsanyi2024} find that clean disentanglement is rarely achieved in
practice, especially under covariate shift, and \citet{bickford2025} argue
the dichotomy should be judged by decision utility rather than theoretical
purity, the operational stance adopted here. \citet{kotelevskii2025b}
respond with vector-valued fusion of complementary UQ measures.
Finance-specific evidence sharpens the aleatoric-floor requirement:
evidential regression with a Normal--Inverse--Gamma prior miscalibrates
systematically under volatility clustering \citep{wong2025}.

GBT-native probabilistic alternatives exist; PGBM's leaf-weight predictive
variance \citep{sprangers2021}, stochastic-ensemble disagreement
\citep{ustimenko2021}, epistemic-aware conformalized quantile regression
\citep{rossellini2024}, and leaf-proximity diagnostics
\citep{sudjianto2026}, but none delivers the formal Bregman excess-risk
decomposition: they produce total variance, disagreement proxies, or
interval-level adjustments (comparison in Table~\ref{tab:uq-methods},
Appendix~\ref{app:extra-floats}). A targeted literature search (2023--2026)
located no published work applying DEUP or direct excess-risk estimation to
cross-sectional stock ranking; the closest finance contribution,
uncertainty-adjusted sorting over rolling residual quantiles
\citep{liu2026}, neither decomposes epistemic from aleatoric uncertainty nor
addresses the ranking-specific $\hat{e}$--score coupling.

\subsection{Regime Detection and Selective Prediction}

Market-side regime detection is mature: Markov-switching models
\citep{hamilton1989}, HMM-gated factor investing \citep{wanglin2020},
hierarchical bull/bear HMMs \citep{oelschlager2023}, jump-model regime
forecasts feeding allocation \citep{shu2024}, deep generative regime models
\citep{nie2026}, Bayesian online change-point detection
\citep{adams2007,habibi2021,huang2025,tsaknaki2025}, factor-premium break
detection \citep{cui2025}, conformal-coupled change points \citep{sun2025},
and similarity-based regime classifiers \citep{mulliner2026}. The recurring
gap is that all condition on market-level observables, returns, volatility,
correlations, and are therefore agnostic to whether any \emph{specific}
downstream model remains valid: a thematic rotation can break a ranker's
feature--label mapping while the VIX stays quiet. \citet{nystrup2017} add
the deployment caution that over-frequent regime switching destroys recovery
convexity, motivating discrete rather than smooth gating;
\citet{pun2023} demonstrate regime-aware allocation value in a
reinforcement-learning setting, again from market observables.

Selective prediction supplies the abstention formalism: optimal
reject-option trade-offs \citep{chow1970,elyaniv2010}, learned selection
functions for deep networks \citep{geifman2017}, dominance of hybrid
abstain-plus-weight rules \citep{franc2021}, joint calibration of aleatoric
and epistemic estimates \citep{chaudhuri2023}, and a five-source
decomposition of the selective-classification gap showing calibration alone
cannot fix ranking errors \citep{rabanser2025}. Existing frameworks operate
at a single granularity (per input), whereas a trading system decides at
(at least) two: which positions to hold, and whether to trade at all today.
Batch-level confidence aggregation \citep{garg2022} and shift-calibration
evidence \citep{ovadia2019} suggest per-item scores can lift to group-level
accuracy estimates in high-$N$, near-i.i.d.\ settings; whether that transfer
survives low-$N$, factor-correlated equity cross-sections is exactly the
question Theorem~\ref{thm:detectability} answers in the negative for
concept shift.

\subsection{Concept Drift and Detectability Limits}
\label{subsec:concept-drift}

The streaming-ML literature distinguishes \emph{real} (or
\emph{concept}) drift---changes in $P(Y \mid X)$---from
\emph{virtual} drift---changes in $P(X)$ with $P(Y \mid X)$
fixed \citep{gama2014}. Real drift is not visible from unlabeled
input marginals alone; virtual drift is, and the canonical survey
therefore separates posterior change (typically requiring outcome
feedback) from drift readable off $P(X)$
\citep{gama2014}. Recent formal work sharpens the limit:
unsupervised drift detection reduces to independence testing
between time and data, for which universal detectors exist under
broad conditions \citep{hinder2020}; supervised drift detection
from model loss alone has no universal detector and can mismatch
real drift systematically \citep{hinder2023}. Theorem~\ref{thm:detectability}
states a closely related impossibility in Le~Cam
total-variation form for fixed cross-sectional error predictors
$g(x)$: under strict concept shift ($\delta_{N}=0$) no score-only
aggregate of $\{g(x_i)\}$ beats chance. The theorem does
\emph{not} claim novelty for that real-versus-virtual separation; the
finance-specific extensions are Corollary~\ref{cor:mean-suppressed}
(correlated low-$N$ double suppression), a fold-conditional reading
for walk-forward-refitted~$g$ (Appendix~\ref{app:detectability}),
the completed $2\times 2$ empirical boundary
(Section~\ref{sec:cifar_external}), and the constructive
partial-horizon $Y$-conditioning escape
(Section~\ref{sec:partial-horizon}).

\subsection{Shift-Robust Conformal Methods and Backtest Inference}

Conformal prediction provides distribution-free coverage
\citep{vovk2005,angelopoulos2023}, extended beyond exchangeability with
shift-weighted guarantees \citep{barber2023}, adaptive online quantile
tracking \citep{gibbs2021,gibbs2024}, multi-horizon time-series variants
\citep{stankeviciute2021,zhang2025}, and epistemic-aware scores
\citep{cabezas2025,cabezas2026,plassier2025}. Coverage degradation under
temporal shift is quantifiable \citep{siahkali2026} and empirically severe
\citep{lee2026}; split conformal on LightGBM is demonstrated in insurance
pricing \citep{manna2025}. Standard conformal outputs remain
undifferentiated intervals rather than decomposed epistemic--aleatoric
pairs, which is why this paper uses $\hat{e}(x)$ as the conformal
normaliser rather than as a replacement for conformal calibration. On the
inference side, the evaluation protocol draws on the Probabilistic and
Deflated Sharpe Ratios \citep{bailey2012,bailey2014}, purged walk-forward
validation \citep{lopezdeprado2018}, Sharpe sampling theory under
non-Normal, serially correlated returns \citep{lopezdeprado2025},
complexity-driven in-sample Sharpe inflation \citep{jacquier2025}, and the
CPCV comparison of \citet{arian2024}; CPCV is statistically stronger but
incompatible with the temporal causality the health index requires, so
strict expanding walk-forward is retained. \citet{barroso2021} establish
that learning from out-of-sample forecast errors carries economic value at
the portfolio level; DEUP can be read as its per-prediction extension.

\subsection{Uncertainty-Aware Exposure Control and Finance Applications}

The exposure-control lineage runs from Kelly sizing \citep{kelly1956} and
its estimation-risk extension \citep{sukhov2026}, through volatility
targeting \citep{harvey2018}, downside-volatility scaling \citep{wang2021},
and risk parity \citep{qian2011}, all of which condition on a single
variance estimate and therefore treat all uncertainty as aleatoric. Model
uncertainty enters with multi-prior ambiguity shrinkage \citep{garlappi2007}
and the smooth-ambiguity framework \citep{klibanoff2005}, operationalised by
BNN-posterior sizing in futures \citep{spears2020}, non-participation
regions for uncertain assets \citep{liao2025}, uncertainty-adjusted sorting
\citep{liu2026}, and subsampling-based model-uncertainty overlays
\citep{buehler2025}. Notably, the published rules converge on
\emph{discrete} exposure reduction rather than smooth inverse-uncertainty
scaling, and hedge-fund due-diligence evidence documents layered controls
operating at distinct organisational levels \citep{cassar2017}. None of
these settings, however, exhibits the structural constraint specific to
ranking: in cross-sectional ranking, extreme scores occupy the rank
positions with the largest permitted displacement, so uncertainty and signal
strength are coupled by construction---a coupling that defeats
inverse-uncertainty sizing. Finance-facing conformal applications are
emerging \citep{schmitt2026,fermanian2025,kaya2025}, and GBT rankers remain
the tabular workhorse \citep{gu2020,leippold2022}.

\subsection{The Gap}

No surveyed method combines GBT compatibility, a formal Bregman excess-risk
decomposition, and cross-sectional ranking validation
(Table~\ref{tab:uq-methods}, Appendix~\ref{app:extra-floats}); only DEUP
approximates the combination, and its only equity-ranking validation is the
author's prior preprint \citep{sanderink2026alpha}. The resulting three-part
gap comprises (i)~ranking-geometry coupling between epistemic uncertainty and
signal strength, (ii)~the failure of per-prediction uncertainty to aggregate
into strategy-level regime signals, and (iii)~model-specific rather than
market-generic regime monitoring. The concept-drift literature
\citep{gama2014,hinder2020,hinder2023} formalises real versus virtual drift
but not correlated low-$N$ cross-sectional rank aggregation or the finance
deployment consequences developed here. These gaps structure the method that follows:
(i) motivates the discrete cap, (ii) is formalised as
Theorem~\ref{thm:detectability} and forces the two-level separation, and
(iii) is answered by a realised-efficacy health index rather than another
market-observable detector.

\section{Method}
\label{sec:method}

\subsection{Notation and Setup}
\label{sec:method-notation}

At each trading date $t$ a universe of $N_t$ stocks is scored by a primary
ranker; portfolio construction consumes only the within-date ordering of the
scores. The prediction target is the benchmark-relative total excess return
of stock~$i$ over horizon $\tau \in \{20, 60, 90\}$ trading days,
\begin{equation}
  r_{i,t}^{(\tau)}
  \;=\; R_{i}(t,\,t+\tau) \;-\; R_{\mathrm{QQQ}}(t,\,t+\tau),
  \label{eq:label}
\end{equation}
where $R_{i}$ is the split-adjusted total return of stock~$i$ (dividends at
ex-date) and $R_{\mathrm{QQQ}}$ the corresponding return of the Invesco QQQ
Trust, a liquid proxy for the AI-sector opportunity set; the
benchmark-adjusted label removes market beta and concentrates the objective
on the cross-sectional component a market-neutral long--short book requires.
Table~\ref{tab:ch3-notation} collects the notation used throughout.

\begin{table}[ht]
  \centering
  \caption{Notation. Equation numbers refer to the definitions in
    Sections~\ref{sec:method-notation}--\ref{sec:method-eval}.}
  \label{tab:ch3-notation}
  \small
  \begin{tabular}{@{}lp{0.72\linewidth}@{}}
    \toprule
    \textbf{Symbol} & \textbf{Definition} \\
    \midrule
    $r^{(\tau)}_{i,t}$ &
      Benchmark-relative excess return of stock $i$ over horizon
      $\tau$ from date $t$ (Eq.~\ref{eq:label}). \\
    $s_{i,t}$ &
      Primary ranker's predicted score for stock $i$ at date $t$. \\
    $\operatorname{rank}\%(\cdot)$ &
      Cross-sectional percentile rank within the universe at date $t$. \\
    $\ell(x_{i,t}; \tau)$ &
      Per-prediction rank-displacement loss
      (Eq.~\ref{eq:rank-displacement}). \\
    $R(f, x)$ &
      Pointwise risk of predictor $f$ at input $x$
      (Eq.~\ref{eq:pointwise-risk}). \\
    $R^{*}(x)$ &
      Bayes-optimal pointwise risk; aleatoric floor
      (Eq.~\ref{eq:bayes-risk}). \\
    $g(x)$ &
      DEUP error predictor: secondary LightGBM model trained
      walk-forward on held-out rank displacements. \\
    $a(x)$ &
      Aleatoric baseline: PIT-safe rolling trailing percentile of matured
      rank-displacement losses. \\
    $\hat{e}(x)$ &
      Estimated epistemic signal, $\max(0,\, g(x) - a(x))$
      (Eq.~\ref{eq:epistemic-signal}). \\
    $H_{\text{real}}(t)$ &
      Trailing realised model efficacy: EWMA of matured RankIC with a
      structural $\tau$-day lag. \\
    $H_{\text{drift}}(t)$ &
      Feature- and score-distribution drift component of the health index. \\
    $H_{\text{disagree}}(t)$ &
      Cross-model disagreement component of the health index. \\
    $H(t)$ &
      Composite continuous health score,
      $H(t)=h\bigl(H_{\text{real}},H_{\text{drift}},H_{\text{disagree}}\bigr)
      \in[0,1]$ (Eq.~\ref{eq:gate}). \\
    $\Gamma(t)$ &
      Continuous regime-trust gate score,
      $\Gamma(t)=\operatorname{clip}\bigl((H(t)-0.3)/(0.7-0.3),\,0,\,1\bigr)$
      (Eq.~\ref{eq:gamma-gate}). \\
    $G(t)$ &
      Binarised deployment gate,
      $G(t)=\mathbf{1}[\Gamma(t)\geq 0.2]\in\{0,1\}$. \\
    $\operatorname{RankIC}(t;\tau)$ &
      Cross-sectional Spearman rank information coefficient at date $t$
      (Eq.~\ref{eq:rankic}). \\
    $\rho_{S}$ & Spearman rank correlation. \\
    $\mathrm{SR}$ & Annualised Sharpe ratio (Eq.~\ref{eq:sharpe}). \\
    $\mathrm{PSR}$ &
      Probabilistic Sharpe ratio (Eq.~\ref{eq:psr};
      \citealp{bailey2012}). \\
    \bottomrule
  \end{tabular}
\end{table}

\subsection{Base Ranker}
\label{sec:method-ranker}

The base ranker is a LightGBM gradient-boosted-tree model. Three arguments
select this backbone: tree ensembles sit at the accuracy frontier for
tabular cross-sectional equity prediction \citep{gu2020,leippold2022}; the
prominent neural UQ stacks (MC~Dropout, Bayesian networks, deep ensembles)
are architecturally incompatible with tree leaf structures; and deep
ensembles additionally incur prohibitive retraining cost under walk-forward
re-estimation (five members over 109 folds would require 545 full fits per
evaluation window). The ranking task, ordering $N_t \approx 84$ assets on
contemporaneous tabular features, also suits GBT leaf partitioning rather
than autoregressive sequence architectures.

The model is trained under an expanding walk-forward protocol spanning
February 2016 to February 2025, producing 109 monthly evaluation folds. At
each fold boundary the training set expands by one calendar month and the
model is re-fitted before scoring the next month out-of-sample; a 90-day
embargo separates the last training observation from the first evaluation
observation, purging overlapping forward-return labels at all three
horizons. Walk-forward is retained over the statistically stronger CPCV
\citep{arian2024} because the health index of
Section~\ref{sec:method-gate} requires strict temporal causality that
combinatorial block recombination would break. The 20-day horizon is
primary: in the development (DEV) period the ranker achieved median RankIC
$0.091$ (a $225\%$ lift over the strongest single-factor baseline, 12-month
momentum at $0.028$), and while the 60- and 90-day horizons showed higher
DEV RankIC ($0.167$, $0.206$), both inverted in the FINAL holdout ($-0.044$,
$-0.052$); the 20-day signal alone retained weak positive directionality
($+0.017$). Features span eight categories (momentum, volatility/liquidity,
drawdown/relative strength, regime/macro, fundamentals, earnings events,
missingness masks, hygiene diagnostics; Table~\ref{tab:feature-categories},
Appendix~\ref{app:extra-floats}), each mapped to its within-date percentile
rank before entering the model. Hyperparameters are deliberately
conservative and fixed across all experiments: base ranker
\texttt{n\_estimators}$\,=100$, \texttt{learning\_rate}$\,=0.05$; error
predictor \texttt{max\_depth}$\,=3$, \texttt{num\_leaves}$\,=8$,
\texttt{n\_estimators}$\,=50$, \texttt{min\_child\_samples}$\,=50$,
\texttt{learning\_rate}$\,=0.05$, row/column subsampling $0.8$,
\texttt{random\_state}$\,=42$. An over-deep base model would compress
residuals toward zero and starve the error predictor of signal.

\subsection{DEUP for Cross-Sectional Ranking}
\label{sec:method-deup}

Following \citet{kotelevskii2025a} and \citet{lahlou2023}, the pointwise
risk of predictor $f$ under loss $\ell$ at input $x$ is
\begin{equation}
  R(f, x) \;=\;
  \mathbb{E}_{P(Y\mid X=x)}\!\bigl[\ell\bigl(Y, f(x)\bigr)\bigr],
  \label{eq:pointwise-risk}
\end{equation}
which decomposes into the Bayes-optimal aleatoric floor
\begin{equation}
  R^{*}(x) \;=\; \inf_{g}\,
  \mathbb{E}_{P(Y\mid X=x)}\!\bigl[\ell\bigl(Y, g(x)\bigr)\bigr]
  \label{eq:bayes-risk}
\end{equation}
and the excess risk $R(f,x)-R^{*}(x)$, the reducible epistemic component.
Unlike posterior variance, excess risk captures model
misspecification, cases where the function class is wrong for the current
input, not merely parameter uncertainty. DEUP estimates it by training a
secondary model on realised held-out losses, with no ensemble or posterior
machinery, which is what makes it GBT-deployable.

DEUP was developed for regression and classification; ranking requires an
adaptation at the loss level. Following \citet{sanderink2026alpha}, the
per-prediction loss is \emph{rank displacement},
\begin{equation}
  \ell\bigl(x_{i,t}; \tau\bigr) \;=\;
     \Bigl| \operatorname{rank}\%\!\bigl(r^{(\tau)}_{i,t}\bigr) -
            \operatorname{rank}\%\!\bigl(s_{i,t}\bigr) \Bigr| \in [0,1],
  \label{eq:rank-displacement}
\end{equation}
the absolute within-date percentile-rank error. Two properties justify the
choice: rank displacement is approximately stock-volatility-invariant (a
stock can occupy any rank percentile regardless of its idiosyncratic
volatility), structurally decoupling the loss target from the volatility
channel that contaminates return-space error; and cross-sectional ordering
is the quantity that determines long--short portfolio construction.

The error predictor $g(x)$ is a second LightGBM regressor trained
walk-forward on held-out rank displacements: for each fold
$k \geq k_0 = 20$, $g$ is fitted on displacement observations from folds
$1,\dots,k-1$ and applied to fold $k$; no fold's predictions condition on
its own residuals. Its fixed fourteen-dimensional feature set comprises the
ranker's \texttt{score} and \texttt{abs\_score}, the within-date
\texttt{cross\_sectional\_rank}, per-stock context (\texttt{vol\_20d},
\texttt{vol\_60d}, \texttt{mom\_1m}, \texttt{adv\_20d}), market state
(\texttt{vix\_percentile\_252d}, \texttt{market\_regime\_enc},
\texttt{market\_vol\_21d}, \texttt{market\_return\_21d}), and three
macro-stress diagnostics (\texttt{credit\_ratio}, \texttt{breadth\_ratio},
\texttt{downside\_rv\_share}). The aleatoric baseline $a(x)$ is a PIT-safe
rolling trailing percentile of \emph{matured} historical rank-displacement
losses, computed only from observations whose forward windows closed
strictly before the as-of date, selected by tiered elimination over
candidate proxies. The epistemic signal is the rectified gap
\begin{equation}
  \hat{e}\bigl(x_{i,t}\bigr) \;=\;
    \max\!\Bigl(0,\; g\bigl(x_{i,t}\bigr) - a\bigl(x_{i,t}\bigr)\Bigr).
  \label{eq:epistemic-signal}
\end{equation}
Because $a(t)$ enters as a date-level constant, cross-sectional ranks of
$\hat{e}$ are preserved: the per-date Spearman correlation between the
PIT-safe and oracle $\hat{e}$ is unity across the sample, so the cap policy
below is deployment-equivalent to its oracle counterpart.

Validation (Section~\ref{sec:res-deup}) also revealed a constraint not
predicted \emph{a priori}: the median per-date Spearman correlation between
$\hat{e}$ and absolute score is
\[
  \rho\bigl(\hat{e},\,|\text{score}|\bigr) \;=\; 0.616
\]
across $1{,}865$ trading dates. Extreme-rank stocks---the model's strongest
directional signals---are simultaneously those at which $\hat{e}$ is
largest, because the geometry of rank displacement permits more displacement
at the boundaries of $[0,1]$ than at the middle. This coupling rules out the
canonical deployment of an epistemic signal: continuous inverse-uncertainty
sizing systematically de-levers precisely the predictions on which the model
is most confident. It is the principal empirical motivation for the discrete
two-level architecture of Section~\ref{sec:method-twolevel}.

The second structural constraint is formal. The following bound
restates, for DEUP-style fixed error predictors on finance
cross-sections, a detectability limit analogous to real versus
virtual drift in the streaming-ML literature
\citep{gama2014,hinder2023} (Section~\ref{subsec:concept-drift}).
It explains why no cross-sectional aggregate of $\{g(x_i)\}$
(mean, quantile, spread, or otherwise) can serve as a regime
detector under concept shift.

\paragraph{Setup and assumptions.}
Let a latent regime $c \in \mathcal{C}$ govern a joint law $P_{c}$ on
$(X, Y)$. Fix a predictor $f$ and a per-prediction loss
$\ell\bigl(f(X), Y\bigr) \in [0, B]$, and let the training law be the
regime mixture $P_{\mathrm{tr}} = \int_{\mathcal{C}} P_{c}\, d\pi(c)$.
Three assumptions govern Theorem~\ref{thm:detectability} and its
corollaries (Table~\ref{tab:assumptions-by-result} records which proof uses
which):
\begin{enumerate}[leftmargin=*,itemsep=2pt]
  \item \textbf{Boundedness.} $\ell,\, g \in [0, B]$.
  \item \textbf{Regime-blind estimator.} $g$ is a fixed measurable function
        of $x$ that does not take $c$ as an argument. By construction the
        DEUP error model targets the mixture-conditional risk and is
        \emph{regime-blind}:
        \begin{equation}
          g(x) \;=\;
          \mathbb{E}_{\mathrm{tr}}\!\bigl[\ell\bigl(f(X), Y\bigr) \,\big|\, X = x\bigr]
          \;=\;
          \int_{\mathcal{C}}
            \mathbb{E}\!\bigl[\ell\bigl(f(X), Y\bigr) \,\big|\, X = x,\, c\bigr]\,
            d\pi(c \mid x).
          \label{eq:g-marginalisation}
        \end{equation}
        In walk-forward deployment the error model is refitted each fold;
        the bound is read conditional on the fitted map $g_{\mathcal{T}}$
        (Appendix~\ref{app:detectability}).
  \item \textbf{Within-regime dependence.} For corollaries only: the average
        pairwise score correlation within regime $c$ is
        \[
          \bar{\rho}_{g}^{(N)}(c)=\tfrac{1}{N(N-1)}\sum_{i\neq j}
          \operatorname{Corr}\!\bigl(g(X_i),g(X_j)\mid c\bigr).
        \]
\end{enumerate}

\begin{table}[ht]
\centering
\small
\caption{Assumptions actually used in each proof.}
\label{tab:assumptions-by-result}
\begin{tabular}{ll}
\toprule
\textbf{Result} & \textbf{Assumptions actually used} \\
\midrule
Theorem~\ref{thm:detectability} & $g$ a fixed $X$-measurable map;
  equal-prior two-regime test \\
Corollary~\ref{cor:concept-shift} & $\delta_N=0$ (joint input invariance) \\
Corollary~\ref{cor:mean-suppressed} & bounded $g\in[0,B]$; conditional
  exchangeability (common $\sigma_g^2(c)$) \\
Corollary~\ref{cor:high-n} & $\eta>0$; regime calibration; bounded
  $\sigma_g^2$; $\bar{\rho}_g^{(N)}\ge 0$ (rate needs
  $\bar{\rho}_g^{(N)}=O(N^{-1})$; variance via
  Cor.~\ref{cor:mean-suppressed}) \\
Corollary~\ref{cor:deployment} & $\delta_N=0$; fixed predictors; stated
  $X$-vs-$Y$ information set \\
\bottomrule
\end{tabular}
\end{table}

\begin{theorem}[Input-law bound for score-only regime detection]
\label{thm:detectability}
Under Assumptions~1--2, at evaluation a cross-section of $N$ inputs
$\mathbf{X} = (X_{1}, \dots, X_{N})$ is drawn under a \emph{single} regime
$c$ (the $X_{i}$ need not be independent). A \emph{score-only detector} is
any measurable map of $g(\mathbf{X}) = (g(X_{1}), \dots, g(X_{N}))$ alone.
Write the total-variation movement of the input law between two regimes
$c_{0}, c_{1}$ at the joint and single-coordinate-marginal levels as
\begin{equation}
  \delta_{N}
    \;=\; d_{\mathrm{TV}}\!\bigl(\mathcal{L}_{c_{0}}(\mathbf{X}),\,
                                 \mathcal{L}_{c_{1}}(\mathbf{X})\bigr),
  \qquad
  \eta
    \;=\; \sup_{c,\, c',\, i}\,
          d_{\mathrm{TV}}\!\bigl(P_{c}^{X_i},\, P_{c'}^{X_i}\bigr).
  \label{eq:tv-movement}
\end{equation}
When coordinates share a common marginal $P_{c}^{X}$, the supremum over
$i$ is redundant. \emph{Concept shift} is marginal invariance per
coordinate ($\eta = 0$; so $P(c \mid X{=}x) \approx P(c)$ by Bayes); its
\emph{strict} form additionally has $\delta_{N} = 0$ (joint input
invariance). For an equal-prior two-point test of $c_{0}$ against $c_{1}$
from $g(\mathbf{X})$ alone, the best achievable balanced accuracy over all
score-only detectors $\phi$ satisfies
\begin{equation}
  \sup_{\phi}\ \mathrm{BAcc}(\phi)
  \;=\;
  \tfrac{1}{2}\Bigl(1 + d_{\mathrm{TV}}\!\bigl(
      \mathcal{L}_{c_{0}}(g(\mathbf{X})),\,
      \mathcal{L}_{c_{1}}(g(\mathbf{X}))\bigr)\Bigr)
  \;\le\;
  \tfrac{1}{2}\bigl(1 + \delta_{N}\bigr).
  \label{eq:detectability-bound}
\end{equation}
\end{theorem}

\begin{remark}[Sequential monitors]\label{rem:sequential}
Any $X$-measurable functional of an input path---including a CUSUM of
cross-sectional aggregates---is bounded by the total variation between path
laws, so multi-date monitors inherit Theorem~\ref{thm:detectability} with
$\delta_{N}$ measuring movement of the full path law.
\end{remark}

\begin{proof}[Proof sketch]
Equation~\eqref{eq:detectability-bound} combines two classical facts. The
equality is Le~Cam's two-point lemma: the Bayes error of an equal-prior
binary test between laws $P_{0}, P_{1}$ is
$\tfrac{1}{2}\bigl(1 - d_{\mathrm{TV}}(P_{0}, P_{1})\bigr)$, and balanced
accuracy is one minus this error. The inequality is the data-processing
inequality for total variation: because $\mathbf{X} \mapsto g(\mathbf{X})$
is a fixed measurable map (Eq.~\ref{eq:g-marginalisation}), the pushforward
laws are no further apart than the input laws. The full proof and the four
corollaries are proved in Appendix~\ref{app:detectability}.
\end{proof}

\begin{corollary}[Concept-shift impossibility]\label{cor:concept-shift}
If the shift is strict concept shift ($\delta_{N} = 0$), then
$\mathcal{L}_{c_{0}}(\mathbf{X}) = \mathcal{L}_{c_{1}}(\mathbf{X})$ and the
pushforward laws coincide for \emph{every} statistic
$T\bigl(g(\mathbf{X})\bigr)$. Consequently the regime ROC curve is the
diagonal and $\mathrm{AUROC}(T) = \tfrac{1}{2}$ exactly, for the mean, the
median, any quantile, the inter-quartile range, the sample variance, or any
other function of $\{g(x_{i})\}$, at \emph{every} cross-section size $N$. No
score-only detector beats chance.
\end{corollary}

\begin{corollary}[Double suppression of the cross-sectional mean]
\label{cor:mean-suppressed}
Suppose the scores $g(X_i)$ are conditionally exchangeable within a regime,
with common regime-conditional variance $\sigma_g^2(c)$ and average pairwise
correlation
$\bar{\rho}_g^{(N)}(c)=\frac{1}{N(N-1)}\sum_{i\neq j}
\operatorname{Corr}\!\bigl(g(X_i),g(X_j)\mid c\bigr)$.
Write $\bar{g}_N=\frac{1}{N}\sum_i g(X_i)$. Then the regime signal in the
mean is bounded by the marginal input movement,
\[
\bigl|\,\mathbb{E}[\bar{g}_N\mid c_0]-\mathbb{E}[\bar{g}_N\mid c_1]\,\bigr|
\;\le\;B\,\eta ,
\]
and its conditional variance is
\[
\operatorname{Var}(\bar{g}_N\mid c)=\sigma_g^2(c)\!\left[\tfrac{1}{N}
+\Bigl(1-\tfrac{1}{N}\Bigr)\bar{\rho}_g^{(N)}(c)\right].
\]
If $\bar{\rho}_g^{(N)}(c)\to\rho_\infty(c)>0$ as $N\to\infty$, the variance
does not vanish and the large-$N$ signal-to-noise ratio is bounded by
$B\,\eta\big/\bigl(\sigma_g(c)\sqrt{\rho_\infty(c)}\bigr)$, independent of
$N$. Under marginal concept shift ($\eta=0$) the signal is zero, so the mean
carries no population regime information; under the finance conditions
($\eta$ small, $\rho_\infty(c)>0$) the mean is suppressed on both axes
simultaneously, negligible signal over a non-vanishing noise floor that
does not improve with universe size.
\end{corollary}

\begin{corollary}[Concentration under covariate-visible shift]
\label{cor:high-n}
Suppose the shift is covariate-visible ($\eta>0$), the error predictor is
regime-calibrated up to a regime-independent constant,
$\mathbb{E}[g(X)\mid c]=R(c)+b$ with $R(c)=\mathbb{E}[\ell\mid c]$, the
conditional variances $\sigma_g^2(c)$ are uniformly bounded, and
$\bar{\rho}_g^{(N)}(c)\ge 0$. Then
\[
\bar{g}_N-b-R(c)\;=\;O_{p}\!\left(\sqrt{\tfrac{1}{N}
+\bar{\rho}_g^{(N)}(c)}\right).
\]
In particular, $\bar{g}_N-b$ consistently estimates the regime risk $R(c)$
whenever $\bar{\rho}_g^{(N)}(c)\to 0$; the parametric rate
$O_{p}(N^{-1/2})$ holds under the stronger condition
$\bar{\rho}_g^{(N)}(c)=O(N^{-1})$, which includes independent or
sufficiently weakly dependent cross-sections. If the regime risks $R(\cdot)$
are separated by a fixed positive gap, the regime AUROC converges to $1$.
This is the high-$N$, near-i.i.d.\ vision regime (CIFAR-10-C,
Section~\ref{sec:cifar_external}).
\end{corollary}

\begin{corollary}[Outcome-conditioning requirement for expert-health
signals]\label{cor:deployment}
Consider a multi-expert allocator that routes capital using a per-expert
health signal $G_i(t)$, and fix the monitoring information set:
contemporaneous statistics are measurable functions of expert $i$'s inputs
$X$ (including cross-sectional aggregates of its own uncertainty scores
$\{g_i(x)\}$, feature-distribution-drift statistics, and disagreement
between fixed predictors), while delayed feedback supplies realised outcomes
$Y$. If $G_i(t)$ is built from contemporaneous, $X$-measurable statistics
alone, then under regime change that is concept-shift dominated for expert
$i$ ($\delta_N\to 0$) its regime-detection balanced accuracy is bounded by
Theorem~\ref{thm:detectability} and is asymptotically at chance.
Consequently, \emph{within this information set}, any health signal
guaranteed to retain regime power under strict concept shift
($\delta_N=0$) must incorporate $Y$-dependent (realised-efficacy)
information. This condition is necessary but not, on its own, sufficient for
reliable regime detection.
\end{corollary}

\paragraph{Interpretation and deployment consequence.}
Theorem~\ref{thm:detectability} depends on the two regimes \emph{only}
through total-variation movement of the input law: per-item signal strength
and cross-section size enter solely through $\delta_{N}$, not through any
property of $g$ on a fixed input distribution; boundedness and dependence
structure enter only in the corollaries (Table~\ref{tab:assumptions-by-result});
Remark~\ref{rem:sequential} extends the bound to path functionals such as
CUSUM monitors. Theorem~\ref{thm:detectability} and
Corollaries~\ref{cor:concept-shift}--\ref{cor:deployment} formalise the
aggregation gap in a finance deployment where $g$ is a
fold-refitted error predictor on correlated low-$N$
cross-sections. The $\delta_{N}=0$ impossibility aligns with the
real-versus-virtual drift separation of Section~\ref{subsec:concept-drift}; the
paper's additions are Corollary~\ref{cor:mean-suppressed}'s
variance--correlation double suppression, the fold-conditional scope
for refitted~$g$, the $2\times 2$ CIFAR/finance boundary test, and
the partial-horizon escape. Score-only summaries of $\{g(x_i)\}$
separate regimes only through input-law movement $\delta_{N}$, so
strict concept shift ($\delta_{N}\to 0$) forces chance detection at
every $N$ regardless of per-item $g$ quality, while cross-section
size and dependence govern concentration
(Corollaries~\ref{cor:mean-suppressed}--\ref{cor:high-n}) rather than
whether a signal exists. Proofs and scope caveats ($\delta_{N}$ versus
marginal $\eta$; bounds on $X$-measurable drift and disagreement) are in
Appendix~\ref{app:detectability}. The operational response is a
strategy-level health index built from matured RankIC, feature drift, and
predictor disagreement; next.

\subsection{Regime-Trust Monitor}
\label{sec:method-gate}

The strategy-level component is a model-specific monitor rather than a
market-observable regime detector: market-level signals infer regime state
from returns, volatility and correlations, and remain agnostic to a base
ranker's internal reliability. The health index aggregates three
independently PIT-safe components. $H_{\text{real}}(t)$ is an exponentially
weighted moving average of \emph{matured} RankIC at the 20-day horizon:
only observations whose forward windows closed strictly before $t$ enter,
so the monitor uses no information unavailable to an operator, at the cost
of a structural 20-day lag at the onset of a break. $H_{\text{drift}}(t)$
is a weighted composite of feature-distribution drift (Population Stability
Index against a 252-day reference), score-distribution drift
(Kolmogorov--Smirnov), and a rank-correlation spike detector.
$H_{\text{disagree}}(t)$ measures within-date Spearman disagreement between
two independently trained rankers (LightGBM and a Rank-Average-2 ensemble).
The composite is
\begin{equation}
  H(t) \;=\; h\!\bigl(H_{\text{real}}(t),\, H_{\text{drift}}(t),\,
                       H_{\text{disagree}}(t)\bigr) \;\in\; [0,1],
  \label{eq:gate}
\end{equation}
where each channel is expanding-$z$-scored (minimum 60 dates, clipped to
$\pm 3$), then combined as
\begin{equation}
  H_{\text{raw}}(t) = z_{\text{real}}(t)
    - 0.3\,z_{\text{drift}}(t) - 0.3\,z_{\text{disagree}}(t),
  \qquad
  H(t) = \sigma\!\bigl(H_{\text{raw}}(t)\bigr),
  \label{eq:ht-combine}
\end{equation}
with $\sigma$ the logistic sigmoid: coefficient $+1$ on normalised realised
efficacy and $-0.3$ on each penalty channel (drift and disagreement), not a
positive-weighted average ($z_{\text{real}}$ is the expanding $z$-score of
the EWMA of matured 20-day RankIC, half-life 30 trading days, minimum 20
observations). A DEV-calibrated monotone rescaling
maps $H(t)$ onto the continuous gate score
\begin{equation}
  \Gamma(t) = \operatorname{clip}\!\bigl((H(t)-0.3)/(0.7-0.3),\,0,\,1\bigr),
  \label{eq:gamma-gate}
\end{equation}
binarised as $G(t)=\mathbf{1}[\Gamma(t)\geq 0.2]$ with the threshold fitted
on the DEV partition. Component weights and which channel drives regime-trust
AUROC are audited in Section~\ref{sec:res-gate} (paragraph below
Table~\ref{tab:regime_auroc}). Binarisation rather than a continuous
multiplier follows \citet{nystrup2017}: smooth regime-driven exposure
destroys recovery convexity, and the continuous-throttle ablation of
Section~\ref{sec:res-gate} confirms it here (FINAL Sharpe $-0.34$ versus
$+0.925$ for the binary variant). The $0.2$ operating point yields a
$47.2\%$ abstention rate at $80.0\%$ bad-day precision over the full
walk-forward sample, with perfectly monotone mean RankIC across
$\Gamma$-buckets (bucket Spearman $\rho_S = 1.0$).

\subsection{Two-Level Deployment}
\label{sec:method-twolevel}

The position-level component is a discrete tail cap. Within active trading
days ($G(t)=1$), weights are adjusted only on the most uncertain quantile:
\begin{equation}
  w_{i,t} \;=\; w_{\text{base},\,i,t} \times
    \Bigl( 1 - \kappa \,\cdot\, \mathbf{1}\!\bigl[\hat{e}(x_{i,t})
           \geq P_{85}\!\bigl(\hat{e}_{t}\bigr)\bigr] \Bigr),
  \label{eq:ecap}
\end{equation}
with $\kappa = 0.30$ and the threshold at the 85th cross-sectional
percentile of $\hat{e}$ on date $t$ (both DEV-calibrated); the base weight
is inverse-volatility sized. Because the cap is an indicator on a single
quantile rather than a continuous function of $\hat{e}$, the
$\rho(\hat{e},|\text{score}|)=0.616$ coupling is immaterial for the bulk of
the book: positions below the 85th percentile keep full base weight, and
because the coupling is positive but imperfect, most extreme-rank
positions---the highest-conviction ideas---fall outside the capped tail. The
cap acts only at the narrow intersection of extreme signal conviction and
extreme estimated displacement risk. Figure~\ref{fig:two-level} summarises
the architecture; the pipeline and gate-detail diagrams are in
Appendix~\ref{app:extra-floats} (Figures~\ref{fig:framework}
and~\ref{fig:regime-gate}).

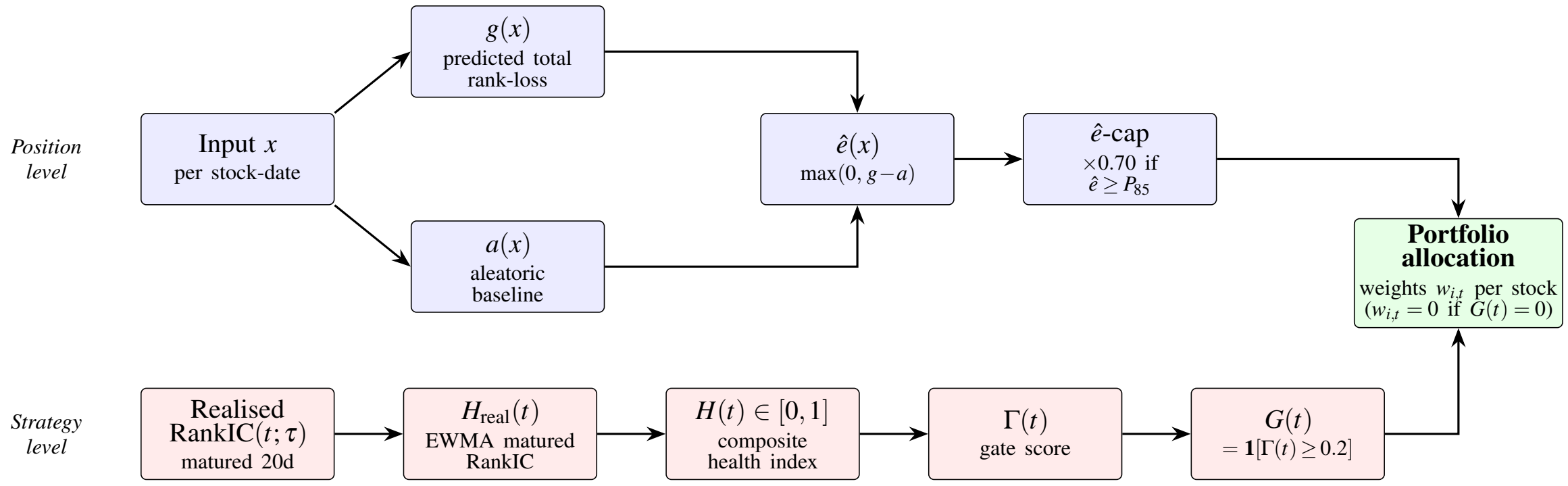
\begin{figure}[ht]
  \centering
  \resizebox{0.95\linewidth}{!}{%
  \begin{tikzpicture}[
      node distance=8mm and 9mm,
      every node/.style={font=\small},
      box/.style={draw, rounded corners=2pt, align=center,
                  text width=24mm, minimum height=12mm, inner sep=2pt,
                  execute at begin node={\baselineskip=9pt\relax}},
      pstock/.style={box, fill=blue!8},
      strat/.style={box, fill=red!8},
      converge/.style={box, fill=green!10, text width=26mm, minimum height=14mm,
                       execute at begin node={\baselineskip=9pt\relax}},
      lbl/.style={font=\scriptsize\itshape, align=center, text width=18mm},
      arr/.style={-{Stealth[length=2.5mm]}, thick},
    ]
    \node[pstock]                       (x)   {Input $x$\\[1pt]
                                               \scriptsize per stock-date};
    \node[pstock, above right=2mm and 10mm of x] (g) {$g(x)$\\[1pt]
                                               \scriptsize predicted total\\[-1pt]
                                               \scriptsize rank-loss};
    \node[pstock, below right=2mm and 10mm of x] (a) {$a(x)$\\[1pt]
                                               \scriptsize aleatoric\\[-1pt]
                                               \scriptsize baseline};
    \node[pstock, right=14mm of x, xshift=42mm] (e) {$\hat{e}(x)$\\[1pt]
                                               \scriptsize $\max(0,\,g\!-\!a)$};
    \node[pstock, right=of e]           (cap) {$\hat{e}$-cap\\[1pt]
                                               \scriptsize $\times 0.70$ if\\[-1pt]
                                               \scriptsize $\hat{e}\!\geq\!P_{85}$};
    \node[strat, below=24mm of x]    (ric)   {Realised\\RankIC$(t;\tau)$\\[1pt]
                                              \scriptsize matured 20d};
    \node[strat, right=of ric]       (hreal) {$H_{\text{real}}(t)$\\[1pt]
                                              \scriptsize EWMA matured\\[-1pt]
                                              \scriptsize RankIC};
    \node[strat, right=of hreal]     (h)     {$H(t)\in[0,1]$\\[1pt]
                                              \scriptsize composite\\[-1pt]
                                              \scriptsize health index};
    \node[strat, right=of h]         (gam)   {$\Gamma(t)$\\[1pt]
                                              \scriptsize gate score};
    \node[strat, right=of gam]       (gate)  {$G(t)$\\[1pt]
                                              \scriptsize $=\mathbf{1}[\Gamma(t)\!\geq\!0.2]$};
    \node[converge, right=18mm of cap, yshift=-15mm] (port)
        {\textbf{Portfolio allocation}\\[1pt]
         \scriptsize weights $w_{i,t}$ per stock\\[-1pt]
         \scriptsize ($w_{i,t}=0$ if $G(t)=0$)};
    \node[lbl, left=2mm of x]   {Position\\level};
    \node[lbl, left=2mm of ric] {Strategy\\level};
    \draw[arr] (x.north east) -- (g.west);
    \draw[arr] (x.south east) -- (a.west);
    \draw[arr] (g.east)       -| (e.north);
    \draw[arr] (a.east)       -| (e.south);
    \draw[arr] (e) -- (cap);
    \draw[arr] (ric)   -- (hreal);
    \draw[arr] (hreal) -- (h);
    \draw[arr] (h)     -- (gam);
    \draw[arr] (gam)   -- (gate);
    \draw[arr] (cap.east) -| (port.north);
    \draw[arr] (gate.east) -| (port.south);
  \end{tikzpicture}
  }
  \caption{Two-level uncertainty architecture: per-prediction
    $\hat{e}$-cap (upper, blue track) and strategy-level regime gate (lower,
    red track). The position-level track converts input $x$ into
    $\hat{e}(x) = \max(0,\, g(x) - a(x))$ and reduces by $30\%$
    ($\times 0.70$, $\kappa = 0.30$) the weight of any stock whose
    $\hat{e}(x)$ exceeds the 85th cross-sectional percentile
    (Eq.~\ref{eq:ecap}). The strategy-level track maps matured realised
    RankIC into the composite health index $H(t)$, rescales it to
    $\Gamma(t)$ (Eq.~\ref{eq:gamma-gate}), and binarises to
    $G(t)=\mathbf{1}[\Gamma(t)\geq 0.2]$. Both tracks converge in the
    allocation block, where $w_{i,t}=0$ whenever $G(t)=0$. The same
    architecture is replayed without modification on CIFAR-10-C in
    Section~\ref{sec:cifar_external}.}
  \label{fig:two-level}
\end{figure}

Seven deployment-policy variants plus two ungated baselines and a na\"{i}ve
kill-switch comparator are evaluated (Table~\ref{tab:policy-variants}); all
numbered variants share the binary gate and differ only in within-day
weighting. Variant~6 (gate + vol-sizing + $\hat{e}$-cap) is the target
policy: the gate eliminates strategy-wide breakdown days, volatility sizing
addresses the aleatoric component within a day, and the cap removes exposure
on the small subset of positions whose predicted displacement risk is
abnormal even within a healthy regime. No single layer subsumes the others:
$G(t)$ is uninformative about within-day position quality, volatility sizing
about predicted ranking error
($\rho(\hat{e},\texttt{vol\_20d}) \leq 0.24$ after orthogonalisation), and
the cap about strategy-wide viability.

\begin{table}[ht]
  \centering
  \caption{Deployment-policy variants evaluated in
    Section~\ref{sec:results}. All numbered variants share the binary
    auto-execution rule ($G(t)=\mathbf{1}[\Gamma(t)\geq 0.2]$) and differ
    only in the within-day weighting rule. Cap parameters $P_{85}$ and
    $\kappa = 0.30$ are DEV-calibrated (Eq.~\ref{eq:ecap}).}
  \label{tab:policy-variants}
  \small
  \begin{tabular}{@{}clllc@{}}
    \toprule
    \textbf{\#} & \textbf{Variant} &
    \textbf{Within-day sizing} &
    \textbf{Position cap} &
    \textbf{Role} \\
    \midrule
    ---  & Ungated Raw          & Equal weight       & None
         & Baseline (no risk control) \\
    ---  & Ungated Vol-Sized    & Inverse volatility & None
         & Baseline (vol only) \\
    \addlinespace
    V1   & Gate~$+$ Raw         & Equal weight       & None
         & Pure $G(t)$ effect \\
    V2   & Gate~$+$ Vol         & Inverse volatility & None
         & $G(t)$ $+$ aleatoric control \\
    V3   & Gate~$+$ UA-Sort     & UA-Sort \citep{liu2026}
         & None & Continuous UQ benchmark \\
    V4   & Gate~$+$ Resid-$\hat{e}$ & Residualised $\hat{e}$ sizing
         & None & Continuous epistemic sizing \\
    V5   & Gate~$+$ $\hat{e}$-Cap & Equal weight
         & $P_{85}$, $\kappa{=}0.30$ & Pure $\hat{e}$-cap effect \\
    \textbf{V6}
         & \textbf{Gate~$+$ Vol $+$ $\hat{e}$-Cap}
         & \textbf{Inverse volatility}
         & $\boldsymbol{P_{85}}$, $\boldsymbol{\kappa{=}0.30}$
         & \textbf{Target policy} \\
    \addlinespace
    K    & Trail-IC kill switch & Equal weight       & None
         & Na\"{i}ve kill-switch comparator \\
    \bottomrule
  \end{tabular}
\end{table}

Table~\ref{tab:monitoring} consolidates every monitoring signal, threshold,
and action the framework emits; the deployment reference for the
practitioner box of Section~\ref{sec:intro}; the per-date wire format
these signals feed to a meta-allocator is specified in
Appendix~\ref{app:allocator}.

\begin{table}[ht]
  \centering
  \caption{Consolidated monitoring and deployment reference. Rows above the
    rule are research-defined and DEV-calibrated on the frozen walk-forward
    panel. Rows below it are \emph{design specifications} for a production
    monitor; they are not validated on a live track record in this paper.}
  \label{tab:monitoring}
  \footnotesize
  \setlength{\tabcolsep}{4pt}
  \renewcommand{\arraystretch}{1.15}
  \begin{tabularx}{\linewidth}{@{}%
    >{\raggedright\arraybackslash}p{2.45cm}%
    >{\raggedright\arraybackslash}X%
    >{\raggedright\arraybackslash}p{2.1cm}%
    >{\raggedright\arraybackslash}p{3.2cm}%
    >{\raggedright\arraybackslash}p{2.2cm}@{}}
    \toprule
    \textbf{Signal} & \textbf{Definition} & \textbf{Threshold} &
    \textbf{Action} & \textbf{Calibration source} \\
    \midrule
    $\Gamma(t)$ &
      DEV-calibrated rescaling of $H(t)$ (Eq.~\ref{eq:gamma-gate}) &
      $\Gamma(t) \geq 0.20$ &
      Trade ($G(t)=1$); apply two-level sizing &
      DEV partition \\
    $\Gamma(t)$ &
      as above &
      $\Gamma(t) < 0.20$ &
      Abstain; zero gross exposure &
      DEV partition \\
    $\hat{e}(x)$ &
      DEUP epistemic signal (Eq.~\ref{eq:epistemic-signal}) &
      $\geq P_{85}$ within date &
      Position weight $\times 0.70$ ($\kappa = 0.30$) &
      DEV partition \\
    $\widehat{P}(\text{good day} \mid H(t))$ &
      Platt-calibrated walk-forward probability
      (Section~\ref{sec:res-gate}) &
      --- (continuous) &
      Meta-allocator sizing input &
      Walk-forward matured labels \\
    $\mathrm{width\_deup}$ / $\mathrm{risk\_score}$ &
      $\hat{e}_{\text{PIT}}$-normalised conformal half-width; within-date
      percentile (Section~\ref{sec:res-conformal}) &
      --- &
      Calibrated per-position risk read; no sizing action (per-date
      Spearman $1.0$ with the $\hat{e}$-cap) &
      Trailing matured losses ($W{=}60$, $h{=}20$) \\
    \midrule
    PSI feature drift &
      Max model-feature PSI vs.\ training baseline (90d recent window) &
      $>0.20$ retrain review;\newline $>0.50$ critical &
      Retrain review;\newline de-risk if critical &
      Production monitor spec \\
    Rolling IC alert &
      4-week rolling avg.\ of weekly realised RankIC (5d horizon) &
      $<0.03$ &
      Alert;\newline retrain review if persistent ($>3$ wk) &
      Production monitor spec \\
    \bottomrule
  \end{tabularx}
\end{table}

\subsection{Evaluation Protocol}
\label{sec:method-eval}

The protocol spans five metric families, signal quality, uncertainty
quality, gate classification, portfolio performance, and statistical
validity, each with acceptance thresholds calibrated on the DEV partition
and \emph{pre-registered} before the FINAL holdout was inspected
(Table~\ref{tab:eval-metrics}); the FINAL partition is evaluated once, with
no post-hoc recalibration. Every metric is reported against explicit
comparators: 12-month momentum at the signal level; the ungated baselines,
the \citet{liu2026} UA-sort, and a trailing-IC kill switch at the portfolio
level (Table~\ref{tab:policy-variants}); a within-date permutation null and
the Fama--French five-factor benchmark at the statistical level.

Signal quality uses the per-date rank information coefficient
\begin{equation}
  \operatorname{RankIC}(t;\tau)
  \;=\;
  \rho_{S}\!\bigl(\, s_{\cdot,t}\,,\; r^{(\tau)}_{\cdot,t} \,\bigr),
  \label{eq:rankic}
\end{equation}
summarised by its median and by IC stability
$\bar{\rho}_{S}/\sigma(\rho_{S})$, plus a cost-survival rate (fraction of
folds whose top-decile median excess return is positive). Uncertainty
quality uses $\rho_{S}(g(x),\ell)$ and quintile monotonicity of
$\hat{e}$-buckets (acceptance: bucket Spearman $=1.0$ at 20 days). Gate
quality scores $H(t)$ and $\Gamma(t)$ by AUROC against
$\mathbf{1}[\text{matured RankIC}>0]$. Portfolio performance uses the
annualised Sharpe ratio on \emph{monthly non-overlapping} returns,
\begin{equation}
  \operatorname{SR} \;=\; \frac{\mu_{\text{m}}}{\sigma_{\text{m}}}\sqrt{12},
  \label{eq:sharpe}
\end{equation}
with a 12-lag Newey--West HAC adjustment of the sampling variance
\citep{newey1987,lopezdeprado2025}.

Two evaluation objects are distinguished to avoid threshold-shopping.
Portfolio-performance and statistical-validity thresholds in
Table~\ref{tab:eval-metrics} (Sharpe, drawdown, PSR/DSR, permutations,
FF5~$\alpha$) were pre-registered for the \emph{ungated equal-weight shadow
portfolio} on the vol-sized LightGBM ranker (Section~\ref{sec:res-base};
Table~\ref{tab:sharpe-canonical})---the alpha-existence question on the
frozen walk-forward panel. The paper's accepted deployable object is the
reliability architecture (gate, cap, monitoring); Variant~6 is judged by
gate-classification quality, the policy ablation of Section~\ref{sec:res-gate},
and the eight-window crisis panel of Section~\ref{sec:res-crisis}, not by
re-applying the ungated Sharpe $\geq 1.5$ bar. Statistical validity combines the
Probabilistic Sharpe Ratio
\begin{equation}
  \operatorname{PSR}(\mathrm{SR}_0)
  \;=\;
  \Phi\!\left(\;
    \frac{(\widehat{\mathrm{SR}} - \mathrm{SR}_0)\sqrt{n-1}}
         {\sqrt{1 - \widehat{\gamma}_{3}\,\widehat{\mathrm{SR}}
                + \tfrac{\widehat{\gamma}_{4}-1}{4}\,
                  \widehat{\mathrm{SR}}^{2}}}
  \;\right)
  \label{eq:psr}
\end{equation}
\citep{bailey2012} ($\mathrm{SR}_0$ is the benchmark Sharpe under the null;
$\widehat{\gamma}_3$ and $\widehat{\gamma}_4$ are sample skewness and
\emph{raw} (non-excess) kurtosis, as in \citealt{bailey2012}), the Deflated Sharpe Ratio over the $N=5$ candidate
fusion family \citep{bailey2014}, White's Reality Check \citep{white2000},
Bonferroni/Benjamini--Hochberg corrections, $5{,}000$-trial within-date
permutation tests, and Fama--French five-factor attribution
\citep{fama2015} with HAC errors. A pre-registered materiality rule governs
feature-augmentation ablations on $g(x)$: $|\Delta\rho| \leq 0.005$ on
held-out $\rho(g,\ell)$ is treated as empirically null, fixed before
inspecting any ablation outcome. Regime robustness uses a retrospective
panel of eight non-overlapping windows, five crisis episodes and three
calm references, 2018--2024, scoring $G(t)$ as correctly active, correctly
abstaining, falsely alarming, or missing (acceptance: $\geq 6/8$). For
allocator-facing sample-size accounting, the average-uniqueness estimator
of \citet{lopezdeprado2018} defines
$\mathrm{eff}\text{-}N = P\,\bar{u}$ over $P$ overlapping position slots,
and the effective episode count $E_{\mathrm{eff}} := \mathrm{eff}\text{-}N$
(Appendix~\ref{app:universe-readiness}).

Reproducibility: all random processes are seeded (\texttt{random\_state}
$=42$), all evaluations run against a frozen read-only DuckDB feature-store
snapshot, and the finance pipeline requires no GPU; the CIFAR-10-C
experiment of Section~\ref{sec:cifar_external} ran on a single rented
NVIDIA~A100 (Appendix~\ref{app:reproducibility}).

\begin{table}[ht]
  \centering
  \caption{Evaluation metrics with formal definitions and acceptance
    thresholds. Thresholds are operating points calibrated on the DEV
    partition; the rightmost column points to the subsection reporting the
    out-of-sample reading.}
  \label{tab:eval-metrics}
  \small
  \begin{tabular}{@{}p{0.27\linewidth}p{0.34\linewidth}cl@{}}
    \toprule
    \textbf{Metric} & \textbf{Formula / Definition} &
    \textbf{Threshold} & \textbf{\S} \\
    \midrule
    \multicolumn{4}{@{}l}{\textit{Signal quality}} \\
    Median RankIC (20d)
      & Median of Eq.~(\ref{eq:rankic}) across evaluation dates
      & $\geq 0.04$ (FINAL) & \ref{sec:res-base} \\
    IC stability
      & $\bar{\rho}_{S}\,/\,\sigma(\rho_{S})$
      & $\geq 0.05$ (FINAL) & \ref{sec:res-base} \\
    Cost survival rate
      & Fraction of folds: top-decile excess $>0$
      & $\geq 60\%$ & \ref{sec:res-base} \\
    \midrule
    \multicolumn{4}{@{}l}{\textit{Uncertainty quality}} \\
    $\rho_{S}\!\bigl(g(x),\ell\bigr)$
      & Per-date Spearman, $g$ vs.\ rank displacement
        (Eq.~\ref{eq:rank-displacement})
      & $\geq 0.15$ (FINAL) & \ref{sec:res-deup} \\
    Quintile monotonicity
      & Bucket Spearman across $\hat{e}$-quintiles
      & $= 1.0$ at 20d & \ref{sec:res-deup} \\
    \midrule
    \multicolumn{4}{@{}l}{\textit{Gate classification}} \\
    $H(t)$ AUROC
      & vs.\ $\mathbf{1}[\text{matured RankIC}>0]$
      & $\geq 0.70$ & \ref{sec:res-gate} \\
    Bucket monotonicity ($\Gamma(t)$-deciles)
      & Bucket Spearman across $\Gamma(t)$-deciles
      & $= 1.0$ & \ref{sec:res-gate} \\
    Abstention rate at $\Gamma(t)\!\geq\!0.2$
      & Fraction of dates with $\Gamma(t) < 0.2$
      & $30$--$50\%$ & \ref{sec:res-gate} \\
    \midrule
    \multicolumn{4}{@{}l}{\textit{Portfolio performance (ungated ranker only)}} \\
    Sharpe (HAC-adjusted)
      & Eq.~(\ref{eq:sharpe}), Newey--West 12-lag on the ungated
        equal-weight shadow series (Table~\ref{tab:sharpe-canonical})
      & $\geq 1.5$ (FINAL) & \ref{sec:res-falsification} \\
    Max drawdown
      & Peak-to-trough monthly return (same shadow series)
      & $\geq -25\%$ & \ref{sec:res-falsification} \\
    \midrule
    \multicolumn{4}{@{}l}{\textit{Statistical validity}} \\
    PSR$(\mathrm{SR}_0=0)$
      & Eq.~(\ref{eq:psr})
      & $\geq 0.95$ & \ref{sec:res-falsification} \\
    DSR ($N_{\text{trials}}=5$)
      & \citet{bailey2014}
      & $\geq 0.95$ & \ref{sec:res-falsification} \\
    Permutation $p$-value
      & $5{,}000$ within-date permutations
      & $\leq 0.05$ & \ref{sec:res-falsification} \\
    FF5 $\alpha$ $t_{\mathrm{HAC}}$
      & Newey--West 12-lag, monthly
      & $\geq 3.0$ & \ref{sec:res-falsification} \\
    \midrule
    \multicolumn{4}{@{}l}{\textit{Multi-crisis robustness}} \\
    $G(t)$ verdict accuracy
      & 8 crisis/calm windows
      & $\geq 6/8$ & \ref{sec:res-crisis} \\
    \bottomrule
  \end{tabular}
  \smallskip
  \par\noindent\footnotesize
  Portfolio-performance and statistical-validity rows (Sharpe through
  FF5~$\alpha$) apply to the shadow series in
  Table~\ref{tab:sharpe-canonical} only. Deployed Variant~6
  (gate\,+\,vol\,+\,$\hat{e}$-cap) is evaluated by gate AUROC, ablation
  Sharpe trade-offs, and multi-crisis verdicts
  (Sections~\ref{sec:res-gate}--\ref{sec:res-crisis}).
\end{table}

\section{Data and Evaluation Protocol}
\label{sec:data}

The universe is a dynamic panel of AI-exposed US equities, at most 100
stocks drawn from ten thematic categories (cloud infrastructure,
semiconductors, AI software, cybersecurity, autonomous systems, data and
analytics, AR/VR, robotics, digital health, fintech). At each monthly
rebalance the eligible set passes a four-condition waterfall: US common
stock on a major exchange; close price $\geq \$5$; 20-day average daily
dollar volume $\geq \$1$ million; membership of a curated thematic
allowlist (a single author-maintained seed list of up to 100 tickers,
frozen as the research snapshot used for all historical dates; manual
updates during research did not produce a date-stamped quarterly
constituent history in the frozen artifact). The first three conditions
are applied point-in-time; \emph{thematic membership is not}
(Section~\ref{sec:res-falsification}). The panel is
listing-survivorship-aware by construction: each of
the 29 stocks that listed after the sample start (e.g.\ ARM, Mobileye)
enters from its first trading date, is never backdated, and is truncated at
delisting rather than backfilled---which addresses delisting dates, not
retrospective inclusion in the AI allowlist. Prices are Polygon.io SIP consolidated
tape (Grouped Daily Bars), ingested once into a frozen read-only DuckDB
store; a data-quality audit found zero close-price discrepancy against an
independent vendor pull on representative tickers. The label is
Eq.~(\ref{eq:label}); labels and features are deliberately
\emph{unwinsorised} (the tree backbone responds to ordinal rank, and
clipping would attenuate the dispersion the ranker exploits), missing
values receive \emph{no imputation} (LightGBM native NaN routing, with
explicit missingness-indicator features), and an automated point-in-time
scanner passes with zero critical violations. A qualitative
signal-hygiene checklist, distilled from universe-churn defects observed
during research signal generation, is provided in
Appendix~\ref{app:hygiene}.

\begin{table}[!htbp]
  \centering
  \caption{DEV/FINAL partition and realised tradable breadth on the frozen
    evaluation window. ``Trading dates'' counts distinct market dates;
    ``Folds'' counts monthly walk-forward evaluations. $N_t$ is the number
    of eligible tickers in the 20-day panel after the \$5 price and
    \$1\,million ADV filters. The DEV/FINAL boundary was fixed at
    1~January~2024.}
  \label{tab:ch4-nt}
  \small
  \resizebox{\linewidth}{!}{%
  \begin{tabular}{@{}llrrrrrrrrrr@{}}
    \toprule
    Partition & Date range & Trading dates & Folds
      & Mean & Median & p10 & p25 & p75 & p90 & Min & Max \\
    \midrule
    ALL & 2016-02-01 -- 2025-02-19 & 2,277 & 109
      & 83.9 & 83 & 71 & 75 & 96 & 98 & 67 & 99 \\
    DEV & 2016-02-01 -- 2023-12-31 & 1,993 &  95
      & 81.8 & 80 & 70 & 74 & 90 & 97 & 67 & 98 \\
    FINAL & 2024-01-01 -- 2025-02-19 &   284 &  14
      & 98.4 & 98 & 98 & 98 & 99 & 99 & 98 & 99 \\
    \bottomrule
  \end{tabular}}
\end{table}

Of $688{,}800$ candidate label triplets, $87.2\%$ were computed; $98.2\%$ of
the drops are pre-IPO observations for late-listed stocks, and the remainder
trace to a single delisting (ABB Ltd.), leaving $99.95\%$ of in-window
(ticker\,$\times$\,date) pairs with a valid 20-day label. One caveat
survives compression: CRSP-style terminal delisting returns are \emph{not}
modelled, observations whose labels cannot mature because a stock exits are
excluded rather than assigned a terminal return, which affects only ABB here
but would require explicit handling in any broader replication. The
training-ready panel comprises $192{,}307$ rows on 39 usable features.

The DEV/FINAL boundary was set as a hard freeze at 1~January~2024
\emph{before} any FINAL-period output was inspected: DEV results were
examined and revised during development, but the FINAL holdout was evaluated
once, with no recalibration of any hyperparameter or threshold.
Table~\ref{tab:ch4-experiment-config} records the frozen experiment frame.
The FINAL partition is materially denser than DEV (mean $N_t$ 98.4 versus
81.8), so DEV--FINAL comparisons coincide with both a temporal regime change
and a breadth increase.

\begin{table}[!htbp]
  \centering
  \caption{Frozen walk-forward experiment configuration.}
  \label{tab:ch4-experiment-config}
  \small
  \begin{tabular}{@{}lp{0.62\linewidth}@{}}
    \toprule
    Parameter & Value \\
    \midrule
    Export snapshot & 21~January~2026 (201,307 rows) \\
    Training-ready rows & 192,307 (20-day primary horizon) \\
    Usable ranker features & 39 (after zero-information elimination) \\
    DEV / FINAL lock & 1~January~2024 (hard freeze) \\
    Fold embargo & 90 trading days (all horizons) \\
    Global random seed & \texttt{random\_state} $= 42$ \\
    Primary ranker & LightGBM: \texttt{n\_estimators}$=100$,
      \texttt{learning\_rate}$=0.05$ \\
    \bottomrule
  \end{tabular}
\end{table}

One exploratory regularity motivates the gate and is worth stating before
the results: the score--return association is regime-stratified
(Table~\ref{tab:ch4-rankic-regime}). At the 20-day horizon the median
within-date rank correlation between score and realised excess return falls
from $0.182$ in the low-VIX third to $0.071$ in the high-VIX third, with
the same monotone ordering at 60 and 90 days, and the per-date association
correlates negatively with the contemporaneous VIX at all horizons. The
predictive content of the feature set is regime-stratified rather than
uniform; exactly the data property a regime-aware trust gate is designed
for, and the reason a single unconditional model under-serves deployment.
Cross-sectional dispersion is in fact \emph{higher} in FINAL than in DEV
(mean 10th--90th spread $0.24$ versus $0.19$;
Figure~\ref{fig:ch4-dispersion}, Appendix~\ref{app:extra-floats}), so the
2024 association loss is not attributable to a narrower target
distribution.

\begin{table}[!htbp]
  \centering
  \caption{Median within-date rank correlation between model score and
    realised excess return, stratified by VIX volatility regime (thirds of
    the 252-day VIX percentile). The final row is the rank correlation
    between the per-date association and the contemporaneous VIX.}
  \label{tab:ch4-rankic-regime}
  \small
  \begin{tabular}{lrrr}
    \toprule
    VIX regime & 20-day & 60-day & 90-day \\
    \midrule
    Low ($\leq$33\%)  & $0.182$  & $0.347$  & $0.298$  \\
    Mid (33--67\%)    & $0.077$  & $0.207$  & $0.246$  \\
    High ($>$67\%)    & $0.071$  & $0.120$  & $0.161$  \\
    \midrule
    Association--VIX $\rho$ & $-0.115$ & $-0.207$ & $-0.181$ \\
    \bottomrule
  \end{tabular}
\end{table}

\section{Results}
\label{sec:results}

Evidence comes from the frozen nine-year walk-forward evaluation of
Table~\ref{tab:ch4-nt}, executed under the PIT-safe protocol of
Section~\ref{sec:method-ranker} and scored against the pre-registered
thresholds of Table~\ref{tab:eval-metrics}. Signal-quality and gate metrics
apply throughout; portfolio-performance thresholds (Sharpe $\geq 1.5$,
PSR/DSR, permutations, FF5~$\alpha$) apply to the \emph{ungated} ranker
(Section~\ref{sec:res-falsification}), while the deployed gated policy is
assessed by ablation and crisis criteria (Section~\ref{sec:res-gate}).
The controlled cross-domain experiment on CIFAR-10-C is reported separately in
Section~\ref{sec:cifar_external} as a mechanism test, not a second
empirical claim about the strategy. Walk-forward tables and figures use
data through February~2025 only; Section~\ref{sec:res-monitoring-corr}
documents post-freeze \emph{monitoring} behaviour illustratively (alert
firing and model lifecycle) without portfolio-performance inference
(Section~\ref{sec:limitations}).

\subsection{Base Ranker Signal Quality}
\label{sec:res-base}

The base ranker must clear a minimum signal bar before any uncertainty layer
can be judged independently. The central finding is asymmetric: FINAL median
RankIC falls below the pre-registered threshold ($+0.017$ at 20 days,
\textbf{FAIL}; Table~\ref{tab:signal_quality_531}) while IC stability, cost
survival, and positive portfolio returns persist, defining the deployment
problem the uncertainty stack addresses.

\paragraph{Factor floor and model lift.}
On the identical 109 folds, the classical factor baselines set the floor:
\texttt{mom\_12m} mean RankIC $0.028$ at 20 days ($0.028$ / $0.016$ at
60/90), a momentum composite $0.025$ / $0.039$ (20/60d), short-term
strength $0.017$ / $0.020$ / $0.009$, and a random-ranking baseline at
zero. A seven-feature momentum-and-volume LightGBM already exceeded the
floor by a wide margin (mean RankIC $0.101$ at 20 days, $+261\%$ over
\texttt{mom\_12m}; $0.128$ at 60 days, $+357\%$; $0.181$ at 90 days,
$+1{,}031\%$ over the 90-day momentum baseline $0.016$); the full production feature set delivered ALL-period mean/median
RankIC of $0.064$/$0.081$ at 20 days. Fusion architectures (learned
stacking, rank-average, regime-blended) added nothing at the primary
horizon, so the production model is the vol-sized LightGBM baseline.

\begin{table}[!htbp]
  \centering
  \caption{Signal quality of the LightGBM ranker versus the acceptance
    thresholds of Table~\ref{tab:eval-metrics}. All RankIC and IC-stability
    figures are \emph{medians}. Thresholds are calibrated for the 20-day
    primary deployment horizon; $^{\dagger}$ indicates the threshold applied
    to all horizons for reference. PASS/FAIL is assessed on FINAL.}
  \label{tab:signal_quality_531}
  \small
  \begin{tabular}{@{}llccccl@{}}
    \toprule
    \textbf{Metric} & \textbf{Hz} &
    \textbf{ALL} & \textbf{DEV} & \textbf{FINAL} &
    \textbf{Threshold} & \textbf{Verdict} \\
    \midrule
    \multicolumn{7}{@{}l}{\textit{Median RankIC
      ($\tilde{\rho}_S$; cf.\ Eq.~\ref{eq:rankic})}} \\[2pt]
    Median RankIC & 20d & 0.081 & 0.091 & $+$0.017
      & $\geq$0.04 & \textbf{FAIL} \\
    Median RankIC & 60d & 0.148 & 0.167 & $-$0.044
      & $\geq$0.04$^{\dagger}$ & FAIL \\
    Median RankIC & 90d & 0.183 & 0.206 & $-$0.052
      & $\geq$0.04$^{\dagger}$ & FAIL \\
    \midrule
    \multicolumn{7}{@{}l}{\textit{IC Stability
      ($\bar{\rho}_S / \sigma(\rho_S)$)}} \\[2pt]
    IC Stability   & 20d & 0.353 & 0.389 & $+$0.069
      & $\geq$0.05 & \textbf{PASS} \\
    IC Stability   & 60d & 0.712 & 0.844 & $-$0.026
      & $\geq$0.05$^{\dagger}$ & FAIL \\
    IC Stability   & 90d & 0.797 & 0.963 & $-$0.129
      & $\geq$0.05$^{\dagger}$ & FAIL \\
    \midrule
    \multicolumn{7}{@{}l}{\textit{Cost Survival Rate
      (fraction of folds: top-10 mean excess $> 0$)}} \\[2pt]
    Cost Survival  & 20d & 69.7\% & 70.5\% & 64.3\%
      & $\geq$60\% & \textbf{PASS} \\
    Cost Survival  & 60d & 80.7\% & 82.1\% & 71.4\%
      & $\geq$60\%$^{\dagger}$ & PASS \\
    Cost Survival  & 90d & 81.7\% & 83.2\% & 71.4\%
      & $\geq$60\%$^{\dagger}$ & PASS \\
    \bottomrule
  \end{tabular}
  \smallskip
  \par\noindent\footnotesize
  $^{\dagger}$~Threshold calibrated for the 20-day deployment horizon;
  applied to all horizons for reference.
\end{table}

\begin{figure}[H]
  \centering
  \includegraphics[width=0.92\linewidth]{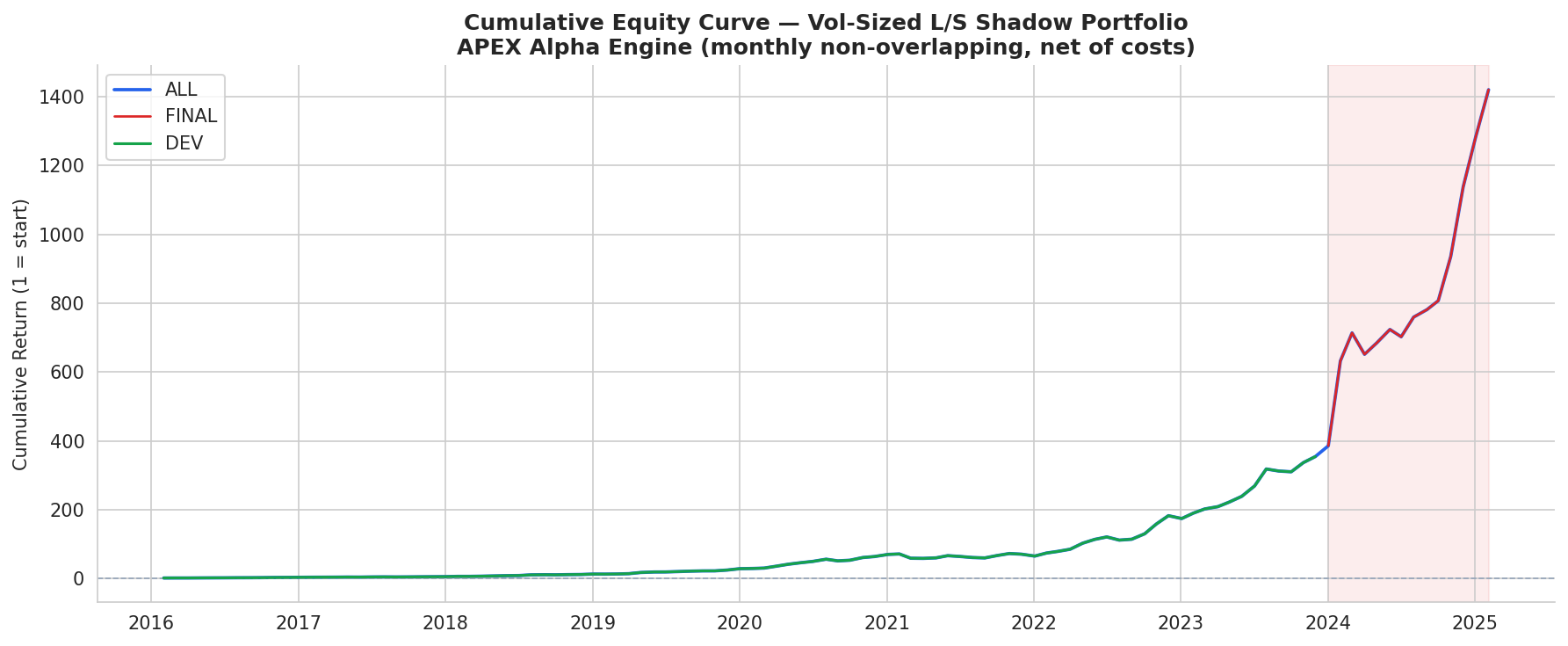}
  \caption{Cumulative equity curve of the ungated equal-weight shadow
    portfolio on the vol-sized LightGBM ranker (monthly non-overlapping,
    net of 10~bps per leg; Table~\ref{tab:sharpe-canonical}).
    Blue: ALL period (February~2016--February~2025, $n = 109$ months);
    green: DEV ($n = 95$); red: FINAL holdout ($n = 14$), chain-linked from
    the DEV level at 1~January~2024 (shaded).}
  \label{fig:equity_curve_52}
\end{figure}

\paragraph{Economic magnitude.}
Portfolio Sharpe figures are reconciled in Table~\ref{tab:sharpe-canonical}.
The headline falsification series is the \emph{ungated equal-weight shadow
portfolio} on the \textbf{vol-sized LightGBM ranker} (top-10/bottom-10,
equal-weight legs, monthly non-overlapping returns from the frozen
walk-forward monthly store; $n=109/95/14$ calendar months). Its annualised
Sharpe is $2.73$ (ALL), $3.12$ (DEV), and $2.34$ (FINAL)---rounded to
$2.730$ on the frozen ungated-raw policy-grid row
(Table~\ref{tab:sharpe-canonical}), a numerical coincidence on different
monthly subsamples ($n=109$ walk-forward store vs.\ $N_{\mathrm{ALL}}=90$
policy store), not evidence of identical return series. Both use
equal-weight top-10/bottom-10 sizing on the vol-sized LightGBM ranker; they
differ from the ungated vol-sized row ($2.614$ ALL), which adds
\emph{inverse-volatility weights within legs} on the policy store. FINAL Sharpe also differs between
the shadow series ($2.34$; 14 months) and the ablation grid ($1.650$ raw /
$1.603$ vol-sized; $N=5$ complete months on the policy store---see
Table~\ref{tab:ablation} caption). All figures are net of a 10~bps-per-leg
cost assumption; at $42.7\%$ mean monthly turnover the annualised drag is
$\approx 1.02\%$ (sensitivity to 50~bps in
Appendix~\ref{app:cost-sensitivity}). Maximum drawdown is $-18.1\%$
full-sample on the shadow curve ($-8.7\%$ in FINAL); monthly hit rate
$82.6\%$ (ALL) and $85.7\%$ (FINAL). The equity curve is
Figure~\ref{fig:equity_curve_52}; the five deepest drawdown episodes,
clustered in high-VIX thematic rotations, are tabulated in
Appendix~\ref{app:extra-floats}
(Table~\ref{tab:drawdown-episodes-531}). Significance testing is deferred
to Section~\ref{sec:res-falsification}.

\begin{table}[H]
  \centering
  \caption{Canonical reconciliation of portfolio Sharpe series (20-day
    horizon, $\times\sqrt{12}$ on monthly non-overlapping returns unless
    noted). ``Vol-sized ranker'' $=$ production LightGBM scored with
    vol-adjusted features; ``equal-weight'' $=$ no inverse-vol weights
    within legs. Prose elsewhere points here.}
  \label{tab:sharpe-canonical}
  \small
  \begin{tabular}{@{}p{0.22\linewidth}p{0.38\linewidth}rrr@{}}
    \toprule
    \textbf{Label} & \textbf{Construction} &
    \textbf{ALL} & \textbf{DEV} & \textbf{FINAL} \\
    \midrule
    Shadow / falsification
      & Vol-sized ranker; equal-weight L/S;
        monthly walk-forward store ($n=109/95/14$)
      & 2.73 & 3.12 & 2.34 \\
    Policy grid: Ungated raw
      & Same ranker; equal-weight; policy store
        (monthly subsample)
      & 2.730 & 3.107 & 1.650$^{\dagger}$ \\
    Policy grid: Ungated vol-sized
      & Same ranker; inverse-vol within-leg weights
      & 2.614 & 2.991 & 1.603$^{\dagger}$ \\
    Policy grid: Variant~6 (gated)
      & Gate $+$ vol $+$ $\hat{e}$-cap (deployed)
      & 1.877 & 1.928 & 0.925$^{\dagger}$ \\
    \bottomrule
  \end{tabular}
  \smallskip
  \par\noindent\footnotesize
  $^{\dagger}$Policy-store FINAL uses $N=5$ complete months
  (Table~\ref{tab:ablation}); shadow FINAL uses 14 months---do not compare
  FINAL levels across rows without this sample difference.
\end{table}

\begin{figure}[H]
  \centering
  \includegraphics[width=0.92\linewidth]{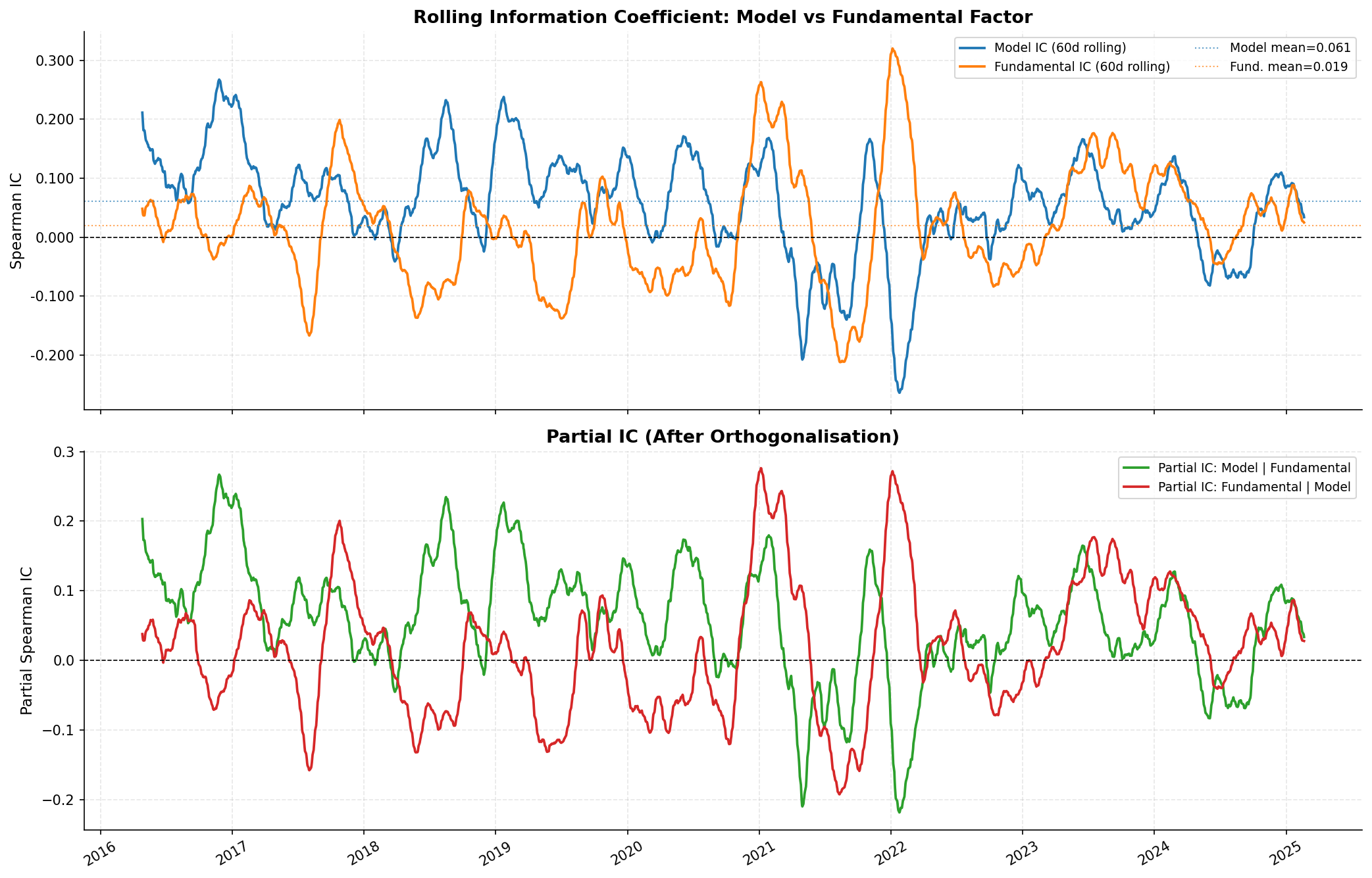}
  \caption{Rolling 60-trading-day mean of daily cross-sectional Spearman IC
    at the 20-day horizon (February~2016--February~2025). Top: model signal
    versus a fundamental-factor benchmark (dotted: full-sample means).
    Bottom: partial ICs after cross-sectional orthogonalisation. The FINAL
    holdout begins January~2024.}
  \label{fig:ic_timeseries_531}
\end{figure}

\paragraph{2024 regime context.}
The FINAL degradation reflects a regime disruption, not a fitting failure:
feature-importance rankings are structurally stable between early
(2016--2019) and late (2022--2023) refit windows (rank correlation $0.976$
at 20 and 60 days, $0.952$ at 90 days, identical top-three features;
Table~\ref{tab:feat-importance-531}, Appendix~\ref{app:extra-floats}), so
previously informative inputs became less predictive rather than the model
overfitting noise. Figure~\ref{fig:ic_timeseries_531} shows the rolling IC
deteriorating from 2024 onwards with a 2021 trough; silent IC degradation
under stable feature rankings is precisely the failure mode the regime-trust
gate targets.

\FloatBarrier
\subsection{DEUP Validation}
\label{sec:res-deup}

\noindent\textit{RQ1 (stock-level separation): per-prediction $\hat{e}(x)$
must rank failure risk distinct from volatility before any strategy-level
claim is meaningful.}

\paragraph{Per-prediction signal quality of $g(x)$.}
Table~\ref{tab:rho-canonical} is the single reference for every
$\rho(\hat{e},\ell)$ and $\rho(g,\ell)$ figure quoted in the paper; rows
differ by predictor ($g$ versus deployable $\hat{e}$), sample, and
post-processing. At 20 days the raw error predictor on the full walk-forward
panel attains $\rho(g,\ell)=0.192$ (row~1); the deployable $\hat{e}$ map
strengthens from DEV to FINAL ($0.142 \to 0.192$; rows~2--3), a different
partition readout than the ALL-pooled $g$ statistic. Longer horizons and
the rank-residualisation diagnostics are in rows~4--7. The dominant feature
of $g(x)$ by gain importance is the within-date \texttt{cross\_sectional\_rank}
of the score, not per-stock volatility or any market-state variable: the
error predictor detects ranking failure, not volatility-aliased noise.

\begin{figure}[!htbp]
  \centering
  \includegraphics[width=0.85\linewidth]{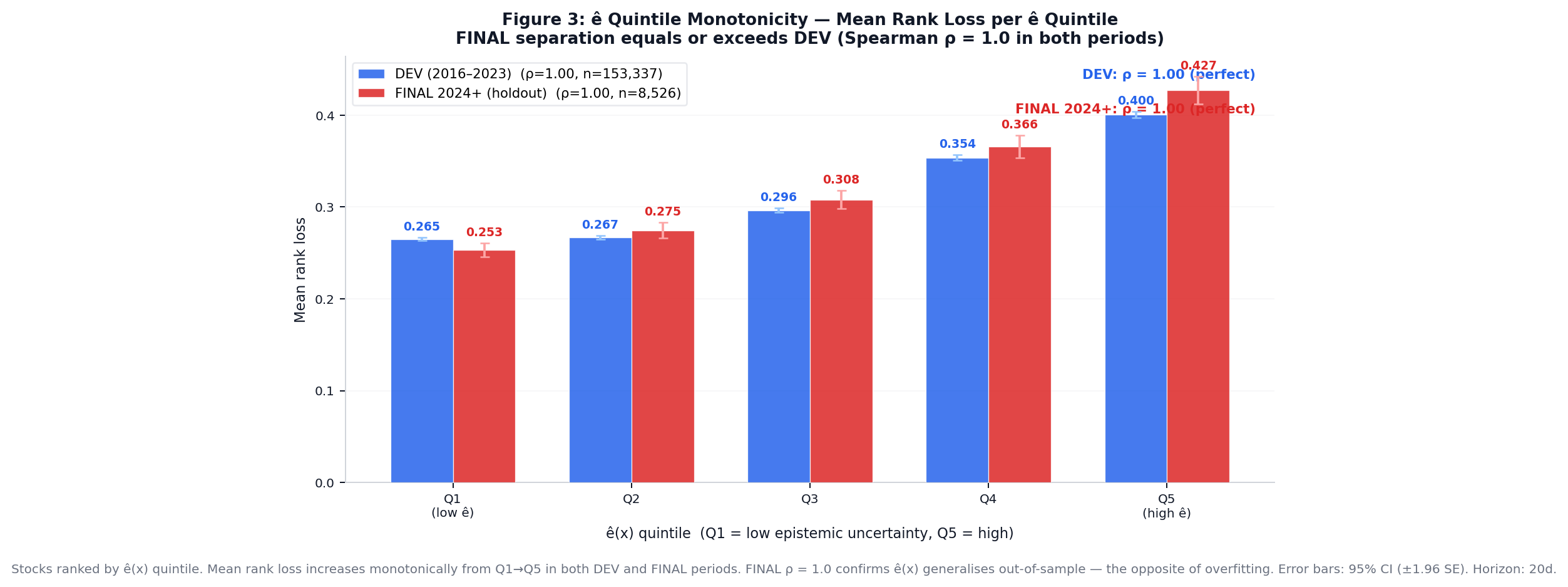}
  \caption{Bucket monotonicity of $\hat{e}(x)$ at the 20-day horizon.
    Observations are assigned to five $\hat{e}$-quintiles by pooled rank
    within each partition; bar height is mean realised rank-displacement
    loss (error bars: 95\% CI). Bucket-Spearman $\rho = 1.0$ in both DEV and
    FINAL (pooled $\rho(\hat{e},\ell)=0.192$ on FINAL); the same pattern
    holds at 60 and 90 days (not shown).}
  \label{fig:quintile_monotonicity_532}
\end{figure}

\paragraph{Holdout generalisation and monotonicity.}
Table~\ref{tab:ehat_holdout_532} reports the holdout rows of
Table~\ref{tab:rho-canonical} with acceptance thresholds; at every horizon
FINAL \emph{exceeds} DEV---the opposite signature of overfitting---with
strict quintile monotonicity (bucket Spearman $=1.0$) throughout
(Figure~\ref{fig:quintile_monotonicity_532}).

The magnitude $\rho \approx 0.19$ on ALL (row~1) is modest by deep-learning
DEUP standards \citep{lahlou2023}, but the rank-displacement target is
bounded in $[0,1]$ with structurally limited variance, and FINAL $\hat{e}$
clears the $0.15$ reference (Table~\ref{tab:ehat_holdout_532}).

\begin{table}[!htbp]
  \centering
  \caption{Canonical reconciliation of $\rho(\hat{e},\ell)$ and related
    association statistics. Each row is a distinct estimand; all prose
    references point here. Horizon $H{=}20$ unless noted; $\ell$ is
    realised rank-displacement loss unless noted.}
  \label{tab:rho-canonical}
  \small
  \begin{tabular}{@{}clllrr@{}}
    \toprule
    &
    \textbf{Predictor} &
    \textbf{Sample} &
    \textbf{Statistic} &
    \textbf{$H$} &
    \textbf{Value} \\
    \midrule
    1 & $g(x)$ error predictor & ALL (pooled) &
       Mean per-date $\rho(g,\ell)$ & 20 & 0.192 \\
    2 & $\hat{e}(x)$ deployable & DEV &
       Pooled $\rho(\hat{e},\ell)$ & 20 & 0.142 \\
    3 & $\hat{e}(x)$ deployable & FINAL &
       Pooled $\rho(\hat{e},\ell)$ & 20 & 0.192 \\
    4 & $g(x)$ before partialisation & ALL &
       Full $\rho(g,\ell)$ (rank-residualisation) & 20 & 0.186 \\
    5 & $g$ residuals vs.\ $\ell$ residuals & ALL &
       Partial $\rho_{\mathrm{partial}}$ & 20 & 0.178 \\
    6 & $\hat{e}$, LGB (RA2 comparison) & ALL &
       Mean per-date $\rho(\hat{e},\ell)$ & 20 & 0.144 \\
    7 & $\hat{e}$, Rank-Avg-2 (same study) & ALL &
       Mean per-date $\rho(\hat{e},\ell)$ & 20 & 0.194 \\
    \midrule
    \multicolumn{6}{@{}l}{\textit{Longer horizons (ALL pooled $g$; DEV/FINAL $\hat{e}$ in
      Table~\ref{tab:ehat_holdout_532})}} \\
    8 & $g(x)$ & ALL & $\rho(g,\ell)$ & 60 & 0.184 \\
    9 & $g(x)$ & ALL & $\rho(g,\ell)$ & 90 & 0.161 \\
    \bottomrule
  \end{tabular}
  \smallskip
  \par\noindent\footnotesize
  Rows~1 vs.\ 2--3: $g$ on ALL ($0.192$) is not contradictory with
  $\hat{e}$ on DEV ($0.142$)---different maps and partitions. Row~6 is the
  LGB arm of the controlled RA2 refit (Section~\ref{sec:res-deup}), not the
  headline ALL $g$ statistic. Rows~4--5: isotonic rank partialisation
  (Section~\ref{sec:res-deup}, robustness paragraph).
\end{table}

\begin{table}[!htbp]
  \centering
  \caption{$\hat{e}(x)$ validation across horizons. Holdout rows are
    Table~\ref{tab:rho-canonical} rows~2--3 (20d) and extensions at 60/90d.
    Bucket monotonicity is the Spearman correlation between $\hat{e}$-quintile
    index and within-quintile mean realised loss (five rank-ordered bucket
    means; acceptance $=1.0$ at 20 days). $\rho(\hat{e},\ell)$ columns are
    pooled cross-sectional Spearman on the frozen $\hat{e}$ store
    (distinct from mean-per-date cross-sectional $\rho$). PASS/FAIL assessed
    on FINAL at 20 days; $^{\dagger}$ same
    reference threshold applied to longer horizons for comparison; $^{\ast}$
    60d FINAL $\rho=0.140$ sits below the $0.15$ reference and passes only
    via bucket monotonicity.}
  \label{tab:ehat_holdout_532}
  \small
  \begin{tabular}{@{}lcccccc@{}}
    \toprule
    \textbf{Horizon} & $\boldsymbol{\rho(\hat{e},\,\ell)}$ \textbf{DEV} &
    $\boldsymbol{\rho(\hat{e},\,\ell)}$ \textbf{FINAL} &
    \textbf{$\Delta$ FINAL$-$DEV} &
    \textbf{Threshold} & \textbf{Bucket~$\rho$} & \textbf{Verdict} \\
    \midrule
    20d & 0.142 & \textbf{0.192} & $+0.050$ & $\geq$0.15 (FINAL)
        & \textbf{1.00} & \textbf{PASS} \\
    60d & 0.103 & \textbf{0.140} & $+0.037$ & $\geq$0.15$^{\dagger}$
        & 1.00 & PASS$^{\ast}$ \\
    90d & 0.138 & \textbf{0.248} & $+0.110$ & $\geq$0.15$^{\dagger}$
        & 1.00 & PASS \\
    \bottomrule
  \end{tabular}
\end{table}

\paragraph{Disentanglement scorecard.}
Table~\ref{tab:deup_scorecard_532} reports six pre-registered failure-mode
tests. Rows~A and~E are the headline separation checks; row~D is an
\emph{expected} negative under Theorem~\ref{thm:detectability}: per-stock
$\hat{e}$ cannot substitute for a strategy-level regime monitor.

\begin{table}[!htbp]
  \centering
  \caption{DEUP $\hat{e}(x)$ disentanglement diagnostic scorecard (six
    pre-registered failure-mode tests). Holdout $\rho(\hat{e},\ell)$ and
    quintile monotonicity are in Table~\ref{tab:ehat_holdout_532}.}
  \label{tab:deup_scorecard_532}
  \small
  \resizebox{\linewidth}{!}{%
  \begin{tabular}{@{}clp{0.34\linewidth}p{0.32\linewidth}c@{}}
    \toprule
    \textbf{ID} & \textbf{Diagnostic} & \textbf{Test} &
    \textbf{Result} & \textbf{Status} \\
    \midrule
    A & Disentanglement
      & $\hat{e}$ residualised on contemporaneous vol and 12m momentum;
        retest vs.\ rank displacement
      & $\rho(\hat{e}_{\text{resid}}, \ell) = 0.11$--$0.24$ across
        horizons; not subsumed by volatility
      & \textbf{PASS} \\
    B & Directional
      & $\hat{e}$ predicts error magnitude, not direction; high-$\hat{e}$
        should yield higher (not lower) realised IC
      & High-$\hat{e}$ quintile $\rightarrow$ elevated IC, consistent
        with rank-geometry coupling
      & \textbf{PASS} \\
    C & Failure detection
      & AUROC of $\hat{e}$ against high-loss-stock indicator
      & AUROC $= 0.53$--$0.61$ (positive separation across horizons)
      & \textbf{PASS} \\
    D & 2024 regime detection
      & Cross-sectional aggregate of $\hat{e}$ during the 2024
        thematic-rotation regime
      & Uniform across stocks; cannot serve as a regime detector
        (empirical realisation of Theorem~\ref{thm:detectability})
      & Expected \\
    E & Baseline comparison
      & $\hat{e}$ vs.\ vol, VIX percentile, $|\text{score}|$; ratio
        of correlations with $\ell$
      & $\hat{e}$ dominates baselines by 3--10$\times$ on held-out
        rank displacement
      & \textbf{PASS} \\
    F & Feature importance
      & Top driver of $g(x)$ should reflect ranking geometry, not
        volatility
      & \texttt{cross\_sectional\_rank} (gain importance 127) is the
        top feature
      & \textbf{PASS} \\
    \bottomrule
  \end{tabular}%
  }
\end{table}

\paragraph{Robustness to rank geometry.}
A sharper objection than volatility aliasing is that $g(x)$ merely
re-encodes the within-date cross-sectional rank (its top feature). A
per-date rank residualisation tests this directly: at each date an isotonic
map is fitted from rank to $g(x)$ and, separately, from rank to realised
displacement, and the partial Spearman between the residual series removes
the rank channel by construction (Table~\ref{tab:rho-canonical}, rows~4--5).
The association is almost undiminished: full $\rho(g,\ell)=0.186$ falls only
to $\rho_{\text{partial}} = 0.178$, a retention of $R = 0.955$
(block-bootstrap 95\% CI $[0.925, 0.985]$); FINAL retains $R = 0.919$, and
the 60/90-day horizons strengthen marginally ($R \approx 1.0$). Roughly
$95\%$ of $g(x)$'s predictive correlation is orthogonal to rank, and the
conclusion is base-model invariant (RA2 refit: $R = 0.85$--$0.87$). A
rank-decoupled variant built from the residual halves the score coupling
(median per-date $\rho(\hat{e},|\text{score}|)$: $0.616 \to 0.317$) while
preserving $92.5\%$ of the loss association, and is available where
position-size neutrality is desired. A complementary ablation adding
PIT-safe covariate-density features (Mahalanobis and nearest-neighbour
novelty) yields no improvement at any horizon
($|\Delta\rho_{\text{partial}}| \leq 0.001$ for LightGBM, $<0.005$ for
RA2): in a deliberately homogeneous walk-forward universe, evaluation
inputs are rarely novel relative to the expanding training set.

\paragraph{Cross-architecture robustness.}
Refitting the entire DEUP pipeline on the Rank-Average-2 (RA2) ensemble
ranker, whose 20-day FINAL median RankIC of $0.033$ is more robust than
LightGBM's $0.017$, yields $\rho(\hat{e}_{\text{RA2}},\ell) = 0.194$ at 20
days against $0.144$ for the LGB-based estimator on the same controlled
comparison (Table~\ref{tab:rho-canonical}, rows~6--7; a $35\%$ uplift;
$0.153$ vs.\ $0.106$ at 60 days, $44\%$), with FINAL strengthening at every
horizon. Portfolio outcomes converge once the
gate is applied: FINAL gated Sharpe $1.017$ (LGB) versus $0.958$ (RA2)
despite divergent raw ranker strength (ALL-period ungated $1.50$ vs.\
$0.62$); the per-stock signal is model-agnostic and the dominant economic
driver is the strategy-level gate. MC~Dropout and deep-ensemble comparators
are not fitted: both require a neural backbone, and a weak MLP comparator
would conflate the uncertainty estimator with the predictive architecture;
the RA2 test is the clean within-design robustness check.

\paragraph{HMM-feature negative ablation.}
A targeted ablation asks whether explicit Markov-switching state features
add information to $g(x)$ that the GBT backbone has not already extracted.
A 3-state Gaussian HMM is fitted per fold on expanding-window market
observables, and its posterior state probabilities, regime duration, and
entropy enter $g(x)$ either replacing the four explicit market-state
features (HMM-replace) or on top of them (HMM-augment), across 92 folds
and $172{,}963$ stock-level predictions at 20 days. Both deltas are
empirically null: $\Delta\rho = -0.00063$ (replace) and $-0.00009$
(augment) at 20 days, and $+0.00208$/$+0.00218$ at 60/90 days, all inside
the pre-registered $|\Delta\rho| > 0.005$ materiality band. LightGBM's
splits on \texttt{market\_vol\_21d} and on a 252-day VIX-percentile
feature (\texttt{vix\_percentile\_252d}) already extract what the HMM
recovers; market-side regime classifiers in
the tradition of \citet{hamilton1989} solve a different problem than DEUP
at this granularity, and the strategy-level comparison in
Section~\ref{sec:res-gate} points the same way.

\paragraph{Limitations and hand-off.}
Two constraints are structural: $\hat{e}$ is uniformly distributed across
stocks during the 2024 rotation (row~D); it detects per-stock displacement
risk \emph{within} a regime, not regime change, and the
$\rho(\hat{e},|\text{score}|)=0.616$ coupling rules out continuous inverse
sizing. $\hat{e}$ therefore cannot be a stand-alone exposure rule; its
deployment forms are the conformal normaliser
(Section~\ref{sec:res-conformal}) and the discrete tail cap inside the
two-level rule (Section~\ref{sec:res-gate}).

\subsection{Conformal Prediction Intervals}
\label{sec:res-conformal}

Normalising split-conformal scores by $\hat{e}(x)$ rather than by
contemporaneous volatility is the position-level test of whether
epistemic separation translates into calibrated risk statements
\citep[cf.][]{plassier2025}. Marginal coverage near the nominal $90\%$
masks severe conditional miscalibration under raw and volatility-normalised
scaling; DEUP normalisation compresses the high-versus-low uncertainty
coverage gap to $0.81\%$ (a $24.9\times$ reduction versus raw) while
narrowing mean width below the raw baseline
(Table~\ref{tab:conformal_533}).

\begin{table}[!htbp]
  \centering
  \caption{Conformal interval quality across normalisation strategies
    (oracle evaluation: DEUP-norm uses the same-date aleatoric floor;
    deployable PIT-safe spreads in Table~\ref{tab:conformal_deployable}).
    Nominal coverage $90\%$ ($\alpha = 0.1$). Conditional spread is the
    empirical coverage gap between highest- and lowest-uncertainty terciles;
    width ratio is top-tercile over bottom-tercile mean width.}
  \label{tab:conformal_533}
  \small
  \begin{tabular}{@{}lccccc@{}}
    \toprule
    \textbf{Variant} & \textbf{Coverage} & \textbf{ECE} &
    \textbf{Mean width} & \textbf{Cond. spread} & \textbf{Width ratio} \\
    \midrule
    Raw (no normalisation)
      & 90.0\% & 0.0001 & 0.675 & 20.2\% & 1.00 \\
    Vol-norm
      & 89.3\% & 0.0074 & 0.839 & 5.9\%  & 1.36 \\
    \textbf{DEUP-norm} ($\hat{e}(x)$)
      & \textbf{89.9\%} & \textbf{0.0009} & \textbf{0.647}
      & \textbf{0.81\%} & \textbf{1.57} \\
    \bottomrule
  \end{tabular}
\end{table}

\paragraph{Deployable PIT-safe variant.}
The oracle result uses the same-date aleatoric floor and is not deployable
without lookahead; a strict point-in-time audit replaces $a(t)$ with a
trailing-window percentile of \emph{matured} losses
($\hat{e}_{\text{PIT}}$, $W = 60$ trading days, $h = 20$ maturation
embargo). The embargo costs only $0.14$ percentage points against the
as-published trailing baseline and yields zero audited look-ahead
violations. Table~\ref{tab:conformal_deployable} implies a $4.94\times$
spread reduction for DEUP-norm versus raw (paired $\Delta = -0.175$, 95\%
CI $[-0.185, -0.164]$; improvement-factor CI $[4.13, 6.10]$), beating
vol-norm ($\Delta = -0.033$, CI $[-0.050, -0.016]$) at narrower mean width
($0.639$ vs.\ $0.842$). The edge is significant across all nine cells of a
calibration-window $\times$ nominal-level grid ($W\in\{40,60,120\}$,
$\alpha\in\{0.05,0.10,0.20\}$), all four aleatoric-baseline windows, a
second base model (RA2), and the 90-day horizon; marginal coverage is
$89.9\%$ on development and $90.6\%$ out-of-sample.

\begin{table}[!htbp]
  \centering
  \caption{Deployable conformal conditional-coverage spread (gap between
    highest- and lowest-$\hat{e}_{\text{PIT}}$ terciles; proportions).
    PIT normaliser: $W = 60$, $h = 20$ maturation embargo; 95\% moving-block
    bootstrap ($N = 1000$, block $= h$).}
  \label{tab:conformal_deployable}
  \small
  \begin{tabular}{@{}lccc@{}}
    \toprule
    \textbf{Variant} & \textbf{Cond.\ spread} & \textbf{95\% CI} &
    \textbf{Marg.\ coverage} \\
    \midrule
    Raw (no normalisation) & $0.2192$ & $[0.207,0.232]$ & $0.900$ \\
    Vol-norm               & $0.0777$ & $[0.062,0.093]$ & $0.893$ \\
    \textbf{DEUP-norm} ($\hat{e}_{\text{PIT}}$)
      & \textbf{0.0444} & $\mathbf{[0.035,0.055]}$ & \textbf{0.900} \\
    \bottomrule
  \end{tabular}
\end{table}

\paragraph{A calibrated risk read, not a sizing lever.}
The deployable half-width is emitted per (date, ticker) as
$\mathrm{width\_deup}$ with a within-date percentile
$\mathrm{risk\_score} \in [0,1]$, both computable at $t$. Because
$\mathrm{width\_deup} = q_{\mathrm{deup}}(t)\,\hat{e}_{\text{PIT}}$ with
the quantile constant within a date, a cross-sectional width cap is
identical to the $\hat{e}$-cap (per-date Spearman $\rho = 1.0$); what the
width adds is the probabilistic guarantee the cap lacks, realised loss
rises monotonically across risk-score quintiles from $0.236$ to $0.409$
(top-minus-bottom $0.173$, CI $[0.159, 0.188]$; per-observation
$\rho = 0.198$, CI $[0.179, 0.217]$). As an economic overlay, however, the
width is null: applied as an exposure throttle on the gated,
$\hat{e}$-capped Variant~6 it changes monthly-annualised Sharpe by
$\Delta = -0.038$ (inside the $0.10$ no-harm band) with a tail-CVaR CI
including zero, and an exhaustive search over eight uncertainty-overlay
levers---continuous $P(\text{good day})$ scaling, inverse-risk position
weighting, meta-RL combinability (scout kill), conjunction de-grossing,
cross-expert width allocation, stress/tail de-gross variants, conformal
risk control (CRC), and Ledoit--Wolf/residual-SNR non-participation---on
an extended dataset
(2017--2026, 15 non-overlapping out-of-sample months) found no lever whose
moving-block-bootstrap Sharpe or tail interval excludes zero. The strongest
candidate, residual-SNR non-participation, delivers out-of-sample
$\Delta\mathrm{Sharpe} = +0.013$ (95\% CI $[-0.19, +0.23]$), FINAL delta
$-0.11$, and width-beyond-score contribution $0.0002$ against a
pre-registered $0.05$ threshold: the score-magnitude filter explains all
apparent gain. A monotone IC-by-quintile mechanism exists in development
($\rho = 0.07$; Q5 signed-return IC $0.035$ vs.\ Q1 $-0.001$) but is too
weak to survive $2\times$ round-trip costs on an honest monthly window.
Split conformal, finally, guarantees marginal coverage under
exchangeability, not instance-conditional coverage
\citep{angelopoulos2023}: the $0.81\%$ spread is empirical, not a
worst-case bound.

\subsection{Regime-Trust Gate, Exposure Ablation, and the Aggregation Gap}
\label{sec:res-gate}

\noindent\textit{RQ1 (strategy-level monitoring and aggregation gap) and
RQ2 (deployment ablation): can realised-efficacy gating add value when
score-only aggregates cannot?}

This subsection reports three interlocking findings: $H(t)$ and $\Gamma(t)$
rank good versus bad IC days above chance and above VIX on the ALL
partition (Table~\ref{tab:regime_auroc}; consolidated map in
Table~\ref{tab:auroc-map}), but the signal is substantially
a persistence measurement of the label's own lagged EWMA; among gated
configurations, the binary gate dominates continuous throttling in the
six-variant ablation, with $\hat{e}(x)$ adding tail protection as a cap
and destroying value as an inverse-sizing signal; and cross-sectional
aggregates of per-prediction uncertainty are near-random regime
signals; the in-sample finance evidence for
Theorem~\ref{thm:detectability}.

\paragraph{$H(t)$ as a regime-trust classifier.}
Table~\ref{tab:regime_auroc} scores each predictor against the PIT-safe
binary target ``matured RankIC $> 0$'' (20-day label lag). $H(t)$ and
$\Gamma(t)$ improve from ALL to FINAL, whereas the VIX percentile is
near-random, market volatility is the best exogenous heuristic but weaker,
and mean stock volatility reverses in FINAL. Bucket analysis confirms perfect
rank-order monotonicity between $\Gamma(t)$ and realised RankIC (Spearman
$\rho = 1.0$ across four $\bar{\Gamma}$ buckets): the bottom bucket
($\bar{\Gamma} \approx 0.006$) has mean RankIC $-0.011$ with $51.2\%$ bad
days, the top ($\bar{\Gamma} \approx 0.939$) $+0.153$ with $11.5\%$. At the
binary operating point the gate achieves precision $80.0\%$, recall
$64.0\%$, abstaining on $47.2\%$ of days, a deliberate
precision-over-recall preference.

\begin{table}[H]
  \centering
  \caption{Regime-trust classifier performance. Binary target: matured
    RankIC $> 0$ (PIT-safe, 20-day lag).
    $N_\mathrm{ALL} = 2{,}218$; $N_\mathrm{FINAL} = 284$; positive-class
    rate 66\%. AUROC point estimates with moving-block bootstrap 95\% CIs
    (block $= 20$ trading days, matching label horizon; $1{,}000$ reps,
    seed~42, recomputed from the frozen \texttt{expert\_health\_lgb\_20d}
    store). Effective sample sizes $N_{\mathrm{eff}} \approx N/20$:
    $\approx 111$ (ALL), $\approx 14$ (FINAL). VIX and volatility predictors
    are inverted so that higher values predict a \emph{good} day.
    Mean-$g(x)$ aggregates appear in Table~\ref{tab:ood_finance}.}
  \label{tab:regime_auroc}
  \footnotesize
  \setlength{\tabcolsep}{4pt}
  \begin{tabularx}{\linewidth}{@{}%
    >{\raggedright\arraybackslash}p{0.20\linewidth}%
    >{\centering\arraybackslash}X%
    >{\centering\arraybackslash}X%
    >{\raggedright\arraybackslash}p{0.16\linewidth}@{}}
    \toprule
    \textbf{Predictor} &
    \textbf{AUROC (ALL) [95\% CI]} &
    \textbf{AUROC (FINAL) [95\% CI]} &
    \textbf{Note} \\
    \midrule
    \textbf{$H(t)$ combined}
      & \textbf{0.721} [0.671, 0.775]
      & \textbf{0.750} [0.624, 0.845]
      & Target system \\
    $H_{\text{realized}}$ only
      & 0.715 [0.665, 0.769]
      & 0.698 [0.549, 0.825]
      & Realised RankIC EWMA$^{\ddagger}$ \\
    \textbf{$\Gamma(t)$ gate score}
      & \textbf{0.710} [0.664, 0.760]
      & \textbf{0.743} [0.635, 0.826]
      & Rescaled $H(t)$ \\
    \midrule
    VIX percentile (252d)
      & 0.449 [0.381, 0.510]
      & 0.504 [0.349, 0.648]
      & Near-random \\
    Market vol (21d)
      & 0.596 [0.532, 0.661]
      & 0.569 [0.406, 0.757]
      & Best heuristic \\
    Mean stock vol (20d)
      & 0.590 [0.526, 0.652]
      & 0.460 [0.249, 0.666]
      & FINAL reversal \\
    \bottomrule
  \end{tabularx}
  \par\smallskip
  {\footnotesize
    $^{\ddagger}$FINAL $H_{\mathrm{realized}}$ AUROC was omitted from the
    frozen \S13.6 classifier JSON; the point estimate and CI above are
    recomputed from the frozen walk-forward health store for this revision.
    \emph{Label} lag-1 autocorrelation $\phi \approx 0.87$ on the same store
    (score-RankIC $\phi \approx 0.847$ in Section~\ref{sec:res-readiness}).
    FINAL CIs rest on $\approx 14$ effective blocks and are treated as
    directional only (moving-block bootstrap under-covers at this block
    count).}
\end{table}

\begin{table}[H]
  \centering
  \caption{AUROC reference map (which number is which). Use this table when
    citing headline discrimination statistics; targets and label maturities
    differ across rows---do not compare AUROCs across rows without matching
    the label column. Mean-$\bar{g}$ rows on matured RankIC use elevated
    mean-$\bar{g}$ $\Rightarrow$ lower matured efficacy (AUROC for good-day
    with inverted sign). $N_{\mathrm{eff}}$ applies block-bootstrap inference
    with 20-day blocks where noted.}
  \label{tab:auroc-map}
  \scriptsize
  \setlength{\tabcolsep}{2.5pt}
  \renewcommand{\arraystretch}{1.08}
  \begin{tabularx}{\linewidth}{@{}%
    >{\raggedright\arraybackslash}p{0.11\linewidth}%
    >{\raggedright\arraybackslash}p{0.24\linewidth}%
    >{\raggedright\arraybackslash}p{0.13\linewidth}%
    >{\centering\arraybackslash}p{0.09\linewidth}%
    >{\centering\arraybackslash}p{0.12\linewidth}%
    >{\centering\arraybackslash}p{0.11\linewidth}%
    >{\centering\arraybackslash}p{0.08\linewidth}@{}}
    \toprule
    \textbf{Signal} & \textbf{Target / label} & \textbf{Maturity} &
    \textbf{Window} & \textbf{$N$} & \textbf{$N_{\mathrm{eff}}$} &
    \textbf{AUROC} \\
    \midrule
    $H(t)$ & matured RankIC${}>0$ & 20d lag & ALL & 2218 & $\approx$111
      & 0.721 \\
    $H(t)$ & matured RankIC${}>0$ & 20d lag & FINAL & 284 & $\approx$14
      & 0.750 \\
    $H_{\text{realized}}$ & matured RankIC${}>0$ & 20d lag &
      ALL/FINAL & 2218/284 & $\approx$111/14 & 0.715/0.698 \\
    $H(t)$ & immature RankIC${}<0$ & same day & ALL & 2218 & ---
      & 0.523 \\
    Mean $\bar{g}$ & matured RankIC${}>0$ & 20d lag &
      ALL/FINAL & 1865/84 & --- & 0.526/0.427 \\
    Mean $\bar{g}$ & bad day (RankIC${}\leq 0$) & same day &
      ALL/FINAL & 1865/84 & --- & 0.55/0.750 \\
    Mean $\bar{g}$ & bad day (RankIC${}\leq 0$) & $+5$d lead & ALL
      & 1865 & --- & 0.55 \\
    $H_{\text{drift}}$ & $H_{\text{real}}<P_{30}$ & $+10$d lead & ALL
      & 2218 & --- & 0.637 \\
    Partial CUSUM ($k{=}5$) & efficacy break (event) & $k{=}5$ partial
      & DEV/ALL & 1865 & --- & 0.796 \\
    VIX pctile & matured RankIC${}>0$ & 20d lag & ALL/FINAL & 2218/284
      & --- & 0.449/0.504 \\
    CIFAR mean $\bar{g}$ & batch BCE broken & batch &
      95$\times$10k & 950k & --- & 0.955 \\
    CIFAR mean $\bar{g}$ & strict concept shift & batch &
      relabelled & 10k & --- & 0.500 \\
    \bottomrule
  \end{tabularx}
\end{table}

\FloatBarrier

\paragraph{AUROC inference and economic validation.}
The gate's headline AUROC is a \emph{persistence measurement}, not a
standalone economic proof: $H_{\mathrm{realized}}$ alone attains
$0.715$/$0.698$ (ALL/FINAL) versus $0.721$/$0.750$ for $H(t)$, and both
series are EWMAs of the same matured-RankIC target whose daily
lag-1 autocorrelation is $\phi \approx 0.87$ (measured on matured
RankIC in the health store). Block-bootstrap CIs therefore use
$N_{\mathrm{eff}} \approx 111$ on ALL, not the nominal $2{,}218$ daily
rows---the same overlap correction applied to the Platt calibration
result ($N_{\mathrm{eff}} \approx 14$ on FINAL; paragraph below). On FINAL,
moving-block bootstrap with $\approx 14$ effective blocks is treated as
directional only (under-coverage at this sample size). On ALL,
$H(t)$'s CI $[0.671, 0.775]$ sits above chance and above the VIX interval
$[0.381, 0.510]$; on FINAL ($N_{\mathrm{eff}} \approx 14$) intervals are
too wide to identify superiority over exogenous heuristics, and the
realised-only ablation overlaps $H(t)$ throughout. The operative economic
tests are therefore the deployment ablation
(Table~\ref{tab:ablation}), the multi-crisis panel
(Section~\ref{sec:res-crisis}), and the gating-cost reconciliation
paragraph below---not the pooled AUROC point estimate in isolation.

\paragraph{$H(t)$ component attribution.}
The gate is not a black-box regime heuristic: it is a fixed-weight composite
of three PIT-safe channels (drift sub-index
$H_{\text{drift,raw}} = 0.4\,\texttt{feat\_drift} + 0.3\,\texttt{score\_drift}
+ 0.3\,\texttt{corr\_spike}$, where \texttt{feat\_drift} is mean absolute
feature $z$-drift vs.\ a trailing 252-day reference,
\texttt{score\_drift} is a Kolmogorov--Smirnov statistic of today's score
distribution vs.\ the trailing 60-day pool, and \texttt{corr\_spike} is mean
pairwise correlation of trailing 20-day returns). Each channel is
expanding-$z$-scored as in Section~\ref{sec:method-gate}, then combined via
Eq.~\ref{eq:ht-combine}; higher drift or cross-expert disagreement
\emph{subtracts} from health; the realised-efficacy channel is the only
additive term. $\Gamma(t)$ and $G(t)$ are monotone transforms of this
scalar (Eqs.~\ref{eq:ht-combine}, \ref{eq:gamma-gate}); they do not introduce
additional learned weights.

Table~\ref{tab:ht-attribution} answers which channel carries the
regime-trust signal. The operative evaluation target matches
Table~\ref{tab:regime_auroc}: a \emph{good day} is matured RankIC $> 0$
(PIT-safe, 20-day label lag). On that target, $H_{\text{realized}}$ alone
achieves AUROC $0.715$/$0.698$ (ALL/FINAL;
Table~\ref{tab:regime_auroc}; FINAL recomputed from the frozen health
store---omitted in the \S13.6 JSON); $H(t)$ reaches $0.721$/$0.750$---gains of
only $+0.006$/$+0.052$ over the realised-only ablation, with overlapping
block-bootstrap CIs on FINAL. Neither $H_{\text{drift}}$ nor
$H_{\text{disagree}}$ appear as standalone good-day predictors in the
frozen component-attribution snapshot. A walk-forward logistic recombination
of the three $z$-channels (M1 fusion experiment) \emph{worsens} calibrated
Brier score
($0.212$ vs.\ $0.207$ at $H{=}20$), confirming that the hand-set
$(\alpha,\beta)=(0.3,0.3)$ weights are not hiding a learnable blend.

The fast channels are not useless---they serve different jobs.
$H_{\text{drift}}$ is a genuine \emph{leading} indicator of slow-channel
deterioration: predicting $H_{\text{realized}} < P_{30}$ at
$+5/+10/+15$ trading days yields AUROC $0.626/0.637/0.640$
(frozen two-channel audit snapshot); $H_{\text{disagree}}$ is noise at
all leads ($0.512/0.497/0.490$). Decomposing $H_{\text{drift}}$ at the
$+10$-day horizon shows opposite OOS stability: \texttt{feat\_drift}
(40\% weight) drops from DEV AUROC $0.655$ to FINAL $0.543$
($-11.2$~pp), while \texttt{score\_drift} rises $0.504 \to 0.650$
($+14.6$~pp) and \texttt{corr\_spike} $0.497 \to 0.582$ ($+8.5$~pp);
the composite $H_{\text{drift,raw}}$ falls $0.654 \to 0.566$. The
feat-drift degradation is consistent with an absolute 252-day
``normality'' reference that is less representative post-2024; the
self-referential score and correlation components remain informative.
Rebalancing drift weights toward score/corr improves lead-time FINAL
AUROC ($0.639$ vs.\ $0.566$) but \emph{degrades} same-day gate inputs
and was not adopted.

For transparency, same-day \emph{immature} RankIC${}<0$ classification
(a \textbf{different} label from the gate's matured good-day target) yields
near-random AUROC for every component on ALL ($H(t)$: $0.523$;
$H_{\text{realized}}$: $0.519$; $H_{\text{drift}}$: $0.524$;
$H_{\text{disagree}}$: $0.535$). This is expected: the gate is a
\emph{regime-level} throttle on matured efficacy, not a same-day tactical
forecaster (within-regime day-to-day $\rho(H,\text{RankIC}) \approx 0.02$
at 20~days). In sum: \textbf{the gate is dominated by the lagged realised
RankIC EWMA}; drift and disagreement provide a small same-day downward
adjustment to $H(t)$ and an operations-grade early-warning on
$H_{\text{drift}}$ spikes ($1.7\times$ precision lift for
$H_{\text{realized}}<P_{30}$ within 10 days), not incremental good-day
discrimination beyond persistence. The headline AUROC therefore validates
monitoring alignment, not incremental forecasting skill; portfolio and
window-level evidence (Tables~\ref{tab:ablation},
\ref{tab:multi_crisis_windows}) carry the economic test.

\begin{table}[!htbp]
  \centering
  \caption{$H(t)$ component attribution ($H{=}20$). Top panel: regime-trust
    target matured RankIC${}>0$ (gate operating label). Middle: association
    with same-day immature RankIC. Bottom: lead-time and drift sub-component
    AUROC for predicting $H_{\text{realized}}<P_{30}$ (research supplement;
    not the gate operating label).}
  \label{tab:ht-attribution}
  \small
  \begin{tabular}{@{}lcccp{3.6cm}@{}}
    \toprule
    Channel / metric & ALL & FINAL & Weight in $H(t)$ & Role \\
    \midrule
    \multicolumn{5}{@{}l}{\textit{Regime-trust: matured RankIC${}>0$ (gate target)}} \\
    $H(t)$ combined & 0.721 & 0.750 & --- & Production gate \\
    $H_{\text{realized}}$ only & 0.715 & 0.698
      & implicit $+1.0$ on $z_{\text{real}}$ & Near-sufficient \\
    $H_{\text{drift}}$, $H_{\text{disagree}}$ & --- & ---
      & $-0.3$ each on $z$ & Not in frozen good-day eval \\
    \midrule
    \multicolumn{5}{@{}l}{\textit{Same-day $\rho(\cdot,\text{RankIC})$ (immature label)}} \\
    $H_{\text{realized}}$ & \multicolumn{2}{c}{$+0.065$}
      & --- & Primary association \\
    $H_{\text{drift,raw}}$ & \multicolumn{2}{c}{$-0.039$}
      & 0.4/0.3/0.3 sub-weights & Mixed \\
    $H_{\text{disagree}}$ & \multicolumn{2}{c}{$-0.086$}
      & $-0.3$ on $z_{\text{disagree}}$ & Noisy \\
    \midrule
    \multicolumn{5}{@{}l}{\textit{Lead-time AUROC $\to H_{\text{real}}<P_{30}$}} \\
    $H_{\text{drift,raw}}$ & --- & ---
      & $-0.3$ on $z_{\text{drift}}$ &
      $+5/+10/+15$d $\to P_{30}$: 0.626/0.637/0.640 \\
    $H_{\text{disagree}}$ & --- & ---
      & $-0.3$ on $z_{\text{disagree}}$ &
      $+5/+10/+15$d: 0.512/0.497/0.490 \\
    \texttt{feat\_drift} & 0.655 & 0.543 & 0.4 in $H_{\text{drift}}$
      & OOS degradation \\
    \texttt{score\_drift} & 0.504 & 0.650 & 0.3 in $H_{\text{drift}}$
      & OOS gain \\
    \texttt{corr\_spike} & 0.497 & 0.582 & 0.3 in $H_{\text{drift}}$
      & OOS gain \\
    \bottomrule
  \end{tabular}
\end{table}

\paragraph{From ranking to a calibrated probability.}
$H(t)$ ranks days well but is severely miscalibrated as a probability (ECE
$\approx 0.27$; Brier skill score $-0.187$ against the $0.66$ base rate;
every reliability bin above the diagonal). A strictly walk-forward Platt
map \citep{platt1999}, trained at each date only on pairs whose 20-day
labels have matured, embargo asserted in code with zero violations across
$1{,}947$ dates, preserves the ranking (AUROC unchanged to $\pm 0.01$)
while reducing the Murphy reliability error from $0.073$ to $0.002$
($97\%$) and the Brier score from $0.266$ to $0.207$; the skill score moves
from $-0.187$ to $+0.106$, with paired
$\Delta\mathrm{Brier} = 0.067$ (moving-block bootstrap 95\% CI
$[0.037, 0.097]$, $p < 0.001$). FINAL reproduces the direction
($\Delta = 0.067$, $[-0.013, 0.135]$, $p = 0.047$) but with
$N_{\mathrm{eff}} \approx 14$ overlapping windows is directional
confirmation only. Isotonic regression \citep{zadrozny2002} attains
marginally lower in-sample error at higher walk-forward variance; the Platt
map is deployed. The gain is robust across good-day thresholds
$\tau \in \{-0.01, 0, +0.02\}$ (all $p < 0.001$) and vanishes only at the
per-period median threshold ($p = 0.18$), where the base rate approaches
one-half \citep[ECE with ten bins;][]{naeini2015}. The meta-layer therefore
consumes $\widehat{P}(\text{good day} \mid H(t))$, a number usable inside
an expected-value sizing rule.

\paragraph{Break taxonomy and partial circularity.}
\label{par:break-taxonomy}
The fast-versus-slow and partial-horizon comparisons condition on a frozen
event taxonomy (parameters locked in the private research archive before
any fusion or partial-horizon evaluation). \emph{Onsets} are defined
from realised efficacy alone: the first day a five-day rolling mean of
\texttt{daily\_rankic} crosses below the development-set 30th percentile,
sustained for at least three trading days, with a 20-day refractory gap
between distinct onsets ($H{=}20$ primary; 39 onsets on ALL). \emph{Typing}
uses only the 20 trading days before each onset $t_0$: an event is
\emph{covariate-visible} if the mean $z$-score of $H_{\text{drift,raw}}$ in
$[t_0{-}20,\,t_0)$ exceeds zero relative to its development mean and
standard deviation; \emph{concept-shift-only} otherwise ($59\%$ of ALL
onsets at $H{=}20$, i.e.\ 23 events; 21 of 33 DEV onsets; a secondary
``strongly covariate-visible'' cut uses mean pre-onset $z > 0.5$). The
partial-horizon RankIC channel and the slow
efficacy EWMA are excluded from this labeling.

This split carries \emph{partial circularity}: typing relies on the same
$H_{\text{drift,raw}}$ composite (feature, score, and correlation
$z$-drifts) that defines the fast channel under test, so concept-shift-only
events are pre-selected as those without a pre-onset elevation in that very
signal. Showing that the fast channel underperforms on concept-shift-only
windows is therefore partly definitional---not an independent falsification
of Corollary~\ref{cor:deployment}. What is not circular: (i)~onsets are
keyed to realised RankIC alone; (ii)~the $Y$-measurable partial-horizon
escape below is excluded from typing; (iii)~the asymmetry remains a
theory-aligned readout under a pre-registered split, not a surprise
discovery. A cleaner label would use ex ante model-feature PSI or KS shifts
that exclude score-drift and gate-score components; we report the
$H_{\text{drift,raw}}$ rule because it is the operational fast monitor.

\paragraph{Can a fast $X$-only channel close the blind window? A test of the
bound.}
$H(t)$ inherits a structural 20-day latency from label maturation. Fusing
the same-day drift channel $H_{\text{drift}}$ (lead AUROC $0.626$--$0.640$
at $+5/+10/+15$-day lags) with the slow realised-efficacy EWMA, via OR
gates, weighted blends, and hysteresis rules over a full hyperparameter
grid, produces a \emph{robust null}: median event-level lead time of $0$
trading days at a fair $10\%$ false-alarm operating point, blend AUROC
within $0.01$ of the slow channel at every horizon; the hysteresis rule at
least does no harm on false alarms ($5.8\%$ vs.\ $11.2\%$) while the OR
gate inflates them to $18.0\%$. The informative split is by the break-taxonomy discussion in
Section~\ref{sec:res-gate}---a direct test of
Corollary~\ref{cor:deployment}, with the circularity caveat noted there.
Primary deployment and gate calibration use the 20-day horizon; the
90-day horizon is reported in the audit because longer-horizon efficacy
breaks are where the fast--slow separation is largest (more concept-shift
events and a wider blind window), while headline gate AUROC remains at
$H{=}20$. At the 90-day horizon the fast channel is indistinguishable from
the slow channel on covariate-visible breaks ($\Delta\mathrm{AUROC} = -0.04$,
CI $[-0.18, +0.13]$) but materially worse on concept-shift-only breaks
($\Delta\mathrm{AUROC} = -0.26$, CI $[-0.38, -0.16]$), which comprise
$59\%$ of breaks at $H{=}20$. The latter gap is partly definitional (typing
uses the same drift composite as the fast channel); it nonetheless confirms
that an $X$-measurable channel carries information where the input law moves
($\delta_{N} > 0$) and is bounded where pre-onset drift did not move---the
detectability ceiling illustrated on financial data.

\paragraph{Binary versus continuous deployment.}
Scaling daily gross exposure by continuous $\Gamma(t)$ reduces crisis
maximum drawdown by roughly $60\%$ but produces a \emph{negative}
FINAL-period Sharpe ($-0.34$): the portfolio remains throttled through the
2024 recovery, destroying the convexity that compensates crisis drawdowns
\citep[cf.][]{nystrup2017}. Table~\ref{tab:ablation} ranks the discrete
policies. Ungated rows calibrate the cost of abstention; among gated
configurations, vol-sizing restores positive FINAL Sharpe relative to raw
sort and the $\hat{e}$-cap trims crisis tail risk. Variant~4 attains the
highest FINAL Sharpe \emph{point estimate} in the grid ($0.953$) via
residualised-$\hat{e}$ sizing, but with only $N=5$ complete monthly
returns the ordering is not identified (Table~\ref{tab:ablation}).
Variant~6 ($0.925$) is retained as the deployed policy for
interpretability and fixed-parameter robustness. Inverse-$\hat{e}$ sizing is
consolidated as null~\ref{sec:nulls} item~1.

\begin{table}[!htbp]
  \centering
  \caption{Deployment-policy ablation (20-day horizon). Sharpe
    $\times\sqrt{12}$ from non-overlapping monthly returns
    ($N_\mathrm{ALL} = 90$; $N_\mathrm{FINAL} = 5$ complete calendar months
    in the policy-evaluation store, versus $N_\mathrm{FINAL} = 14$ in the
    headline ungated series of Section~\ref{sec:res-base}; FINAL levels
    are not comparable across the two series). FINAL Sharpe$^\dagger$:
    point estimates only ($N=5$); no bootstrap CI is reported because
    five monthly blocks do not support stable inference---do not infer
    ordering between Variant~4 ($0.953$) and Variant~6 ($0.925$).
    Use ALL and crisis MaxDD for relative variant assessment. All
    parameters fitted on DEV and frozen. Crisis MaxDD from monthly series
    (Mar--Jul 2024). Literature comparators (V3, K) and the full variant
    grid appear in Table~\ref{tab:comparators} and
    Appendix~\ref{app:experiment-registry}.}
  \label{tab:ablation}
  \small
  \begin{tabular}{@{}llrrr@{}}
    \toprule
    & Variant & Sharpe (ALL) & Sharpe (FINAL)$^\dagger$ & Crisis MaxDD \\
    \midrule
    & Ungated raw           & 2.730 & 1.650 & $-7.4\%$ \\
    & Ungated vol-sized     & 2.614 & 1.603 & $-8.8\%$ \\
    \midrule
    1 & Gate + Raw sort     & 1.928 & $-0.322$ & $-7.4\%$ \\
    2 & Gate + Vol-sizing   & 1.886 &  0.375 & $-8.4\%$ \\
    4 & Gate + Resid-$\hat{e}$ & 1.987 & 0.953 & $-6.0\%$ \\
    \textbf{6} & \textbf{Gate + Vol + $\hat{e}$-Cap (P85)}
      & \textbf{1.877} & \textbf{0.925} & $\mathbf{-6.7\%}$ \\
    \bottomrule
  \end{tabular}
\end{table}

\paragraph{Gating cost versus validity alignment.}
The gate is scored against \emph{model health} (matured RankIC), not
realised P\&L; Table~\ref{tab:ablation} states the portfolio trade-off on
the frozen policy-grid store ($N_{\mathrm{ALL}}=90$ months;
Table~\ref{tab:sharpe-canonical}), distinct from the $n=109$ shadow
falsification series. Ungated raw and vol-sized policies dominate
Variant~6 in risk-adjusted return in both partitions
($2.730$/$1.650$ and $2.614$/$1.603$ versus $1.877$/$0.925$ ALL/FINAL),
while crisis maximum drawdown improves only modestly
($-7.4\%$/$-8.8\%$ to $-6.7\%$; complete grid in
Table~\ref{tab:experiment-policy-grid}). Variant~6's FINAL Sharpe would
fail the pre-registered $\geq 1.5$ bar in Table~\ref{tab:eval-metrics};
that bar was registered for the ungated ranker
(Section~\ref{sec:res-falsification}), and the deployed policy is judged
here and in Section~\ref{sec:res-crisis} instead. The pattern is coherent
with the 2024 confound flagged in Section~\ref{sec:res-base}: IC
deterioration under stable feature rankings while shadow returns remain
strong. What the gate
\emph{buys} is abstention aligned with validity rather than luck
($80.0\%$ bad-day precision at $47.2\%$ abstention;
Table~\ref{tab:regime_auroc}), a modest tail trim, and---in multi-book
deployments---a redeployable expert-health signal: Figure~\ref{fig:regime-gate}
specifies routing notional to active strategies rather than literal
cash-holding. What it \emph{costs} is carry whenever model-validity and
P\&L diverge---the IC-bad, P\&L-good episodes an allocator must either
accept as insurance or hedge elsewhere. Variant~6 is retained despite lower
Sharpe than ungated baselines because interpretability, fixed-parameter
robustness, and alignment with the validated monitoring object outweigh
incremental point-estimate optimisation on five monthly policy returns
(Table~\ref{tab:ablation}; Variant~4's $0.953$ FINAL Sharpe is not
statistically separated from Variant~6's $0.925$).

\paragraph{Literature comparators.}
\citet{liu2026} UA-Sort (V3) and a Trail-IC kill rule (K) are the
pre-registered portfolio benchmarks of Section~\ref{sec:method-eval}.
Table~\ref{tab:comparators} gives frozen policy-grid head-to-head Sharpe:
Trail-IC matches Gate~$+$~Raw to four decimals because
$w(t)=\min(1, c_{\mathrm{trail}}\,H_{\mathrm{realized}})$ is a \emph{single
scalar per date} that multiplies all scores before sort selection: rank order
is unchanged, so the long/short book and leg returns coincide with V1. On
gated days $\Gamma(t)\geq 0.20$ typically implies
$H_{\mathrm{realized}}\gtrsim 0.38$, so with $c_{\mathrm{trail}}\approx 8.30$
the weight saturates at $w(t)=1$ anyway; the exact Sharpe match is the
rank-invariance property, not exposure scaling. As implemented, Trail-IC is
\emph{degenerate} as an exposure rule: a date-level scalar that multiplies
all scores before sort is rank-preserving, so book identity with
Gate~$+$~Raw is guaranteed \emph{a priori} and K4 is a sanity check, not a
discriminating experiment. A non-degenerate exposure-scaling comparator is
the continuous-$\Gamma(t)$ throttle null (FINAL Sharpe $-0.34$; intro null
list). Kill criterion K4 is recorded as null~\ref{sec:nulls}
item~\ref{null:k4}. Definitions and the complete policy grid are in
Appendix~\ref{app:experiment-registry}.

\begin{table}[!htbp]
  \centering
  \caption{Head-to-head: literature comparators vs.\ Gate~$+$~Raw and the
    deployed policy (20-day horizon). Sharpe $\times\sqrt{12}$ from
    non-overlapping monthly returns; crisis MaxDD from the Mar--Jul~2024
    monthly subsample ($N = 3$). FINAL Sharpe$^\dagger$: $N=5$ months;
    point estimates only (Table~\ref{tab:ablation}). Source: frozen policy
    evaluation archive (February~2025 snapshot).}
  \label{tab:comparators}
  \small
  \begin{tabular}{@{}llrrrr@{}}
    \toprule
    & Variant & Sharpe (ALL) & Sharpe (DEV) & Sharpe (FINAL)$^\dagger$ & Crisis MaxDD \\
    \midrule
    V1 & Gate + Raw (benchmark)        & 1.928 & 2.039 & $-0.322$ & $-7.4\%$ \\
    V3 & Gate + UA-Sort \citep{liu2026}
       & 1.933 & 2.046 & $-0.322$ & $-7.4\%$ \\
    K  & Gate + Trail-IC (K4 kill)      & 1.928 & 2.039 & $-0.322$ & $-7.4\%$ \\
    \addlinespace
    \textbf{V6} & \textbf{Gate + Vol + $\hat{e}$-Cap (deployed)}
       & \textbf{1.877} & \textbf{1.928} & \textbf{0.925} & $\mathbf{-6.7\%}$ \\
    \bottomrule
  \end{tabular}
\end{table}

\paragraph{The aggregation gap.}
\label{sec:agg_gap}

\noindent\textit{RQ1 (strategy-level): pooling per-stock scores is the
natural aggregation test; Theorem~\ref{thm:detectability} predicts failure
under strict concept shift.}

If $g(x)$ correctly ranks which stocks the model is most wrong about, why
not aggregate the per-stock scores into a date-level regime signal?
Table~\ref{tab:ood_finance} fixes the evaluation convention for $X$-only
aggregates here: target $\mathbb{1}[\text{daily RankIC} < 0]$, higher score
$=$ more likely bad day (elevated mean-$g$ $\Rightarrow$ elevated expected
error). Under that orientation the long-panel answer is near chance, with a
short-window covariate caveat below.

First, the daily cross-sectional median, 90th percentile, and inter-quartile
spread of $g(x)$ and $\hat{e}(x)$ achieve bad-day AUROC $\approx
0.48$--$0.52$, with Spearman correlations to future RankIC within $\pm 0.03$
of zero. Second, the PIT-safe daily mean $\bar{g}_{\mathrm{agg}}$ attains
same-day bad-day AUROC $\approx 0.55$ over ALL and $0.750$ in FINAL
(Table~\ref{tab:ood_finance}); a $+5$-day lead bad-day AUROC remains
$\approx 0.55$. The ALL-panel reading is the operative one for deployment:
$0.55$ is indistinguishable from chance over $N_{\mathrm{ALL}} = 1{,}865$
days. The FINAL $0.750$ coincides with concurrent covariate shift
($\delta_N > 0$) and, on that $N = 84$ slice, $\bar{g}_{\mathrm{agg}}$ even
edges $\Gamma(t)$ ($0.710$). In real markets breaks are often
covariate-accompanied, so $X$-only aggregates may carry incidental signal;
Theorem~\ref{thm:detectability} says you cannot \emph{rely} on that channel
under strict concept shift. Third, the named OOD-accuracy predictors (AC,
ATC, DoC, GDE) adapted to the daily equity cross-section remain near chance
on ALL (Table~\ref{tab:ood_finance}); only efficacy channels built from
matured outcomes ($\Gamma(t)$ and realised RankIC) clear the matured
good-day gate in Table~\ref{tab:regime_auroc}. \textbf{Label discipline}
(Table~\ref{tab:auroc-map}): on the gate's operating target (matured
good-day), $H(t)$ reaches $0.721$ while $\bar{g}_{\mathrm{agg}}$ reaches
only $0.526$ on the overlapping $\hat{e}$ store ($N_{\mathrm{ALL}} =
1{,}865$); no $X$-only aggregate exceeds $\approx 0.55$ on any label
tested (same-day bad-day $\bar{g}\approx 0.55$; matured good-day
$\bar{g}\approx 0.53$). On the \emph{same-day} bad-day target,
$H(t)$ itself is near chance ($0.523$), statistically tied with
$\bar{g}_{\mathrm{agg}}$---the separation is therefore
\emph{label-dependent}: efficacy channels win on matured outcomes; score
aggregates do not beat chance on any target, and the gate does not beat
$\bar{g}$ on the $X$-only label. The mechanism is
Eq.~(\ref{eq:g-marginalisation}): $g$ marginalises over the latent regime,
so under concept shift its cross-sectional mean is approximately invariant
by construction. The RA2 replication reproduces the pattern, indicating the
gap is architectural rather than a LightGBM artefact.

\begin{table}[!htbp]
  \centering
  \caption{Named OOD-accuracy predictors as bad-day detectors in the equity
    cross-section. Target: $\mathbb{1}[\text{daily RankIC} < 0]$ (higher
    score $=$ more likely bad day; the orientation used throughout
    Section~\ref{sec:agg_gap} for $X$-only aggregates). $N_{\mathrm{ALL}} =
    1{,}865$ trading days (2016-12 to 2024-05); $N_{\mathrm{FINAL}} = 84$
    (Jan--May~2024), capped by the $\hat{e}$ prediction store, shorter than
    Table~\ref{tab:regime_auroc}. Confidence proxy
    $\mathrm{conf}_{i} = 1 - \hat{e}_{\mathrm{pctile},i}$;
    $\bar{g}_{\mathrm{agg}}$ is the daily mean of $g$. COT and nuclear norm
    lack a faithful one-dimensional finance analogue; entropy has no
    PIT-safe date overlap.}
  \label{tab:ood_finance}
  \small
  \begin{tabular}{@{}lccl@{}}
    \toprule
    Predictor & AUROC (ALL) & AUROC (FINAL) & Channel \\
    \midrule
    AC$_{\mathrm{fin}}$  & 0.435 & 0.487 & $X$-only \\
    ATC$_{\mathrm{fin}}$ & 0.474 & 0.473 & $X$-only \\
    DoC$_{\mathrm{fin}}$ & 0.435 & 0.487 & $X$-only \\
    GDE$_{\mathrm{fin}}$ & 0.527 & 0.636 & $X$-only \\
    $\bar{g}_{\mathrm{agg}}$ & 0.554 & 0.750 & $X$-only \\
    \midrule
    $H_{\mathrm{drift}}$ (feature drift) & 0.517 & 0.588 & $\delta_N > 0$ \\
    $\Gamma(t)$ & 0.507 & 0.710 & efficacy ($Y$) \\
    Matured RankIC ($H_{\mathrm{realized}}$) & 0.508 & 0.836
      & efficacy ($Y$) \\
    \bottomrule
  \end{tabular}
\end{table}

\subsection{Partial-Horizon RankIC Monitoring}
\label{sec:partial-horizon}

\noindent\textit{RQ1 (constructive escape): a $Y$-measurable channel outside
the score-only ceiling mandated by Corollary~\ref{cor:deployment}.}

The partial-horizon channel is the paper's constructive answer to
Theorem~\ref{thm:detectability}: a $Y$-measurable statistic excluded from
the break-taxonomy discussion in Section~\ref{sec:res-gate} and therefore
outside the score-only ceiling.

\paragraph{Definition and PIT timing.}
For horizon $h$ and partial window $k<h$, define
\begin{equation}
  \texttt{partial\_rankic}_{k}(t)
  \;=\;
  \operatorname{Spearman}\!\Bigl(
    \{\text{score}_{i,t}\},\,
    \{r^{(k)}_{i,t}\}\Bigr),
  \label{eq:partial-rankic}
\end{equation}
the within-date Spearman correlation between scores and realised $k$-day
forward returns. The statistic is observable at $t{+}k$ trading days---$h{-}k$
days before the matured $h$-day RankIC that feeds $H_{\text{realized}}$---so
it supplies early efficacy information without using contemporaneous
$X$-only aggregates.

\paragraph{Implementation (frozen on DEV).}
The production-grade detector applies three layers, all calibrated on DEV
only:
\begin{enumerate}[leftmargin=*,itemsep=2pt]
  \item \textbf{Partial window} $k=5$ at $h=20$ (15-day lead over matured
        RankIC; AUROC stable on a $k\in\{3,5,10\}$ grid).
  \item \textbf{Smoothing:} EWMA with half-life $10$ trading days on the
        reindexed partial stream (robust to HL $\in\{3,5,10,15\}$;
        AUROC span $<0.08$).
  \item \textbf{Sequential detector:} Page CUSUM on the daily partial stream,
        $C_t = \max\{0,\, C_{t-1} + (\mu_0 - x_t) - \kappa\}$ with
        $\kappa = 0.5\sigma$ (DEV std of the partial stream) and alarm
        boundary set to a $10\%$ DEV false-alarm rate (matched to the slow
        channel audit).
\end{enumerate}

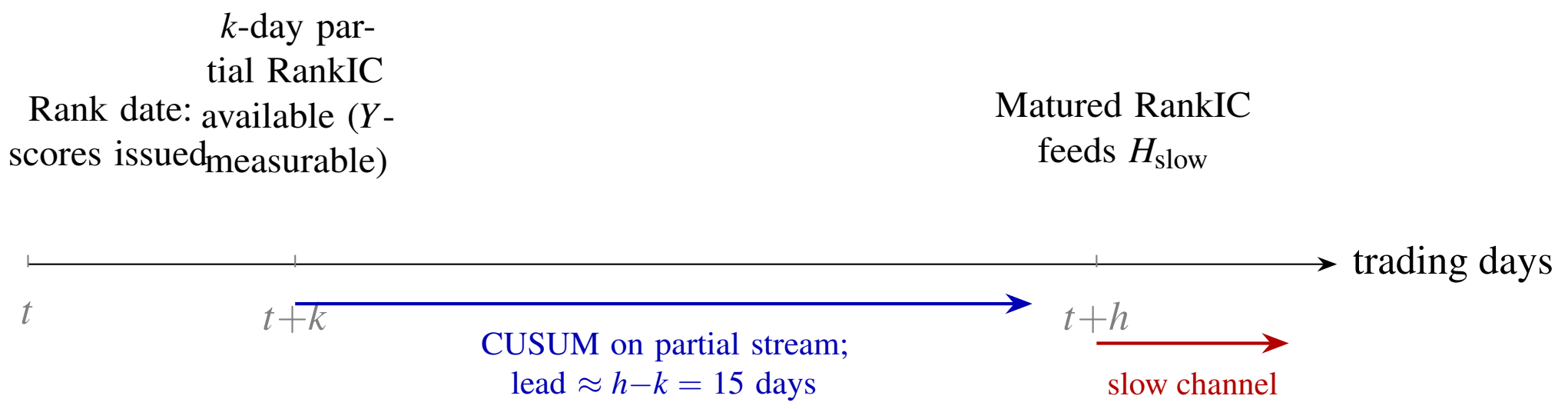
\begin{figure}[!htbp]
  \centering
  \begin{tikzpicture}[x=0.46cm,y=1cm,>=Stealth,
    stage/.style={font=\footnotesize, align=center, text width=2.55cm}]
    \draw[->] (0,0) -- (24.5,0) node[right,font=\small] {trading days};
    \foreach \x/\lab in {0/$t$, 5/$t{+}k$, 20/$t{+}h$} {
      \draw[dashed,gray] (\x,0.08) -- (\x,-0.08)
        node[below=3pt,font=\small] {\lab};
    }
    \node[stage] at (1.5,1.15) {Rank date:\\scores issued};
    \node[stage] at (5,1.45) {$k$-day partial RankIC\\available ($Y$-measurable)};
    \node[stage] at (20.5,1.15) {Matured RankIC\\feeds $H_{\text{slow}}$};
    \draw[blue!70!black,thick,->] (5,-0.34) -- (18.8,-0.34)
      node[midway,below=3pt,font=\scriptsize,align=center,text width=3.4cm]
      {CUSUM on partial stream;\\lead $\approx h{-}k = 15$ days};
    \draw[red!70!black,thick,->] (20,-0.68) -- (23.6,-0.68)
      node[midway,below=3pt,font=\scriptsize] {slow channel};
  \end{tikzpicture}
  \caption{Partial-horizon monitoring timeline ($h=20$, $k=5$). The CUSUM
    fires on the partial-efficacy stream up to $h{-}k$ trading days before
    the matured label that drives $H_{\text{realized}}$, realising the
    $Y$-conditioning escape from Theorem~\ref{thm:detectability}.}
  \label{fig:partial-horizon-timeline}
\end{figure}

\paragraph{Results.}
On DEV/ALL the partial CUSUM attains date-level AUROC $0.796$ (CI
$[0.747, 0.841]$) against the slow channel's $0.714$ (paired
$\Delta{=}+0.079$ ($[0.030, 0.132]$)) at a $9.9\%$ false-alarm rate
versus $8.6\%$. On concept-shift-only windows---the non-tautological half
of the taxonomy, because typing did not use partial RankIC---AUROC is
$0.760$ versus $0.393$ for the $X$-only fast channel (paired
$\Delta{=}+0.367$ ($[0.194, 0.537]$)). At equal false-alarm budget the
partial channel detects $10$ of $21$ concept-shift-only \emph{DEV}
events versus the slow channel's $5$, with median lead $+25$ trading days ($[18, 32]$) on
jointly detected events. On covariate-visible breaks the channels are
indistinguishable ($\Delta = -0.002$). The direction holds out-of-sample
(FINAL AUROC $0.775$ vs.\ $0.685$). Table~\ref{tab:auroc-map} situates
this AUROC against the gate's matured good-day statistics.

\subsection{Multi-Crisis Robustness}
\label{sec:res-crisis}

\noindent\textit{RQ3 (regime generalisation): model-health versus market-fear
across eight frozen crisis and calm windows.}

A pooled-sample AUROC cannot exclude over-fitting to the single prominent
2024 episode; the eight-window panel of
Table~\ref{tab:multi_crisis_windows}, five stress episodes and three calm
references, 2018--2024, on frozen walk-forward outputs with no
window-specific recalibration, tests whether the gate's discrimination
generalises.

\begin{table}[!htbp]
  \centering
  \caption{Multi-crisis window definitions and per-window results (20-day
    horizon). IC (20d) $=$ mean daily matured RankIC; $\bar{\Gamma}$ $=$
    mean continuous gate score; \%Abs $=$ share of days with
    $\Gamma < 0.20$; VIX\% $=$ mean 252d VIX percentile. $G$ verdicts:
    CA $=$ Correctly Active; Abs $=$ Correctly Abstaining; FA $=$ False
    Alarm; MC $=$ Missed Crisis (from window-mean $\bar{\Gamma}$ and
    window-mean IC, not day-level majority votes: a window can be
    ``correctly active'' while abstaining on most days, e.g.\ Calm~2018:
    CA verdict but $61.8\%$ daily abstention). Day-level gate accuracy
    appears in the \%Abs column. VIX benchmark: abstain if the VIX percentile
    exceeds the 67th on more than half of window days.}
  \label{tab:multi_crisis_windows}
  \small
  \begin{tabular}{@{}llrrrrll@{}}
    \toprule
    Window & Period & IC (20d) & $\bar{\Gamma}$ & \%Abs & VIX\%
      & G verdict & VIX verdict \\
    \midrule
    \multicolumn{8}{@{}l}{\textit{Crisis episodes}} \\
    COVID recovery      & Jun--Dec 2020 & $+0.062$ & 0.375 & 47.3 & 52.0
      & CA & CA \\
    Meme mania          & Jan--Sep 2021 & $-0.040$ & 0.210 & 73.4 & 95.8
      & \textbf{MC} & Abs \\
    Inflation shock     & Jan--Jun 2022 & $-0.024$ & 0.077 & 85.5 & 71.8
      & Abs & Abs \\
    Late rate-hiking    & Jul--Dec 2023 & $+0.034$ & 0.381 & 39.7 & 94.3
      & CA & \textbf{FA} \\
    AI rotation         & Mar--Jul 2024 & $-0.013$ & 0.123 & 76.2 & 95.1
      & Abs & Abs \\
    \midrule
    \multicolumn{8}{@{}l}{\textit{Calm reference periods}} \\
    Calm 2018           & Jan--Dec 2018 & $+0.088$ & 0.323 & 61.8 & 63.6
      & CA & CA \\
    Calm 2019           & Jan--Dec 2019 & $+0.122$ & 0.566 & 14.7 & 83.2
      & CA & \textbf{FA} \\
    Calm H1 2023        & Jan--Jun 2023 & $+0.104$ & 0.486 & 10.5 & 97.1
      & CA & \textbf{FA} \\
    \midrule
    \multicolumn{6}{@{}l}{Score} & \textbf{7/8 (87.5\%)} & 5/8 (62.5\%) \\
    \bottomrule
  \end{tabular}
\end{table}

$G(t)$ scores $7/8$ against $5/8$ for the VIX benchmark (crisis episodes
alone: $4/5$ each). Verdicts are window-mean summaries: Calm~2018 is
``correctly active'' because mean $\bar{\Gamma}=0.323>0.20$ despite
abstaining on $61.8\%$ of days; day-level precision/recall differ from the
window scorecard. The decisive divergence is on profitable high-VIX
windows, where the VIX gate false-alarms three times. In H2~2023 the mean
VIX percentile was $94.3\%$ yet the model generated 20-day IC $+0.034$
(rising to $+0.108$ / $+0.159$ at 60/90 days); $G(t)$ stayed active
($\bar{\Gamma} = 0.381$) while a VIX gate would have suppressed one of the
strongest periods. The same pattern recurs in 2019 (IC $+0.122$, VIX
percentile $83.2\%$, $\bar{\Gamma} = 0.566$) and H1~2023 (IC $+0.104$, VIX
percentile $97.1\%$, $\bar{\Gamma} = 0.486$). Conversely $G(t)$ correctly
abstains in both unambiguous deterioration episodes: the 2022 inflation
shock ($\bar{\Gamma} = 0.077$, $85.5\%$ abstention; 90-day IC $-0.118$)
and the 2024 AI rotation ($\bar{\Gamma} = 0.123$, $76.2\%$ abstention),
where short-horizon pricing broke (20d IC $-0.013$) while longer horizons
stayed marginally positive (60d $+0.070$, 90d $+0.142$), a
horizon-specific precision unavailable to a single VIX signal.

The single failure is the 2021 meme-stock dislocation: mean
$\bar{\Gamma} = 0.210$ barely exceeds the threshold while the 20-day IC of
$-0.040$ is marginally negative. Two mitigations: the gate was already
abstaining on $73.4\%$ of individual days, and the deterioration is mild
($-0.040$ at 20d is within sampling noise at $N \approx 100$; 60/90-day
ICs are near zero at $-0.009$/$-0.010$). The VIX gate abstained correctly
here, but through a mechanism that does not generalise, retail
speculation inflating market-wide implied volatility without systematic
factor failure. With eight windows, a Fisher exact test does not reject
equal accuracy on the five crisis windows ($p > 0.20$); the panel's value
is primarily qualitative: the 2024 behaviour is not an isolated artefact,
the structural mechanism (model health versus market anxiety) is
documented, and the one failure mode (slow EWMA in margin-call-style
dislocations) is identified.

\subsection{Factor Attribution and Statistical Falsification}
\label{sec:res-falsification}

\noindent\textit{RQ3 (falsification): data-mining, serial dependence, fat
tails, and factor-mimicking objections on the frozen walk-forward panel.}

An allocator's first objection to a reported Sharpe of $2.73$ is not
scepticism of the point estimate but of its \emph{generators}: serial
dependence, fat tails, model mining, and factor replication. The
\emph{ungated} headline series of Section~\ref{sec:res-base}---not the
deployed gated policy (Variant~6 at $0.925$ FINAL Sharpe on the policy
store)---is subjected to the pre-registered portfolio-performance battery of
Table~\ref{tab:eval-metrics}
(Table~\ref{tab:falsification_536}). Gate adoption trades headline Sharpe for
validity alignment; Variant~6 is evaluated by gate AUROC, ablation trade-offs,
and the multi-crisis panel instead (Sections~\ref{sec:res-gate}
and~\ref{sec:res-crisis}).

\begin{table}[H]
  \centering
  \caption{Statistical falsification scorecard for the ungated equal-weight
    shadow portfolio on the vol-sized LightGBM ranker (109 non-overlapping
    monthly returns, February~2016--February~2025;
    Table~\ref{tab:sharpe-canonical}). Portfolio thresholds in
    Table~\ref{tab:eval-metrics} apply to this series only; deployed
    Variant~6 ($0.925$ FINAL Sharpe on five policy-store months) is assessed
    in Sections~\ref{sec:res-gate} and~\ref{sec:res-crisis}. HAC/PSR/DSR per
    \citet{bailey2012,bailey2014}; permutations are PIT-safe within-date
    shuffles ($5{,}000$ trials); White RC uses $10{,}000$ centred-bootstrap
    resamples ($K = 5$).}
  \label{tab:falsification_536}
  \footnotesize
  \setlength{\tabcolsep}{3pt}
  \renewcommand{\arraystretch}{1.05}
  \begin{tabularx}{\linewidth}{@{}%
    >{\raggedright\arraybackslash}p{0.28\linewidth}%
    >{\centering\arraybackslash}p{0.07\linewidth}%
    >{\raggedright\arraybackslash}X%
    >{\centering\arraybackslash}p{0.14\linewidth}%
    >{\centering\arraybackslash}p{0.07\linewidth}@{}}
    \toprule
    \textbf{Test} & \textbf{Period} & \textbf{Realised statistic} &
    \textbf{Threshold} & \textbf{Verdict} \\
    \midrule
    HAC-adjusted Sharpe (Newey--West, 12 lags)
      & ALL    & $\mathrm{SR}_{\mathrm{HAC}} = 2.79$
      & $\geq 1.5$ & \textbf{PASS} \\
    HAC-adjusted Sharpe
      & FINAL  & $\mathrm{SR}_{\mathrm{HAC}} = 2.46$
      & $\geq 1.5$ & \textbf{PASS} \\
    Probabilistic Sharpe (PSR)
      & ALL    & $z = 6.95$, $\mathrm{PSR} > 99.999\%$
      & $\geq 0.95$ & \textbf{PASS} \\
    Probabilistic Sharpe (PSR)
      & FINAL  & $z = 5.08$, $\mathrm{PSR} > 99.999\%$
      & $\geq 0.95$ & \textbf{PASS} \\
    Deflated Sharpe ($N = 5$, $\mathrm{SR}^{*} = 0.469$)
      & ALL    & $\mathrm{DSR} > 99.999\%$
      & $\geq 0.95$ & \textbf{PASS} \\
    Deflated Sharpe ($N = 5$)
      & FINAL  & $\mathrm{DSR} = 99.995\%$
      & $\geq 0.95$ & \textbf{PASS} \\
    Permutation null (IC)
      & ALL    & $0/5{,}000$, $p < 0.0002$
      & $\leq 0.05$ & \textbf{PASS} \\
    Permutation null (IC)
      & FINAL  & $77/5{,}000$, $p = 0.0154$
      & $\leq 0.05$ & \textbf{PASS} \\
    Permutation null (gross Sharpe)
      & ALL    & $0/5{,}000$, $p < 0.0002$
      & $\leq 0.05$ & \textbf{PASS} \\
    Permutation null (gross Sharpe)
      & FINAL  & $43/5{,}000$, $p = 0.0086$
      & $\leq 0.05$ & \textbf{PASS} \\
    White's Reality Check ($10{,}000$ bootstrap, $K = 5$)
      & ALL    & $V_{n} = 8.24$, $0/10{,}000$, $p < 0.0001$
      & $\leq 0.05$ & \textbf{PASS} \\
    Bonferroni (FWER, $K = 5$ candidates)
      & ALL    & all five at $p < 0.0001$
      & $\leq 0.05$ & \textbf{PASS} \\
    FF5 alpha $t_{\mathrm{HAC}}$ (long--short)
      & ALL    & $\alpha = +87.4\%$~p.a., $t_{\mathrm{HAC}} = 9.34$
      & $\geq 3.0$ & \textbf{PASS} \\
    FF5 + AI-thematic factor
      & ALL    & $\alpha = +78.4\%$~p.a., $t_{\mathrm{HAC}} = 10.28$;
        $\beta_{\mathrm{AI}} = 0.956$, $t = 1.65$
      & $\geq 3.0$ ($\alpha$) & \textbf{PASS} \\
    \bottomrule
  \end{tabularx}
\end{table}

\FloatBarrier

\paragraph{Provenance of the headline Sharpe.}
To prevent misreading, the provenance of the headline figure is stated
explicitly (Table~\ref{tab:sharpe-canonical}). The Sharpe ratio of
$2.73$ (ALL) / $2.34$ (FINAL) is a \emph{monthly, non-overlapping
long--short Sharpe} on the \emph{ungated equal-weight shadow portfolio}
driven by the \textbf{vol-sized LightGBM ranker} (long leg minus short leg
over 109 monthly folds; 14 FINAL months). It matches the policy-grid
ungated raw row at ALL ($2.730$) and is \emph{not} the inverse-vol-within-leg policy
($2.614$ ALL). It is not Variant~6, not a daily deployed-capital Sharpe,
and not a factor-hedged P\&L series. The HAC-adjusted FINAL reading
($2.34 \to 2.46$) clears the pre-registered $\geq 1.5$ bar; Variant~6
($0.925$ FINAL on five policy-store months) would fail that same bar by
design. The Fama--French five- and six-factor regressions reported below are
\emph{attribution} tests that decompose this return stream into
factor-spanned and residual ($\alpha$) components for inference; they are
not hedges applied to the returns, and the residual $\alpha$ they recover
is not a separately traded factor-neutral portfolio. Annualisation and
serial-dependence corrections for this monthly series are handled by the
HAC, PSR, and DSR analysis that follows.

\paragraph{Sharpe robustness: HAC, PSR, DSR.}
Newey--West HAC adjustment (12 monthly lags) moves the Sharpe marginally
\emph{up} ($2.73 \to 2.79$ ALL; $2.34 \to 2.46$ FINAL), indicating mild
negative rather than inflating positive serial dependence (ACF diagnostics:
lag-1 $+0.147$, Ljung--Box $p = 0.12$; lag-2 $-0.20$, $p = 0.03$); the
naive Sharpe also compresses from $2.73$ (monthly) to $1.71$ (annual)
purely through serial-dependence misspecification, while the HAC bars stay
at $2.79$--$2.93$ across aggregation frequencies
(Figure~\ref{fig:horizon_stab_52}, Appendix~\ref{app:extra-floats}). The
FINAL return distribution has skewness $+2.46$ and excess kurtosis $+7.82$
(best month $+64.0\%$; Figure~\ref{fig:return_dist_52},
Appendix~\ref{app:extra-floats}), which the PSR handles explicitly
($z = 6.95$ ALL, $5.08$ FINAL; both $\mathrm{PSR} > 99.999\%$). The DSR
discounts selection across the $N = 5$ fusion family at
$\mathrm{SR}^{*} = 0.469$ and still exceeds $99.99\%$ in both partitions.
Because labels and features are deliberately unwinsorised, headline
performance retains sensitivity to extremes; the fat-tail-aware battery
conditions inference on that preserved distribution.

\paragraph{Permutation null and family-wise selection bias.}
Two PIT-safe permutation tests, an IC permutation shuffling realised
returns across tickers within each date, and a Sharpe permutation drawing
random $K{=}10$/$K{=}10$ books on the same dates, yield zero matching
permuted samples in $5{,}000$ trials at the ALL level ($p < 0.0002$; the
realised IC is $5.8\times$ the null 99th percentile), and remain
significant in the 14-month FINAL ($p = 0.0154$ IC; $p = 0.0086$ Sharpe).
White's Reality Check ($V_n = 8.24$, $0/10{,}000$ bootstrap exceedances)
and Bonferroni/Benjamini--Hochberg corrections across the $K = 5$
candidates reject the family-wise null at $p < 0.0001$; bootstrap
selection-stability shows the deployed vol-sized model winning $71.2\%$ of
resamples (baseline LGB $25.5\%$; each fusion variant $\leq 1.7\%$), so
the published figure is the dominant rather than marginal choice. The scope
is deliberately conditional: none of these tests corrects for the choice of
the AI-exposed universe itself, for retrospective construction of the
thematic allowlist, nor for the counterfactual that many thematic universes
could have been inspected and only the strongest reported. The battery
supports ``not a noise, autocorrelation, or within-family search artefact
\emph{conditional on the universe being given}''; it does not establish that
the alpha below survives arbitrary universe selection, allowlist hindsight,
or a broad-market panel.

\paragraph{Fama--French factor attribution.}
The long--short FF5 regression \citep{fama2015}, HAC errors at twelve
monthly lags, delivers annualised $\alpha = +87.4\%$~p.a.\ at
$t_{\mathrm{HAC}} = 9.34$ (long-only $+66.1\%$, $t = 8.74$); adjusted
$R^{2} = 0.051$, so $94.9\%$ of variance is idiosyncratic; bootstrap 95\%
CI on $\alpha$: $[+67.5\%, +107.4\%]$. FF6 momentum control:
$\beta_{\mathrm{MOM}} = 0.772$ ($t = 3.02$) with residual $\alpha$ at
$t_{\mathrm{HAC}} = 9.01$. All overlapping 24-month windows from
February~2018 show positive FF5 $\alpha$ (weakest $+35.3\%$~p.a.; windows
overlap and are not independent). \emph{This alpha is conditional on the
single curated AI-thematic universe} (see the scope paragraph above and
Section~\ref{sec:nulls-limits}); no claim is made that it survives
broad-market or sector-neutral construction.

\paragraph{Thematic membership and selection bias.}
Thematic \emph{membership} is not point-in-time. At each monthly rebalance,
tradability filters (price, ADV, listing status) use information available
on that date; membership in the ten-category AI allowlist does \emph{not}.
It is a single author-curated seed list of up to 100 tickers (frozen
research archive allowlist), assembled with post-2022/2024 thematic knowledge and held fixed across the 2016--2025
backtest. Quarterly review was a governance intent during research, not a
frozen, date-stamped constituent history in the artifact. Accordingly,
the reported FF5 $\alpha$ ($+87.4\%$~p.a., HAC $t = 9.34$) may be
\emph{upward-biased} by classic thematic-universe hindsight---stocks are
partly in the panel because they later became AI winners---in addition to
being conditional on a single thematic sector. None of the falsification
tests in this subsection detect that mechanism; they are conditional on the
universe being given (listing survivorship is handled; allowlist selection
is not). A conservative read treats $\alpha$ as an existence demonstration
on a post-hoc-curated panel, not transportable skill; point-in-time
eligibility rules (GICS/NAICS revenue exposure, third-party AI indices with
historical constituents, or a 2016-fixed negative-control subset) are
deferred to future replication work.

\paragraph{AI-thematic factor stress test.}
Adding an equal-weight AI-thematic factor (active-universe absolute return
minus QQQ over the same horizon) compresses alpha by about 10 points to
$+78.4\%$~p.a.\ while the thematic loading stays insignificant
($\beta_{\mathrm{AI}} = 0.956$, $t = 1.65$); the alpha $t$-statistic
\emph{rises} to $10.28$ because the factor absorbs residual noise rather
than alpha. FF6$+$AI leaves $\alpha = +76.9\%$~p.a.\
($t_{\mathrm{HAC}} = 8.82$; $\beta_{\mathrm{AI}} = 0.859$, $t = 1.54$).
This does not substitute for a second-universe replication; the factor is
built from the same panel and therefore does \emph{not} validate
point-in-time thematic membership---only whether the long--short spread
coincides with the equal-weight AI-basket return.

\begin{figure}[!htbp]
  \centering
  \includegraphics[width=\linewidth]{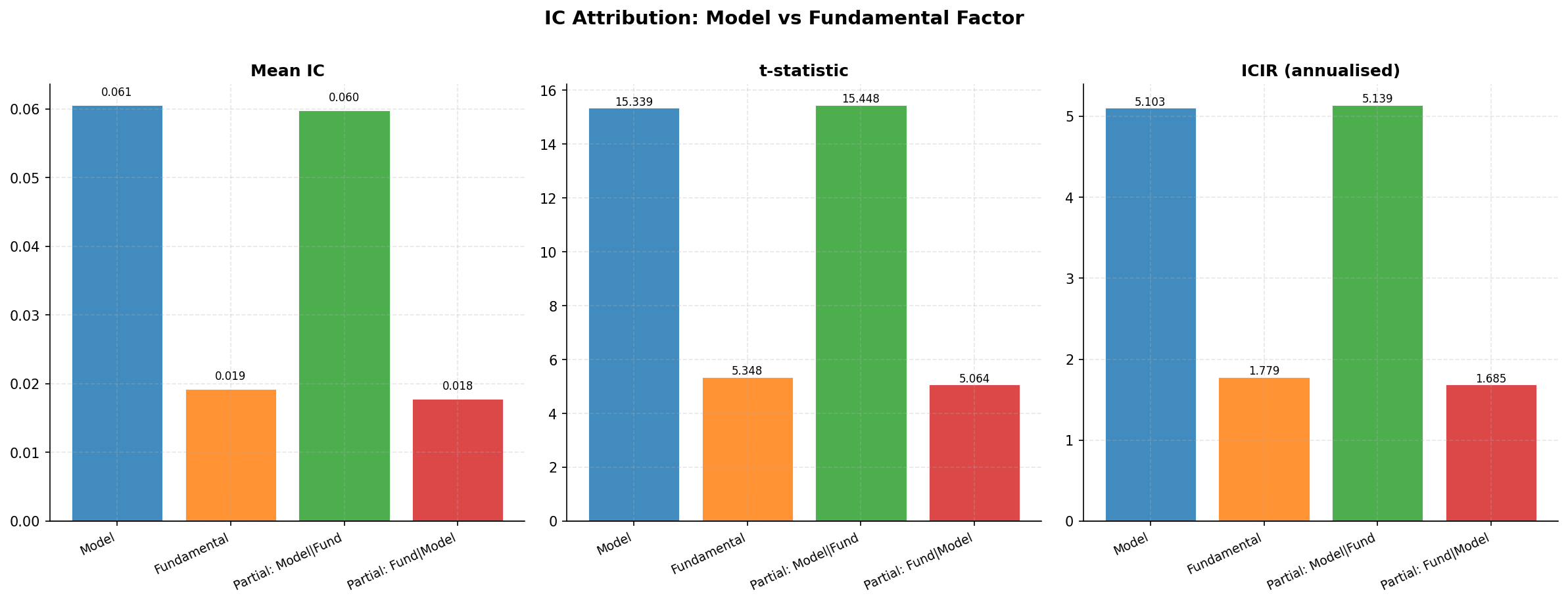}
  \caption{Information-coefficient attribution: model versus fundamental
    factor. Panels report mean IC (left), $t$-statistic (centre), and
    annualised ICIR (right) for the full ICs and both partial ICs, computed
    from daily cross-sectional Spearman ICs across 2{,}277 evaluation
    dates.}
  \label{fig:ic_comparison_52}
\end{figure}

\paragraph{Fundamental-quality orthogonality.}
A proprietary quarterly-filing fundamental composite (86 of 100 tickers)
has standalone power (daily Spearman IC $0.0192$, $t = 5.35$,
$n = 2{,}277$ dates) against the model's $0.0606$ ($t = 15.34$). The
model's partial IC conditional on the fundamental factor retains
$\rho_{S} = 0.0598$ ($t = 15.45$) (a loss of only $1.32\%$) and an
augmented time-series regression adding a fundamental-mimicking portfolio
as a sixth regressor retains $99.8\%$ of HAC alpha
($+89.3\% \to +89.1\%$~p.a.\ on the reduced 86-ticker universe;
$\beta_{\mathrm{fund}} = 0.025$, $t = 0.21$), with shared cross-sectional
$R^{2}$ of only $0.5\%$. Monthly model and fundamental ICs are moreover
uncorrelated across the 109 evaluation months (Pearson $r = -0.01$), and a
year-by-year cross-sectional $R^{2}$ decomposition shows negligible shared
variance in every year. Figure~\ref{fig:ic_comparison_52} summarises the
full and partial IC comparison.

\paragraph{Counter-cyclical convexity diagnostic.}
Regressing the 109 monthly net long--short returns on market return and its
square (HAC lag 2) gives $\hat{\beta}_1 = -0.117$
($t_{\mathrm{HAC}} = -0.75$, $p = 0.456$), minimal directionality,
consistent with the low factor $R^{2}$, and a marginally positive squared
term $\hat{\beta}_2 = 2.38$ ($t_{\mathrm{HAC}} = 1.91$, $p = 0.057$): mild
positive convexity, pre-empting the concern that the documented bear-market
outperformance (regime-conditional panels in
Figure~\ref{fig:stress_scorecard_52}, Appendix~\ref{app:extra-floats})
masks implicit short-volatility exposure. The result is marginal and should
not be over-read.

\paragraph{Limitations and hand-off.}
Conditional on the frozen universe and model family, the alpha is not a
noise, autocorrelation, multiple-testing, academic-factor, or
fundamental-factor artefact. The tests do not de-bias the thematic universe
\emph{choice} or the retrospective allowlist construction (paragraph
above), model borrow availability, or estimate institutional capacity, and
three external-validity concerns remain open. First, the FINAL partition is
only 14 months and is not a strict pre-registration of the entire
methodology: the gate weights and threshold are DEV-calibrated, and the
methodology was developed after observing the \emph{kind} of failure that
appears in FINAL; FINAL statistics are a frozen-rule evaluation, not
unconditional post-design evidence. Second, the universe is a single
thematic sector, and the 2024 rally simultaneously degraded daily IC and
amplified monthly Sharpe, a documented confound. Third, all results to this
point are walk-forward shadow returns. The cross-domain boundary test
(Section~\ref{sec:cifar_external}) therefore carries more external-validity
weight than the headline FINAL Sharpe alone. Illustrative post-freeze
monitoring behaviour is logged in Section~\ref{sec:res-monitoring-corr}
without portfolio-performance inference.

\subsection{Universe-Readiness Gate}
\label{sec:res-readiness}

\noindent\textit{RQ3 (scope boundary): an explicit \emph{stop} verdict on
downstream allocator and alpha-timing claims for this panel.}

Before deployment conclusions are drawn, the development panel is scored
against a pre-registered \emph{universe-readiness gate} asking whether
breadth, IC persistence, and effective sample size are jointly adequate for
alpha-timing or allocator meta-layers (derivation and tier ladder in
Appendix~\ref{app:universe-readiness}). On the frozen panel (median breadth
$N \approx 86$; first-order score-RankIC autocorrelation
$\phi \approx 0.847$; peak-to-trough IC shift to detect
$\Delta\mathrm{IC} \approx 0.028$; average pairwise idiosyncratic
correlation $\rho_e \approx 0.004$), the univariate minimum detection delay
is $D^{*} \approx 722$ trading days at $z_{\alpha} = 2$ against a 30-day
detection budget, and even at the autocorrelation-free floor
($\phi \to 0$), $D^{*} \approx 60$ days still exceeds the budget at this
breadth. A joint checkpoint pooling score, $\hat{e}$, $g$, and $G$ at the
10-day mid-hold mark shortens the delay roughly sevenfold
($D^{*}_{\mathrm{joint}} \approx 100$ days) but still exceeds it. The
panel carries only $\mathrm{eff}\text{-}N \approx 84$ concurrent
independent positions (from $P \approx 7{,}400$ overlapping holds,
$\bar{u} \approx 0.011$), hence $E_{\mathrm{eff}} \approx 84$ allocation
episodes. Against the tier ladder (Table~\ref{tab:universe-tiers}) the
configuration passes Tier~A (signal existence) and fails Tiers~B--E; the
composite verdict is \emph{stop} for downstream alpha-timing and allocator
programmes. Scenario prechecks at $N = 150$, $500$, $800$ (holding
$\Delta\mathrm{IC}$ and $\phi$ fixed) project $D^{*} \approx 414$, $124$,
$78$ days, breadth expansion alone is insufficient without remeasuring
$\phi$ on a rebuilt, lower-autocorrelation universe. These failures bound
the downstream programmes, not the single-expert reliability primitive
itself.

\subsection{Live Monitoring Corroboration (Illustrative)}
\label{sec:res-monitoring-corr}

The frozen walk-forward evaluation ends in February~2025; a separate
post-freeze engineering period (March--April~2026) asks a narrower
question: \emph{did the monitoring stack fire as designed when confronted
with a structural break outside the backtest window?} This subsection
answers that question only. It does \textbf{not} validate trading alpha,
gate AUROC, or execution quality: the observation window spans four weekly
signals ($N = 1$ tariff episode), broker reconciliation failed mid-April,
and the health-board UI remained immature---so no Sharpe, P\&L, or
fill-quality metrics from the post-freeze audit enter the evidence base
(Section~\ref{sec:limitations}). The statistical anchor for
$G(t)$ remains the eight-window multi-crisis panel
(Section~\ref{sec:res-crisis}).

\paragraph{Monitoring trigger taxonomy.}
Table~\ref{tab:monitoring-triggers} records the alert channels wired into
the weekly drift-check pipeline. Thresholds match
Table~\ref{tab:monitoring} (design specifications).

\begin{table}[!htbp]
  \centering
  \caption{Production monitoring triggers (design specification). Actions
    are alerts and retrain reviews unless noted; the research gate
    $G(t)=\mathbf{1}[\Gamma(t)\geq 0.20]$ (Table~\ref{tab:monitoring}) is
    evaluated only on the frozen walk-forward panel.}
  \label{tab:monitoring-triggers}
  \footnotesize
  \setlength{\tabcolsep}{3.5pt}
  \renewcommand{\arraystretch}{1.08}
  \begin{tabularx}{\linewidth}{@{}%
    >{\raggedright\arraybackslash}p{2.0cm}%
    >{\raggedright\arraybackslash}X%
    >{\raggedright\arraybackslash}p{2.3cm}%
    >{\raggedright\arraybackslash}p{2.5cm}@{}}
    \toprule
    \textbf{Channel} & \textbf{Definition} & \textbf{Threshold} &
    \textbf{Action} \\
    \midrule
    PSI drift &
      Max model-feature PSI vs.\ training baseline (90d window) &
      $>0.20$ review;\newline $>0.50$ critical &
      Retrain review;\newline de-risk haircut \\
    Rolling IC &
      4-week mean of weekly realised RankIC (5d horizon) &
      $<0.03$ &
      Alert;\newline retrain if persistent \\
    $\Gamma(t)$ advisory &
      Walk-forward $H(t)$ gate score (Eq.~\ref{eq:gamma-gate}) &
      $<0.30$ for $\geq 2$ wk &
      Low-confidence alert \\
    $G(t)$ binary &
      $\mathbf{1}[\Gamma(t)\geq 0.20]$ &
      $<0.20$ &
      Research abstention rule (not hard production gate in v4) \\
    \bottomrule
  \end{tabularx}
\end{table}

\paragraph{Model lifecycle driven by alerts.}
After the February~2025 freeze, production models advanced v1$\to$v4 in
response to monitoring alerts rather than on a fixed calendar
(Table~\ref{tab:lifecycle-corr}). Promotion used shadow scoring and
retrain-gate checks; backtest Sharpe figures in the Notes column are
calibration outputs, not live track-record inference.

\begin{table}[!htbp]
  \centering
  \caption{Post-freeze model lifecycle (monitoring-corroboration log).
    Production windows are approximate cutover dates. All versions: 13-feature
    LightGBM ranker; $n_{g}$ = feature count in $g(x)$.}
  \label{tab:lifecycle-corr}
  \footnotesize
  \setlength{\tabcolsep}{3pt}
  \renewcommand{\arraystretch}{1.05}
  \begin{tabularx}{\linewidth}{@{}%
    >{\raggedright\arraybackslash}p{1.55cm}%
    >{\raggedright\arraybackslash}p{1.35cm}%
    >{\raggedright\arraybackslash}p{2.0cm}%
    c
    >{\raggedright\arraybackslash}X
    >{\raggedright\arraybackslash}X@{}}
    \toprule
    \textbf{Version} & \textbf{Train} & \textbf{Production} & $n_{g}$ &
    \textbf{Retrain trigger} & \textbf{Notes} \\
    \midrule
    v1\_2026Q1 &
      2024-01-01 & 03-10 -- 03-30 & 11 &
      PSI on \texttt{vol\_20d}, \texttt{adv\_20d} &
      Initial deployment \\
    v2\_2026Q1 &
      2025-10-01 & 03-30 -- 04-06 & 11 &
      Q1 mean RankIC $\approx -0.145$; PSI on vol features &
      Promoted after IC inversion alert \\
    v3\_2026Q2\_A &
      2026-01-01 & 04-06 -- 04-07 & 11 &
      Feature lookback bug (\texttt{mom\_12m}/\texttt{beta\_252d} NaN);
      shadow IC sign-flip ($p<0.0001$) &
      One-day shadow; not retained \\
    v4\_2026Q2\_A &
      2026-02-01 & 04-07 -- present & 11 &
      Tariff regime in train window; compound IC\,+\,PSI\,+\,$\Gamma$
      alerts (below) &
      4-fold rolling CV (backtest calibration only) \\
    \bottomrule
  \end{tabularx}
\end{table}

\paragraph{Compound alert sequence (April~2026 tariff episode; $N=1$).}
On 2~April~2026 a broad tariff announcement hit a universe already flagged:
the 13~March PSI check recorded $\mathrm{PSI}=1.93$ on \texttt{vol\_20d}
and $1.85$ on \texttt{adv\_20d} (both $\gg 0.20$). On 6~April (first
post-shock signal week) $\Gamma(t)=0.330$ remained marginally above the
$0.30$ advisory line while that week's realised long--short RankIC was
$-0.604$. By 13~April, as the outcome entered the trailing EWMA,
$\Gamma(t)$ fell to $0.247$, model-feature max PSI reached $0.645$ on
\texttt{adv\_20d} (above the $0.50$ critical band), and the automated
de-risk protocol cut gross exposure to $\approx 0.15\times$ (IC history
$1/4$ positive weeks combined with PSI\,${>}\,0.50$). On 20~April
$\Gamma(t)=0.249$ with a two-week rolling IC of $-0.359$, firing the
second consecutive sub-$0.30$ advisory alert. The formal research abstention
threshold $\Gamma<0.20$ was never breached (minimum $0.247$); production
v4 treated $\Gamma$ as a monitoring and sizing signal rather than a hard
cash gate---but the de-risk haircut achieved a near-abstention in gross
terms.

\paragraph{What this corroborates---and what it does not.}
\emph{Corroborated:} (i)~PSI and rolling-IC channels fired \emph{before}
and during a ranker failure the backtest had not seen; (ii)~retrain
promotion followed alert logic rather than a fixed schedule; (iii)~low
$\Gamma(t)$ aligned directionally with negative realised IC in every week
of the episode. \emph{Not corroborated:} gate discrimination AUROC (no matured-label panel),
strategy profitability (paper-trading L/S $-13.7\%$ vs.\ SPY $+12.1\%$
over Mar~30--Apr~20 with non-decision-grade $N_{\mathrm{days}}=15$, after
a mid-episode de-risk to $\approx 0.15\times$ gross exposure),
execution fidelity (mid-April reconcile failures), or data hygiene
(delisted/acquisition-pending tickers JNPR, CYBR, INFA;
Appendix~\ref{app:hygiene}). Full audit prose is retained in the private
post-freeze monitoring archive; the drift-triage protocol is an internal
operations document, not part of the statistical evidence base.

\section{Cross-Domain Boundary: CIFAR-10-C}
\label{sec:cifar_external}

\noindent\textit{RQ1 (cross-domain): the bound predicts aggregation succeeds
where $\delta_{N}>0$ and fails where $\delta_{N}=0$; CIFAR-10-C tests that
dichotomy under the same mean-$g$ rule.}

\begin{figure}[!htbp]
  \centering
  \includegraphics[width=\linewidth]{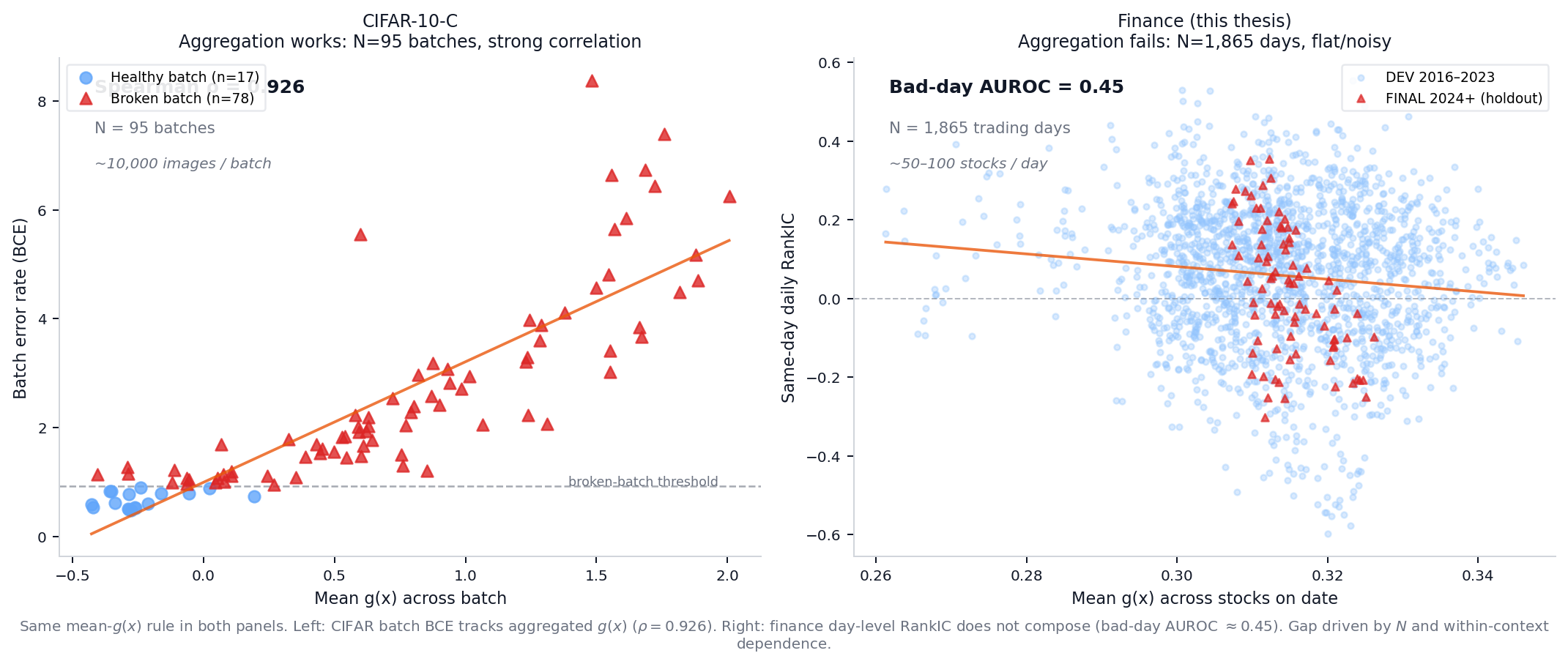}
  \caption{Aggregation boundary across domains under an identical
    mean-$g(x)$ rule. \textbf{Left:} CIFAR-10-C (mean $g(x)$ per batch
    versus batch BCE across 95 corruption batches of 10{,}000 images each
    (Spearman $\rho = 0.926$). \textbf{Right:} finance (mean $g(x)$ across
    the daily cross-section versus same-day RankIC across 1{,}865 trading
    days (scatter illustrative; deployable bad-day summary: same-day AUROC
    $\approx 0.55$ on ALL; Section~\ref{sec:agg_gap}). Per-item $g(x)$ ranks
    errors within both contexts; batch/day-level composition succeeds only
    when $N$ is large and within-context dependence is weak.}
  \label{fig:agg_boundary}
\end{figure}

Theorem~\ref{thm:detectability} predicts that the aggregation failure of
Section~\ref{sec:agg_gap} is a structural consequence of three environment
properties, small per-context sample size, within-context conditional
covariance, and concept shift in which $P(Y \mid X)$ moves through a regime
variable absent from the inputs. If so, the identical estimator family
applied in a domain violating the small-$N$ and non-i.i.d.\ conditions
should aggregate successfully. This section reports that controlled check.
Figure~\ref{fig:agg_boundary} contrasts the two regimes under the same
mean-$g(x)$ rule.

\paragraph{Design.}
CIFAR-10-C \citep{hendrycks2019} applies 19 corruption types at five
severities to the CIFAR-10 test set, producing 95 batches of 10{,}000
images. A ResNet-18 trained on clean CIFAR-10 following the recipe of
\citet[Appendix~C]{lahlou2023} reaches clean top-1 accuracy $93.88\%$; the
DEUP $g(x)$ estimator uses the original five-fold cross-pretraining with
density, variance, and Bayesian features. The single deviation from the
original protocol is the density estimator: the Masked Autoregressive Flow
produced NaNs on one fold and was replaced with a diagonal-covariance
Gaussian Mahalanobis-style density on the penultimate embeddings. Five
sanity checks pass: clean accuracy; mean $g(\text{corrupted}) >$ mean
$g(\text{clean})$ with $\Delta = +3.74$; mean within-batch $\rho = 0.655$
at severity 3; permutation-null $|\rho| = 0.002$; oracle AUROC $1.000$.

\begin{figure}[!htbp]
  \centering
  \includegraphics[width=0.82\linewidth]{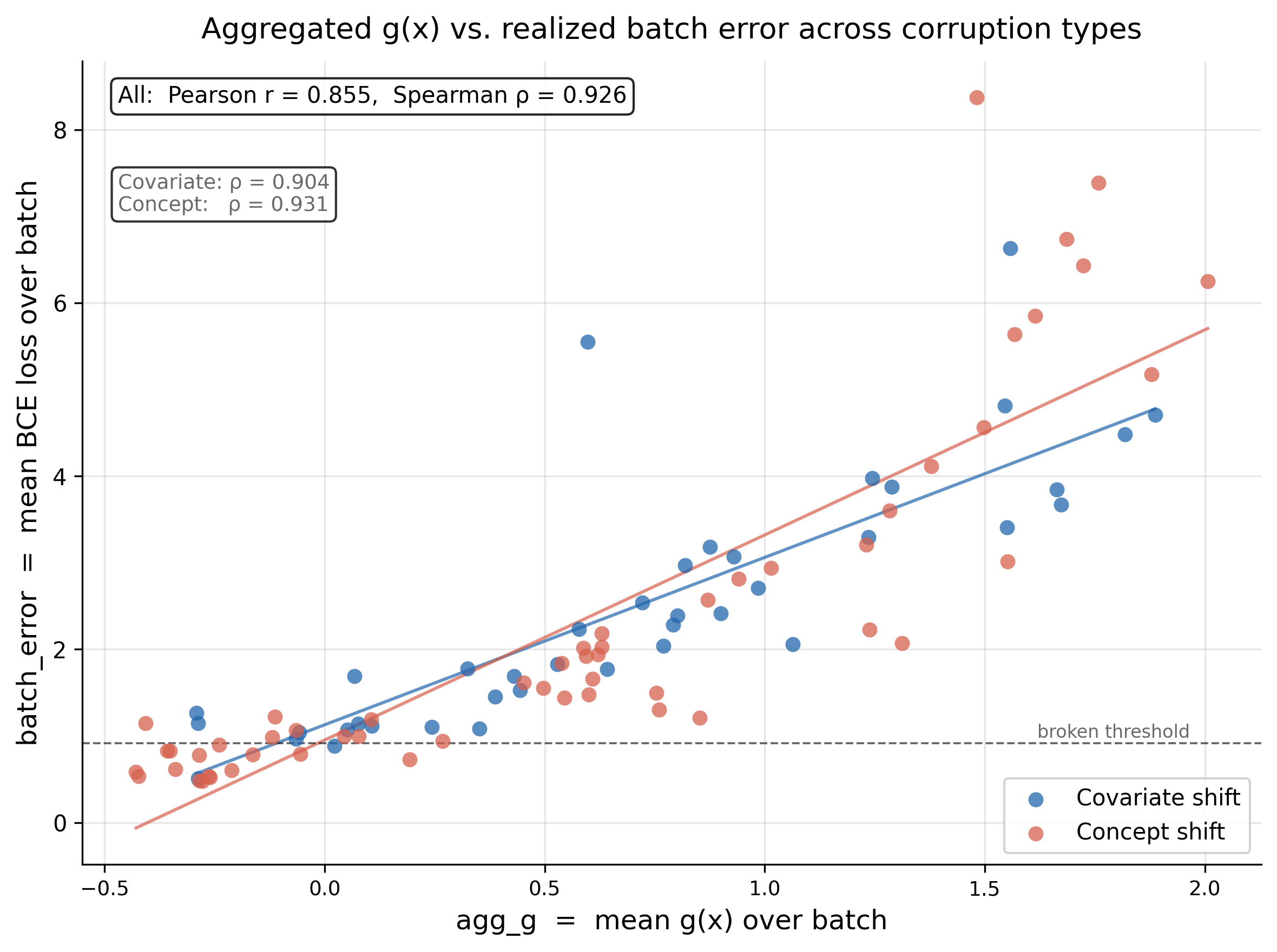}
  \caption{CIFAR-10-C aggregate-level validation by corruption family:
    mean $g(x)$ versus batch BCE (95 batches $\times$ 10{,}000 images).
    Covariate-shift (blue, $\rho = 0.904$) and concept-shift (red,
    $\rho = 0.931$); dashed: $2\times$ clean-BCE threshold
    ($\mathrm{BCE} = 0.916$).}
  \label{fig:cifar_agg}
\end{figure}

\paragraph{Results.}
Per image, $g(x)$ ranks losses well within batches: median within-batch
Spearman $\rho = 0.712$ across the 95 batches, 93 of 95 positive
(sign-test $p = 1.15 \times 10^{-25}$), 82 of 95 at or above
$\rho \geq 0.30$; per-family medians are $0.627$ (covariate shift) and
$0.732$ (label-altering perturbations). At the aggregate level, mean batch
$g(x)$ tracks batch BCE ($r = 0.855$, $\rho = 0.926$; AUROC $0.955$
against a $2\times$ clean-BCE ``broken batch'' threshold, 78 of 95 batches
broken), with both corruption families composing strongly
(Figure~\ref{fig:cifar_agg}: $\rho = 0.904$ and $0.931$). A
Gaussian-density baseline saturates at AUROC $0.961$.

\begin{table}[!htbp]
  \centering
  \caption{Cross-domain DEUP aggregation under the same estimator and
    mean-$\bar{g}$ rule. Finance mean-$\bar{g}$: bad-day RankIC
    ($N = 1{,}865$; Section~\ref{sec:agg_gap}); composite finance: $H(t)$
    good-day FINAL ($N = 284$; Table~\ref{tab:regime_auroc}).}
  \label{tab:cross_domain}
  \small
  \begin{tabular}{@{}lp{0.38\linewidth}p{0.38\linewidth}@{}}
    \toprule
    Metric & Finance & CIFAR-10-C \\
    \midrule
    Per-context $\rho(g,\,\ell)$ &
      $0.192$ (ALL pooled; Table~\ref{tab:rho-canonical}) &
      $0.712$ (median within-batch) \\
    Mean $\bar{g}$ AUROC &
      $0.55$ same-day\,/\,$0.55$ lead $+5$d (bad day) &
      $0.955$ (broken batch) \\
    Composite AUROC &
      $0.75$ [$H(t)$, FINAL] &
      $0.961$ (density baseline) \\
    \bottomrule
  \end{tabular}
\end{table}

\paragraph{Mechanism.}
The contrast in Table~\ref{tab:cross_domain} is exactly the one
Theorem~\ref{thm:detectability} isolates: detectability from any score-only
aggregate is bounded by the input-law movement $\delta_{N}$, and $N$
governs only how fast a covariate-visible signal concentrates. CIFAR-10-C
corruptions alter pixels, so they \emph{move} $P(X)$
($\delta_{N} > 0$), including the batches labelled ``concept shift'',
whose label-altering perturbations still change the image and therefore
remain covariate-visible ($\eta > 0$), which is why they aggregate just as
well. At $N = 10{,}000$ images sharing only the corruption operator,
within-batch dependence is weak and Corollary~\ref{cor:high-n} predicts
$O_{p}(N^{-1/2})$ concentration around the regime risk. At
$N \approx 50$--$100$ stocks sharing factor exposures driven by the regime
itself, Corollary~\ref{cor:mean-suppressed} predicts double suppression:
the regime signal in the mean is bounded by $B\eta$ with $\eta \approx 0$
under finance concept shift, while the correlated noise floor caps the
effective sample size at $N_{\mathrm{eff}} \to 1/\bar{\rho}_{F}$.

\paragraph{Strict concept shift: the relabelled-CIFAR test.}
The main corruption batches exemplify $\delta_{N} > 0$; strict concept
shift ($\delta_{N} = 0$) is tested directly by holding the same 10{,}000
test images fixed and applying $K = 95$ random label permutations, each a
regime that changes $\mathbb{E}[\ell \mid c]$ while leaving every
pixel (and hence every $g(x_i)$) identical. Joint input invariance is
confirmed empirically: $\mathrm{std}(\bar{g}_{\mathrm{agg}}) = 0.000$
across permutations. Corollary~\ref{cor:concept-shift} predicts
$\mathrm{AUROC} = \tfrac{1}{2}$ for all score-only aggregates; the result
is $\mathrm{AUROC} = 0.500$ for the mean and $0.500$ for the IQR spread,
against a batch-BCE standard deviation of $2.225$ across permutations
($\mathrm{BCE} \in [0.458, 16.260]$); the regime risks vary substantially
and are nonetheless invisible. The result is mechanically
guaranteed ($g(x)$ never accepts the label as input), which is precisely
the theorem's content: a concept shift that changes $P(Y \mid X)$ but not
$P(X)$ is invisible to any regime-blind score-only detector.

\begin{figure}[!htbp]
  \centering
  \includegraphics[width=\linewidth]{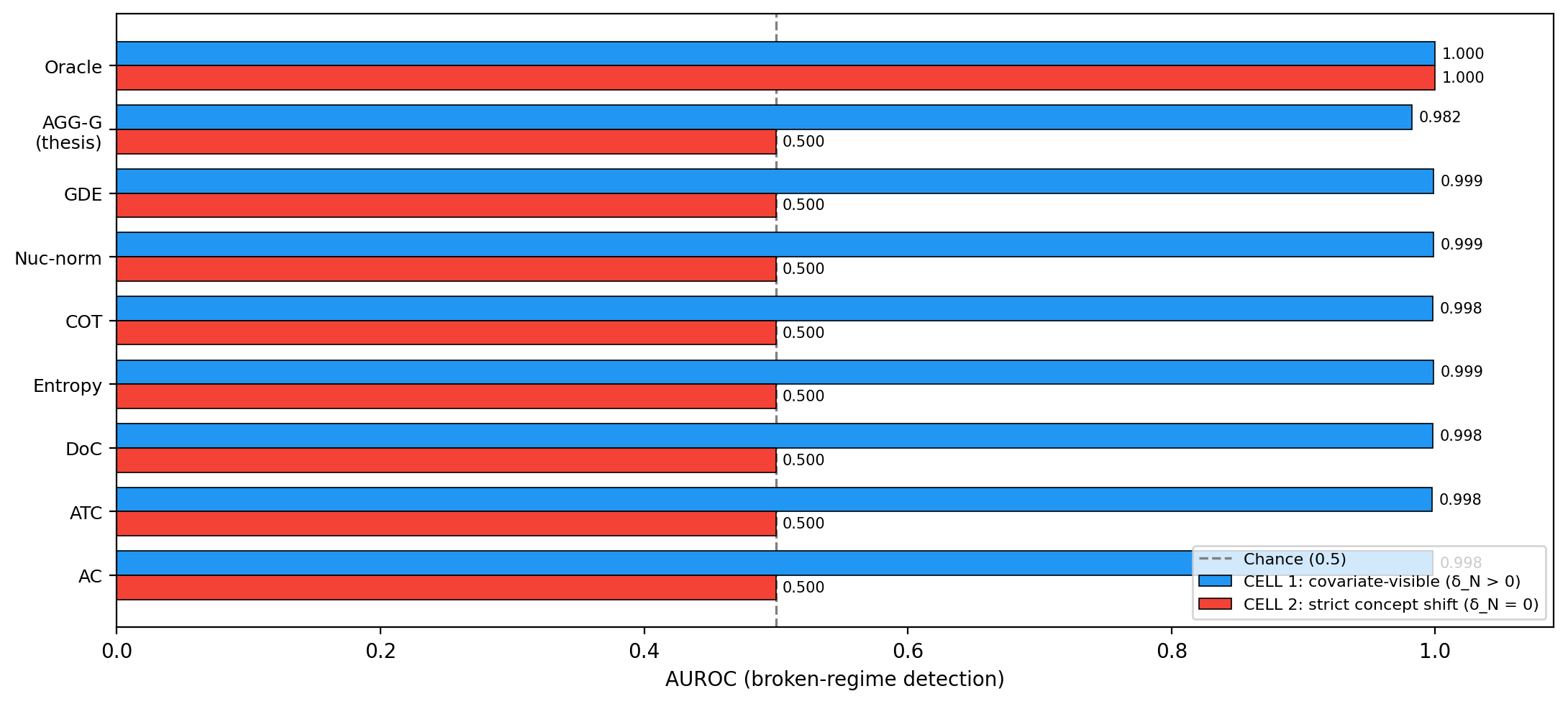}
  \caption{OOD-accuracy predictors as regime detectors on CIFAR-10-C
    (Table~\ref{tab:ood_predictors}). Blue: covariate-visible batches
    ($\delta_{N} > 0$); red: strict concept shift ($\delta_{N} = 0$).
    $X$-only methods track batch accuracy in CELL~1 and collapse to chance
    in CELL~2; the realised-error oracle (uses $Y$) is unaffected.}
  \label{fig:ood_predictors}
\end{figure}

\paragraph{The OOD-accuracy-prediction toolbox under both regimes.}
Because the data-processing argument bounds \emph{any} $X$-measurable
statistic, the relabelled design also benchmarks the unsupervised
OOD-accuracy-prediction literature as regime detectors on the identical
classifier and batches: Average Confidence \citep{hendrycks2017}, Average
Thresholded Confidence \citep{garg2022}, Difference of Confidences
\citep{guillory2021}, mean predictive entropy, Confidence Optimal
Transport \citep{lu2023}, the softmax nuclear norm \citep{deng2023}, and
two-model disagreement \citep[GDE;][]{jiang2022}. Under covariate-visible
shift every method tracks batch accuracy ($\mathrm{AUROC} \geq 0.982$;
ATC accuracy-MAE $0.061$ against AC/DoC $0.226$/$0.177$ under classifier
overconfidence). Under $\delta_{N} = 0$ every $X$-only method, including
GDE, collapses to $\mathrm{AUROC} = \tfrac{1}{2}$ exactly
(cross-permutation std $< 10^{-16}$); only the realised-error oracle,
which uses $Y$, is unaffected (Table~\ref{tab:ood_predictors},
Figure~\ref{fig:ood_predictors}). The GDE collapse is the empirical
content of Corollary~\ref{cor:deployment}'s sharpening: cross-model
disagreement is itself $X$-measurable and equally powerless under strict
concept shift.

\begin{table}[!htbp]
  \centering
  \caption{OOD-accuracy-prediction methods as regime detectors on CIFAR
    ($N = 10{,}000$). AUROC against batch top-1 accuracy as the regime
    label, hence $\bar{g}_{\mathrm{agg}} = 0.982$ here differs from the
    $0.955$ broken-batch target in the text. CELL~1: 95 CIFAR-10-C
    corruption batches ($\delta_{N} > 0$). CELL~2: 95 label permutations of
    the fixed clean pool ($\delta_{N} = 0$).}
  \label{tab:ood_predictors}
  \small
  \begin{tabular}{@{}lccl@{}}
    \toprule
    Method & CELL~1 AUROC & CELL~2 AUROC & Channel \\
    \midrule
    AC (avg.\ confidence)          & 0.998 & 0.500 & $X$-only \\
    ATC (avg.\ thresholded conf.)  & 0.998 & 0.500 & $X$-only \\
    DoC (diff.\ of confidences)    & 0.998 & 0.500 & $X$-only \\
    Entropy                        & 0.999 & 0.500 & $X$-only \\
    COT (conf.\ optimal transport) & 0.998 & 0.500 & $X$-only \\
    Nuclear norm                   & 0.999 & 0.500 & $X$-only \\
    GDE (two-model disagreement)   & 0.999 & 0.500 & $X$-only \\
    $\bar{g}_{\mathrm{agg}}$ (this paper) & 0.982 & 0.500 & $X$-only \\
    \midrule
    Realised-error oracle          & 1.000 & 1.000 & uses $Y$ \\
    \bottomrule
  \end{tabular}
\end{table}

\paragraph{Scope.}
This is a controlled check at a single high-$N$ i.i.d.\ point, not a
population study; only one $N$ is tested, and a low-$N$ i.i.d.\ ablation
that would isolate cross-section size from dependence is future work. The
completed $2\times 2$ boundary reads: covariate-visible, high-$N$
(corruptions): $\mathrm{AUROC} = 0.955$; concept-shift, high-$N$
(relabelled): $0.500$; concept-shift, low-$N$, $\bar{\rho}_{F} > 0$
(finance, ALL bad-day): $\approx 0.55$, all consistent with the detectability
bound (FINAL covariate-accompanied lift noted in Section~\ref{sec:agg_gap}).
Aggregation failure in the equity setting is a structural property of the
data-generating environment, not a deficiency of the estimator; the
two-level separation of $G(t)$ from $\hat{e}(x)$ addresses exactly the
failure mode the boundary characterises.

\section{What Does Not Work, and Limitations}
\label{sec:nulls-limits}

\subsection{Negative Results}
\label{sec:nulls}

Seven nulls are consolidated here; items~1--6 point to the introductory
summary box and the subsection cited; item~\ref{null:k4} records the
literature comparators in full.
\begin{enumerate}[leftmargin=*,itemsep=3pt]
  \item \emph{Continuous inverse-uncertainty sizing.} See intro null~1 and
  Section~\ref{sec:res-gate} (Table~\ref{tab:ablation}).
  \item \emph{Cross-sectional $g(x)$ aggregates as regime detectors.}
  See intro null~2 and Section~\ref{sec:agg_gap}.
  \item \emph{VIX-percentile gating.} See intro null~3 and
  Section~\ref{sec:res-crisis}.
  \item \emph{HMM regime features.} See intro null~4 and
  Section~\ref{sec:res-deup}.
  \item \emph{Continuous $\Gamma(t)$ throttling and fast-channel fusion.}
  See intro null~5; continuous $\Gamma$ and same-day drift fusion in
  Section~\ref{sec:res-gate}.
  \item \emph{Conformal-width economic overlays and $H(t)$ recombination.}
  See intro null~6; conformal overlays in Section~\ref{sec:res-conformal};
  walk-forward logistic/isotonic/rolling $H(t)$ recombination failed the
  materiality gate ($\Delta\mathrm{Brier} \leq 0.005$).
  \item \label{null:k4} \emph{Literature UA-Sort and Trail-IC (K4).}
  Under the shared binary $G(t)$ gate, \citet{liu2026} UA-Sort (V3,
  $\lambda = 0.05$ DEV-calibrated) and Trail-IC (date-level
  $w(t) = \min(1, c_{\mathrm{trail}} H_{\mathrm{realized}})$,
  $c_{\mathrm{trail}} = 8.30$) do not beat Gate~$+$~Raw on FINAL Sharpe
  ($-0.322$); Trail-IC matches V1 because date-level score scaling is
  rank-preserving (Section~\ref{sec:res-gate}, literature comparators) and
  therefore \emph{degenerate}: equality with V1 is guaranteed by
  construction, not learned from data; non-degenerate exposure scaling is
  covered by the continuous-$\Gamma(t)$ null (intro null~2). Neither
  comparator approaches Variant~6
  ($0.925$ FINAL). \textbf{Kill criterion K4} triggered: IC-based or
  literature-UQ \emph{sorting} is not a substitute for the $\hat{e}$-cap
  tail guard (Table~\ref{tab:comparators};
  Appendix~\ref{app:experiment-registry}).
\end{enumerate}

\subsection{Limitations}
\label{sec:limitations}

\begin{enumerate}[leftmargin=*,itemsep=4pt]
  \item \textbf{Universe selection.} The study uses one curated panel of
  $\approx 100$ AI-exposed US equities (median tradable breadth
  $N \approx 86$). Tradability filters and listing dates are applied
  point-in-time, but the thematic allowlist is a post-hoc-curated seed
  snapshot (Section~\ref{sec:res-falsification}), not a historically
  re-derived membership rule. Every statistical test in
  Section~\ref{sec:res-falsification} is conditional on that universe being
  given; none corrects for the counterfactual of choosing among many
  possible thematic universes or for hindsight in allowlist construction. A broad-market, sector-neutral, or
  multi-sector replication is required before the trading alpha can be
  called portable. The breadth shortfall is quantified by the pre-registered
  universe-readiness gate (verdict \emph{stop}; $D^{*} \approx 722$ trading
  days against a 30-day budget, $\mathrm{eff}\text{-}N \approx 84$;
  Section~\ref{sec:res-readiness}). The detectability dichotomy,
  $D^{*}$ formula, minimum-breadth inversion, and tier ladder (Tiers
  A--E) are derived in Appendix~\ref{app:universe-readiness}
  (Eqs.~\eqref{eq:dichotomy}--\eqref{eq:dicmin};
  Tables~\ref{tab:dstar-grid} and~\ref{tab:universe-tiers}). Separately, the deliberate
  non-winsorisation of labels and features leaves performance exposed to a
  small number of extreme outcomes, a methodological choice stress-tested
  by the fat-tail-aware PSR/DSR/permutation battery.
  \item \textbf{Calibration boundary; FINAL is not a pre-registration.}
  The DEV/FINAL split is frozen and PIT-safe, but the architecture was
  developed with knowledge of the \emph{type} of failure that appears in
  the 2024 FINAL period. FINAL Sharpe, PSR, DSR, and attribution results on
  the \emph{ungated} ranker are frozen-rule evaluations, not strict
  pre-registered evidence for the whole research programme; deployed
  Variant~6 is judged by gate and crisis criteria
  (Sections~\ref{sec:res-gate}--\ref{sec:res-crisis}), not the ungated
  Sharpe $\geq 1.5$ bar.
  \item \textbf{No live or paper-trading performance validation.}
  Walk-forward shadow returns (through February~2025) are the only
  portfolio-performance evidence base. Post-freeze materials in the private
  research archive cover roughly four weekly signals ($\sim$15 daily P\&L
  observations): Sharpe intervals are wide and include zero; execution and
  hygiene defects (broker reconcile failures, delisted tickers) invalidate
  any alpha inference. Section~\ref{sec:res-monitoring-corr}
  documents \emph{monitoring alert firing} and model lifecycle from that
  period as illustrative engineering corroboration only---it does not
  upgrade the statistical claims of Sections~\ref{sec:res-gate}
  and~\ref{sec:res-crisis}.
  \item \textbf{Execution, borrow, and capacity.} The 10~bps-per-leg cost
  model with sensitivity to 50~bps (Appendix~\ref{app:cost-sensitivity})
  is adequate for comparing research variants but is not a capacity study:
  hard-to-borrow constraints, locate failures, short-borrow fees (estimated
  at 3--8\% annually for hard-to-borrow AI names), crowding, and
  order-book impact at institutional AUM are unmodelled. Portfolio
  performance is shadow evidence for the framework, not an underwriting
  case for allocation.
  \item \textbf{Factor coverage.} FF5, FF6, the fundamental composite, and
  the AI-thematic factor do not exhaust institutional controls; industry,
  crowding, short-interest, borrow-fee, idiosyncratic-volatility, and
  options-implied factors remain untested. Strong bear-market performance
  in a concentrated universe may reflect implicit short-beta or
  short-squeeze exposure not fully spanned by FF5.
  \item \textbf{Modest validation panels.} The multi-crisis analysis covers
  eight hand-selected windows (Fisher exact vs.\ the VIX gate not
  significant); CIFAR-10-C is a single high-$N$ mechanism check; the FINAL
  partition spans 14 months.
  \item \textbf{Novelty scope of the architecture.} The two-level risk
  structure---strategy-level kill switch plus position-level outlier
  cap---is a rigorous account and mechanistic justification of controls
  familiar in institutional multi-book practice, not a structural
  invention. The cap's incremental portfolio value is not established by
  the FINAL point estimate on $N=5$ monthly returns (Variant~6 at $0.925$
  vs.\ Variant~4 at $0.953$; intervals not identified);
  its retained role is interpretability and discrete tail-risk
  control. This does not imply a degenerate signal: $\approx 95\%$ of
  $g(x)$'s predictive content is orthogonal to cross-sectional rank
  (Section~\ref{sec:res-deup}).
  \item \textbf{Theorem scope.} Strong concept-shift impossibility requires
  \emph{joint} input invariance ($\delta_{N} = 0$), stricter than marginal
  concept shift; the second-moment escape hatch is closed here empirically
  (IQR aggregate at AUROC $\approx 0.50$), not by the theorem. The bounds
  are population-level, with finite-$N$ fluctuation visible in
  Section~\ref{sec:cifar_external}; any $X$-measurable detector, including
  cross-model disagreement, is equally bounded under strict concept shift,
  so the surviving channel is realised-outcome conditioning
  (Appendix~\ref{app:detectability}).
\end{enumerate}

\paragraph{Quarantining the alpha: what transports and what does not.}
Because the empirical study is conducted on a single curated universe, it
matters which contributions are logically independent of that universe.
Three tiers separate cleanly. \emph{(i) Universe-independent within stated
assumptions:} Theorem~\ref{thm:detectability} and its corollaries are
mathematical statements about score aggregates, and the CIFAR-10-C boundary
involves no equity universe; these claims cannot be ``an AI-sector bet''
because they contain no sector and no returns. \emph{(ii) Mechanistic
claims instantiated in (but not logically dependent on) the AI
universe:} the rank-geometry coupling ($0.616$), the label-asymmetric
aggregation gap (matured-efficacy gate $\approx 0.72$ vs.\ $X$-only
aggregates $\lesssim 0.55$ on every label tested; Table~\ref{tab:auroc-map}),
the model-health versus market-regime dissociation
(HMM null; $7/8$ vs.\ $5/8$ verdicts), and the conformal
conditional-coverage improvement ($20.2\% \to 0.81\%$ oracle, $4.44\%$
deployable) are architecture properties hypothesised to recur in
comparable low-$N$, factor-correlated long--short books, transportability
untested. \emph{(iii) Universe-dependent:} exactly one headline quantity
does not transport without replication; the trading-performance magnitude
(FF5 $\alpha = +87.4\%$~p.a., $t_{\mathrm{HAC}} = 9.34$), which is
explicitly conditional on the curated universe. Downstream alpha-timing
and allocator meta-layer claims are additionally ruled out on this panel by
the readiness tiers (Appendix~\ref{app:universe-readiness}): breadth and
overlap dependence leave $D^{*} \gg H_{\mathrm{det}}$ and
$\mathrm{eff}\text{-}N \ll 500$ even under optimistic $\phi$ scenarios.
On the most adversarial
reading---that the entire $+87.4\%$ is a pure AI-sector exposure---no
tier-(i) or tier-(ii) contribution is weakened, because none is derived
from that return.

\paragraph{A predicted null, not a discovered one.}
The near-chance performance of the disagreement channel (finance GDE
$\mathrm{AUROC} \approx 0.527$; relabelled-CIFAR GDE exactly $0.500$) is
the kind of result usually reported post hoc as ``one configuration that
did not help.'' Here it is consistent with a prediction derivable in
advance: Corollary~\ref{cor:deployment} designates every $X$-measurable
health channel, confidence aggregates, drift statistics, cross-model
disagreement, as bounded by $\delta_{N}$ and powerless under strict
concept shift. The two observations differ in status: the finance
near-null is a genuine empirical measurement, while the relabelled-CIFAR
$0.500$ is mechanically exact ($g(x)$ never sees the label) and serves as
a consistency check of the corollary's logic rather than an independent
measurement. Together they license the architectural redirect, toward a
realised-efficacy ($Y$-conditioning) monitor, as a consequence of the
proof rather than an empirical accident, and they make the detectability
bound, rather than the trading result, the sharp contribution of this
paper.

\section{Conclusion}
\label{sec:conclusion}

Cross-sectional rankers can fail silently: per-stock error structure may
look stable while portfolio-level deployment becomes unsafe under regime
shift. Standard volatility controls track aleatoric noise; they do not
test whether the \emph{model} still ranks correctly. As posed in
Section~\ref{sec:intro} and Table~\ref{tab:contribution-map}, this paper
evaluates three research questions under Theorem~\ref{thm:detectability}
as the formal spine.

\paragraph{What is new.}
Relative to \citet{sanderink2026alpha} (consolidated items C1--C2), the
increment is N1--N4 in Table~\ref{tab:contribution-map}: a detectability
bound with finance-specific corollaries, a CIFAR-10-C boundary test,
partial-horizon outcome monitoring, and falsification plus
universe-readiness gates that bound transport claims. Reused gate and DEUP
statistics confirm the bound-compatible architecture under stronger
tests---not as independent discoveries.

\paragraph{Answers to the research questions.}
As stated in Section~\ref{sec:intro}:
\textbf{RQ1 (Separation).} \textbf{Yes} at stock level; \textbf{no} at
strategy level---and the negative result is now structural, not empirical
noise. Per-stock $\hat{e}(x)$ ranks failure risk distinct from volatility
(holdout strengthening, quintile monotonicity, RA2 replication;
Section~\ref{sec:res-deup}). Date-level aggregates of $\{g(x_i)\}$ remain
near chance on the long ALL panel, as in \citep{sanderink2026alpha};
Theorem~\ref{thm:detectability} and Corollary~\ref{cor:concept-shift}
identify the ceiling under strict concept shift ($\delta_{N}=0$).
Partial-horizon monitoring (Section~\ref{sec:partial-horizon}) is the
principled $Y$-measurable escape.

\textbf{RQ2 (Decision value).} \textbf{Yes, within scope.} Strategy
abstention must read realised efficacy
(Corollary~\ref{cor:deployment}); the frozen two-level stack---gate on
$H(t)$ plus discrete $\hat{e}$-cap, not inverse-uncertainty sizing---adds
value in component ablations and an eight-window crisis panel
(Sections~\ref{sec:res-gate}, \ref{sec:res-crisis}). Gating dominates;
the cap trims tail exposure without penalising score tails. Reported
Sharpes are conditional existence checks on a universe whose base ranker
failed its pre-registered holdout bar
(Table~\ref{tab:signal_quality_531}), not a claim of transportable alpha.

\textbf{RQ3 (Robustness).} \textbf{Yes, within stated boundaries.}
Multi-crisis verdicts favour model-health over VIX-style proxies
(Section~\ref{sec:res-crisis}); a five-test falsification battery addresses
data-mining and factor-mimicking objections
(Section~\ref{sec:res-falsification}); the universe-readiness diagnostic
withholds allocator-grade claims on this panel
(Section~\ref{sec:res-readiness}). All three answers are conditional on a
curated $\lesssim 100$-name universe and a 14-month FINAL holdout.

\paragraph{Synthesis.}
The prior working paper observed that per-stock uncertainty does not
aggregate into regime trust; this paper supplies the ceiling that
explains that observation, confirms it outside finance, and evaluates an
architecture forced by the proof rather than chosen by design preference.
Tier-(i) claims---the bound and the CIFAR boundary---do not depend on the
equity universe; tier-(ii) mechanistic claims---coupling, gate AUROC,
ablation ordering---are instantiated on the panel but align with the
theorem's predictions (Section~\ref{sec:nulls-limits}). The validated
object is the reliability boundary and the monitoring stack that respects
it; headline trading performance is secondary evidence.

\paragraph{Future work.}
Theory should isolate cross-section size from within-date dependence
(low-$N$ i.i.d.\ vision ablation; synthetic aggregate-AUROC surfaces over
$(N,\eta)$). Empirical work should add universes, neutralisation, borrow
and locate modelling, and AUM-scaled impact before leading with Sharpe;
the universe-readiness gate specifies a concrete rebuild target (breadth
$N \approx 300$--$800$, lower residual-IC autocorrelation).
Partial-horizon $Y$-monitoring is the principled route to shortening the
abstention blind window; a multi-year live or paper-trading record with
full cost capture would address questions retrospective simulation cannot.

\section*{Acknowledgements}
The author thanks colleagues at Quant Signal Lab for research feedback and
infrastructure support.

\section*{Disclosure}
The author maintains a private systematic research stack and
household-affiliated paper-trading infrastructure used only for
post-freeze monitoring corroboration (Section~\ref{sec:res-monitoring-corr}).
No live or paper-trading performance figures enter the empirical claims of
this paper. The research was conducted independently without external funding
or paid client mandates.


\appendix

\section{Detectability Bound: Proofs and Scope}
\label{app:detectability}

This appendix gives the complete proofs and scope of
Theorem~\ref{thm:detectability}. The mathematics is deliberately
elementary: Le~Cam two-point testing and the data-processing inequality
suffice. The core $\delta_{N}=0$ impossibility for
$X$-only detectors is folklore in the concept-drift literature
(real versus virtual drift; Section~\ref{subsec:concept-drift}).
The appendix's value lies in (i)~the precise
covariate-visible versus concept-shift dichotomy, (ii)~the honest
separation of joint from marginal input invariance, (iii)~the
fold-conditional reading for walk-forward-refitted~$g$, and
(iv)~the architectural consequence for constructing per-expert
health signals. It is not claimed as a hard impossibility result
beyond the stated assumptions.

\subsection{Notation and result-specific assumptions}
Let $c \in \mathcal{C}$ be a latent regime. For each $c$ let $P_{c}$ be a
joint law on $(X, Y) \in \mathcal{X} \times \mathcal{Y}$. Fix a predictor
$f$ and a per-prediction loss $\ell\bigl(f(X), Y\bigr) \in [0, B]$. The
training law is the regime mixture
$P_{\mathrm{tr}} = \int_{\mathcal{C}} P_{c}\, d\pi(c)$, and the error
predictor is the mixture-conditional risk
$g(x) = \mathbb{E}_{\mathrm{tr}}[\ell(f(X), Y) \mid X = x]$. At evaluation a
cross-section of size $N$ is drawn under a single regime~$c$:
$\mathbf{X} = (X_{1}, \dots, X_{N}) \sim P_{c}^{X,(N)}$, the $N$-marginal of
$P_{c}$ over inputs; the $X_{i}$ need not be independent. Write
$g(\mathbf{X}) = (g(X_{1}), \dots, g(X_{N}))$. A \emph{score-only detector}
is any measurable map $\phi : \mathbb{R}^{N} \to \{0, 1\}$ (or score
$T : \mathbb{R}^{N} \to \mathbb{R}$) acting on $g(\mathbf{X})$ alone.
Throughout, $d_{\mathrm{TV}}(P, Q) = \sup_{A} |P(A) - Q(A)| \in [0, 1]$.
Three assumptions are used (Table~\ref{tab:assumptions-by-result} records
which proof uses which):
\begin{enumerate}[leftmargin=*,itemsep=3pt]
  \item \textbf{Boundedness.} $\ell,\, g \in [0, B]$.
  \item \textbf{Regime-blind estimator.} $g$ is a fixed measurable function
        of $x$ that does not take $c$ as an argument; it estimates the
        mixture-conditional risk, not the regime. (This is the defining
        property of a DEUP-style per-item error predictor.) In the
        walk-forward deployment the error model is refitted each fold, so
        the bound is read \emph{conditional on the fitted map}
        $g_{\mathcal{T}}$: given the training information $\mathcal{T}$,
        $g_{\mathcal{T}}$ is held fixed under both regime hypotheses and the
        result applies fold-by-fold. The pooled cross-fold AUROC reported in
        Section~\ref{sec:results} is therefore an application of the bound
        across a sequence of fitted maps rather than a single literal
        instantiation: the corresponding pooled statement replaces the input
        law of $\mathbf{X}$ with the joint law of $(\mathcal{T}, \mathbf{X})$,
        so any across-fold detectability exceeding the per-fold $\delta_{N}$
        must be carried by movement in the fitted-model state $\mathcal{T}$
        itself.
  \item \textbf{Within-regime dependence.} The average pairwise correlation
        of the scores is
        \[
          \bar{\rho}_{g}(c) = \tfrac{1}{N(N-1)} \sum_{i \neq j}
          \mathrm{Corr}\bigl(g(X_{i}), g(X_{j}) \mid c\bigr),
        \]
        with $\bar{\rho}_{g} \approx 0$ for near-i.i.d.\ vision batches and
        $\bar{\rho}_{F} > 0$ for equity cross-sections sharing factor
        exposures.
\end{enumerate}
The movement quantities $\delta_{N}$ (joint) and $\eta$ (supremum over
regimes and coordinates of single-coordinate marginal TV) are as defined in
Eq.~\eqref{eq:tv-movement}. Table~\ref{tab:assumptions-by-result} in
Section~\ref{sec:method-deup} records which assumptions each result uses.

\subsection{Proof of Theorem~\ref{thm:detectability}}
Write $P_{j} = \mathcal{L}_{c_{j}}(g(\mathbf{X}))$ for $j \in \{0, 1\}$. For
a two-point test with equal priors the Bayes-optimal rule is the
likelihood-ratio test, whose error is
$\tfrac{1}{2}\bigl(1 - d_{\mathrm{TV}}(P_{0}, P_{1})\bigr)$ (Le~Cam's
two-point lemma); since balanced accuracy under equal priors equals one
minus this Bayes error,
$\sup_{\phi} \mathrm{BAcc}(\phi) =
 \tfrac{1}{2}\bigl(1 + d_{\mathrm{TV}}(P_{0}, P_{1})\bigr)$. The map
$\mathbf{X} \mapsto g(\mathbf{X})$ is a fixed measurable function
(Assumption~2), hence a deterministic channel, so the data-processing
inequality for total variation gives
$d_{\mathrm{TV}}(P_{0}, P_{1}) \le
 d_{\mathrm{TV}}\bigl(\mathcal{L}_{c_{0}}(\mathbf{X}),
 \mathcal{L}_{c_{1}}(\mathbf{X})\bigr) = \delta_{N}$. Combining the two
yields Eq.~\eqref{eq:detectability-bound}; the equivalent minimax statement
is that the sum of Type-I and Type-II errors of any score-only test is at
least $1 - \delta_{N}$. $\hfill\square$

\subsection{Proof of Corollary~\ref{cor:concept-shift}}
If $\delta_{N} = 0$ then
$\mathcal{L}_{c_{0}}(\mathbf{X}) = \mathcal{L}_{c_{1}}(\mathbf{X})$, so for
any measurable $T \circ g$ the pushforwards coincide,
$\mathcal{L}_{c_{0}}(T(g(\mathbf{X}))) =
 \mathcal{L}_{c_{1}}(T(g(\mathbf{X}))) =: \mu$. For independent draws
$T_{0}, T_{1} \sim \mu$, symmetry gives $P(T_{1} > T_{0}) = P(T_{0} >
T_{1})$, and with $2P(T_{1} > T_{0}) + P(T_{1} = T_{0}) = 1$,
\[
  \mathrm{AUROC}(T)
  = P(T_{1} > T_{0}) + \tfrac{1}{2} P(T_{1} = T_{0})
  = \tfrac{1 - P(\text{tie})}{2} + \tfrac{P(\text{tie})}{2}
  = \tfrac{1}{2}.
\]
This holds for the mean, median, any quantile, the IQR, the sample
variance, or any other score-only statistic, at every $N$. $\hfill\square$

\subsection{Proof of Corollary~\ref{cor:mean-suppressed}}
\emph{Signal.}
$\mathbb{E}[\bar{g}_N\mid c]=\mathbb{E}[g(X)\mid c]=\int g\,dP_c^{X}$.
Centring $\tilde{g}=g-\tfrac{B}{2}\in[-\tfrac{B}{2},\tfrac{B}{2}]$,
\[
\Bigl|\textstyle\int g\,d(P_{c_0}^{X}-P_{c_1}^{X})\Bigr|
=\Bigl|\textstyle\int \tilde{g}\,d(P_{c_0}^{X}-P_{c_1}^{X})\Bigr|
\le \tfrac{B}{2}\,\bigl\|P_{c_0}^{X}-P_{c_1}^{X}\bigr\|_1
=\tfrac{B}{2}\cdot 2\,d_{\mathrm{TV}}\!\bigl(P_{c_0}^{X},P_{c_1}^{X}\bigr)
\le B\,\eta,
\]
which proves the (absolute) signal bound. \emph{Noise.} Under conditional
exchangeability with common variance $\sigma_g^2(c)$ and average pairwise
correlation $\bar{\rho}_g^{(N)}(c)$,
$\operatorname{Var}\!\bigl(\sum_i g(X_i)\mid c\bigr)
=N\sigma_g^2+N(N-1)\bar{\rho}_g^{(N)}\sigma_g^2$, hence
$\operatorname{Var}(\bar{g}_N\mid c)
=\sigma_g^2[\tfrac1N+(1-\tfrac1N)\bar{\rho}_g^{(N)}]$.
\emph{SNR.} If $\bar{\rho}_g^{(N)}(c)\to\rho_\infty(c)>0$, the variance
tends to $\rho_\infty(c)\sigma_g^2(c)$; dividing the signal bound by its
square root gives the stated $N$-independent ratio. Conditional
exchangeability is an idealisation for a heterogeneous equity
cross-section; without it the exact identity is
$\operatorname{Var}(\bar{g}_N\mid c)=N^{-2}\bigl[\sum_i \sigma_i^2(c)
+2\sum_{i<j}\operatorname{Cov}(g(X_i),g(X_j)\mid c)\bigr]$, of which the
common-variance expression is the equal-variance special case, and the
signal bound holds as
$|\mathbb{E}[\bar{g}_N\mid c_0]-\mathbb{E}[\bar{g}_N\mid c_1]|\le
B\,d_{\mathrm{TV}}(\bar{P}^{X}_{c_0},\bar{P}^{X}_{c_1})\le B\eta$ for the
average marginal law $\bar{P}^{X}_{c}=\tfrac1N\sum_i P^{X}_{c,i}$, with
$\eta=\sup_{c,c',i} d_{\mathrm{TV}}(P_{c}^{X_i},P_{c'}^{X_i})$ as in
Eq.~\eqref{eq:tv-movement}, without any exchangeability assumption. \qed

\subsection{Proof of Corollary~\ref{cor:high-n}}
By the calibration assumption $\mathbb{E}[\bar{g}_N\mid c]=R(c)+b$, so
$\bar{g}_N-b-R(c)=\bar{g}_N-\mathbb{E}[\bar{g}_N\mid c]$. Chebyshev's
inequality gives, for every $t>0$,
$\mathbb{P}\bigl(|\bar{g}_N-\mathbb{E}[\bar{g}_N\mid c]|\ge t\mid c\bigr)
\le \operatorname{Var}(\bar{g}_N\mid c)/t^2$. By
Corollary~\ref{cor:mean-suppressed},
$\operatorname{Var}(\bar{g}_N\mid c)
=\sigma_g^2(c)\bigl[\tfrac1N+(1-\tfrac1N)\bar{\rho}_g^{(N)}(c)\bigr]
\le \sigma_g^2(c)\bigl[\tfrac1N+\bar{\rho}_g^{(N)}(c)\bigr]$ with
$\sigma_g^2(c)$ bounded, hence
\[
\bar{g}_N-b-R(c)=O_{p}\!\left(\sqrt{\tfrac1N+\bar{\rho}_g^{(N)}(c)}\right).
\]
The special cases follow: $\bar{\rho}_g^{(N)}(c)\to 0$ gives consistency;
$\bar{\rho}_g^{(N)}(c)=O(N^{-1})$ gives the parametric $O_{p}(N^{-1/2})$
rate; and a fixed positive separation of $R(\cdot)$ drives the regime AUROC
to $1$. \qed

\subsection{Proof of Corollary~\ref{cor:deployment}}
Apply Theorem~\ref{thm:detectability} and
Corollary~\ref{cor:concept-shift} to expert $i$'s scores $\{g_i(x)\}$ under
its concept-shift regime. Feature-distribution drift is a functional of
$P(X)$, and cross-model disagreement
$D(X)=\mathbf{1}[\hat{f}_A(X)\neq \hat{f}_B(X)]$ is a deterministic
function of $X$ for fixed predictors; both are therefore $X$-measurable
and, by the data-processing inequality applied to the input-law movement
$\delta_N$, the pushforward total variation between regime laws---and hence
the excess balanced accuracy $\mathrm{BAcc}(\phi)-\tfrac12$ over chance---is
at most $\tfrac12\delta_N$, which is zero at $\delta_N=0$. Realised efficacy is a functional of the realised
$Y$ and is not bounded by Theorem~\ref{thm:detectability}. Hence, among the
channels in the stated information set, only the realised-$Y$ channel
retains power at strict concept shift; this is a necessity claim within
that information set; it does not assert sufficiency, nor does it exclude
exogenous side information lying outside the set. The relabelled-CIFAR
experiment (Section~\ref{sec:cifar_external}) is a controlled illustration:
holding the images fixed under label permutation forces every
$g(x_i)$ (and hence the disagreement statistic) to be identical across
the two regimes, so GDE AUROC $=0.500$ by construction. \qed

\subsection{Scope and limitations}
\begin{enumerate}[leftmargin=*,itemsep=3pt]
  \item \textbf{Strong impossibility needs joint, not marginal,
        invariance.} Corollary~\ref{cor:concept-shift} requires
        $\delta_{N} = 0$, stronger than concept shift $\eta = 0$. If a
        regime changes the \emph{dependence} structure among inputs (e.g.\
        crisis factor-correlation spikes) while leaving marginals fixed,
        then $\delta_{N} > 0$ and a second-moment aggregate (sample
        covariance or IQR of $\{g(x_{i})\}$) is \emph{not} excluded; there
        is residual detectability with excess balanced accuracy up to
        $\tfrac{1}{2}\delta_{N}$. The
        empirical finding that the IQR/spread aggregate also attains
        $\mathrm{AUROC} \approx 0.50$ (Section~\ref{sec:agg_gap}) is what
        closes this gap in the finance data; the theorem alone does not.
  \item \textbf{Every $X$-measurable detector is bounded.} The result says
        nothing against detectors that condition on realised outcomes
        (matured RankIC). It \emph{does} bound feature-distribution
        statistics and cross-model disagreement: both are deterministic
        functions of the inputs $X$, hence bounded by $\delta_{N}$ and
        powerless at strict concept shift. They recover regime power only
        under covariate-visible movement ($\delta_{N} > 0$), the
        sharpening illustrated by the GDE collapse in
        Section~\ref{sec:cifar_external}.
  \item \textbf{Population, not finite-sample.} The bounds are
        population-level; at finite $N$ the realised AUROC fluctuates
        around them.
  \item \textbf{Not DEUP-specific.} Assumption~2 is the only structural
        input; any regime-blind per-item error or uncertainty estimator
        targeting $\mathbb{E}[\ell \mid X]$ inherits the result.
  \item \textbf{Relation to \citet{singh2024}.} A conceptual cousin
        (aggregation can erase task-relevant information), not an
        application: \citet{singh2024} bound bag-to-instance generalisation
        under covariate shift, whereas
        Theorem~\ref{thm:detectability} bounds instance-to-context
        detection under concept shift, opposite direction, different shift
        type.
\end{enumerate}

\section{Experiment Registry}
\label{app:experiment-registry}

Table~\ref{tab:experiment-registry} maps each major falsification experiment
to the claim tested, the verdict, and the location in the main text. Rows
are limited to tests that change what a reader should believe about the
theorem, the deployment architecture, or a named comparator; engineering
grids (calibrator choice, drift sub-weight sweeps, model-version lifecycle)
are documented in the private research archive but omitted here.

\begin{table}[ht]
  \centering
  \caption{Experiment registry. Policy Sharpes are monthly
    $\times\sqrt{12}$ from the frozen policy evaluation archive
    (February~2025 snapshot).}
  \label{tab:experiment-registry}
  \footnotesize
  \setlength{\tabcolsep}{3pt}
  \renewcommand{\arraystretch}{1.12}
  \begin{tabularx}{\linewidth}{@{}%
    >{\raggedright\arraybackslash}p{2.4cm}%
    >{\raggedright\arraybackslash}X%
    >{\raggedright\arraybackslash}p{1.6cm}%
    >{\raggedright\arraybackslash}p{2.6cm}@{}}
    \toprule
    \textbf{Experiment} & \textbf{Claim / hypothesis} & \textbf{Verdict}
      & \textbf{Where} \\
    \midrule
    Gate + Raw (V1) & Binary $G(t)$ abstention alone & Baseline gated
      & \S\ref{sec:res-gate}, Tab.~\ref{tab:ablation} \\
    Gate + Vol (V2) & Aleatoric vol-sizing restores FINAL Sharpe
      & Partial ($0.375$ FINAL) & Tab.~\ref{tab:experiment-policy-grid} \\
    Gate + UA-Sort (V3) & Liu \citep{liu2026} UQ sort beats raw gate
      & \textbf{Null} (FINAL $-0.322$) & Tab.~\ref{tab:comparators} \\
    Gate + Resid-$\hat{e}$ (V4) & Residualised continuous $\hat{e}$ sizing
      & Strong FINAL point est.\ ($0.953$); not deployed
      & Tab.~\ref{tab:experiment-policy-grid} \\
    Gate + $\hat{e}$-Cap (V5) & Tail cap without vol-sizing
      & Partial ($0.731$ FINAL) & Tab.~\ref{tab:experiment-policy-grid} \\
    Gate + Vol + $\hat{e}$-Cap (V6) & Three-layer deployed stack
      & \textbf{Adopted} ($0.925$ FINAL point est.) & Tab.~\ref{tab:ablation} \\
    Trail-IC (K) & Trailing IC sizing beats raw gate (K4)
      & \textbf{Null} ($\equiv$ V1) & Tab.~\ref{tab:comparators} \\
    Continuous $\Gamma(t)$ throttle & Smooth exposure beats binary gate
      & \textbf{Null} ($-0.34$ FINAL) & \S\ref{sec:res-gate} \\
    Inverse / raw $\hat{e}$ sizing & $w \propto 1/\hat{e}$ adds IR
      & \textbf{Null} ($\rho = 0.616$) & \S\ref{sec:nulls} item~1 \\
    $g(x)$ / $\hat{e}$ aggregates & Cross-sectional UQ detects regimes
      & \textbf{Null} (AUROC $\approx 0.5$) & \S\ref{sec:agg_gap} \\
    VIX-percentile gate & Market vol proxy beats $H(t)$
      & \textbf{Null} (3 FA on high-VIX windows) & \S\ref{sec:res-crisis} \\
    HMM features in $g(x)$ & Explicit regime state improves $\hat{e}$
      & \textbf{Null} ($|\Delta\rho| < 0.001$) & \S\ref{sec:res-deup} \\
    Fast drift $\oplus$ slow EWMA & Same-day drift closes blind window
      & \textbf{Null} (lead $0$d) & \S\ref{sec:res-gate} \\
    Conformal-width overlays & Width-based sizing / abstention
      & \textbf{Null} ($\Delta$Sharpe CI inc.\ 0)
      & \S\ref{sec:res-conformal} \\
    $H(t)$ recombination (M1--M3) & Learned blend beats Platt $H(t)$
      & \textbf{Null} & \S\ref{sec:nulls} item~6 \\
    Partial-horizon RankIC CUSUM & $Y$-channel escapes detectability bound
      & \textbf{Positive} (+25d lead) & \S\ref{sec:partial-horizon} \\
  \bottomrule
  \end{tabularx}
\end{table}

\begin{table}[ht]
  \centering
  \caption{Complete deployment-policy grid (frozen policy evaluation archive,
    February~2025 snapshot). Same metrics as
    Table~\ref{tab:comparators}. FINAL Sharpe$^\dagger$: point estimates
    on $N=5$ complete months only; not for inferential variant ordering
    (cf.\ Table~\ref{tab:ablation} caption).}
  \label{tab:experiment-policy-grid}
  \small
  \begin{tabular}{@{}llrrrr@{}}
    \toprule
    & Variant & Sharpe (ALL) & Sharpe (DEV) & Sharpe (FINAL)$^\dagger$ & Crisis MaxDD \\
    \midrule
    --- & Ungated raw           & 2.730 & 3.107 &  1.650 & $-7.4\%$ \\
    --- & Ungated vol-sized     & 2.614 & 2.991 &  1.603 & $-8.8\%$ \\
    \midrule
    V1 & Gate + Raw             & 1.928 & 2.039 & $-0.322$ & $-7.4\%$ \\
    V2 & Gate + Vol-sizing      & 1.886 & 1.971 &  0.375 & $-8.4\%$ \\
    V3 & Gate + UA-Sort         & 1.933 & 2.046 & $-0.322$ & $-7.4\%$ \\
    V4 & Gate + Resid-$\hat{e}$ & 1.987 & 2.041 &  0.953 & $-6.0\%$ \\
    V5 & Gate + $\hat{e}$-Cap   & 1.702 & 1.750 &  0.731 & $-6.3\%$ \\
    \textbf{V6} & \textbf{Gate + Vol + $\hat{e}$-Cap}
       & \textbf{1.877} & \textbf{1.928} & \textbf{0.925} & $\mathbf{-6.7\%}$ \\
    K  & Gate + Trail-IC        & 1.928 & 2.039 & $-0.322$ & $-7.4\%$ \\
    \bottomrule
  \end{tabular}
\end{table}

\section{Universe-Readiness Diagnostic: Derivation and Tier Ladder}
\label{app:universe-readiness}

This appendix develops the universe-readiness diagnostic applied in
Section~\ref{sec:res-readiness}. It combines two standard results from
active-portfolio theory and time-series statistics into a single readiness
criterion; every numerical value is computed on the frozen
development-panel metrics.

The first ingredient is a \emph{detectability dichotomy}. Write the
cross-sectional return of stock $i$ on date $t$ as
$r_{i,t} = \beta_i F_t + e_{i,t}$, with common factor $F_t$, idiosyncratic
$e_{i,t}$ of variance $\sigma_e^2$ and average pairwise correlation
$\rho_e$, and standardised scores $z_i$ ($\sum_i z_i = 0$,
$\sum_i z_i^2 = N$). The quantity to monitor for skill decay is the
dollar-neutral slope statistic $S_t = \tfrac{1}{N}\sum_i z_i r_{i,t}$
(which coincides with the realised RankIC numerator when scores and returns
are mapped to within-date ranks), not a level statistic
$M_t = \tfrac{1}{N}\sum_i r_{i,t}$. When scores are factor-neutral
($\mathrm{Cov}_{\mathrm{cs}}(z,\beta)=0$) the common factor is annihilated
from $S_t$, so that
\begin{equation}
  \mathrm{Var}(S_t) = \frac{\sigma_e^2(1-\rho_e)}{N}
  \xrightarrow{N\to\infty} 0,
  \qquad
  \mathrm{Var}(M_t) \xrightarrow{N\to\infty}
  \sigma_F^2\,\mu_\beta^2 + \sigma_e^2\rho_e > 0,
  \label{eq:dichotomy}
\end{equation}
where $\sigma_F^2$ is the common-factor variance and $\mu_\beta$ the mean
loading. The slope statistic's noise vanishes with breadth; the level
statistic retains a common-factor floor. Factor-neutrality is equally
load-bearing: if the scores leak factor exposure
($\mathrm{Cov}_{\mathrm{cs}}(z,\beta)=\gamma\neq 0$) the slope statistic
inherits its own floor $\mathrm{Var}(S_t)\to\sigma_F^2\gamma^2>0$ and
ceases to benefit from breadth.

The second ingredient turns Eq.~\eqref{eq:dichotomy} into a delay. Let $N$
denote median breadth, $\phi$ the first-order autocorrelation of the daily
RankIC series, $\Delta\mathrm{IC}$ the peak-to-trough IC shift to detect,
$H_{\mathrm{det}}$ the detection budget in trading days
($H_{\mathrm{det}} = 30$ here, aligned with the null-hold detection
comparison; distinct from the production hold $H_{\mathrm{prod}} = 20$),
and $z_{\alpha}$ the one-sided normal multiplier ($z_{\alpha} = 2$,
$\alpha \approx 0.025$). With per-stock idiosyncratic variance
$v = \sigma_e^{2}(1-\rho_e)$, the per-day detection signal-to-noise ratio
is $\mathrm{SNR}_1 = \Delta\mathrm{IC}\sqrt{N}/(\sigma_e\sqrt{1-\rho_e})$,
the $\sqrt{N}$ breadth scaling of the fundamental law of active management
\citep{grinold1989}. Serial dependence inflates the running-mean variance
by the AR(1) long-run factor $(1+\phi)/(1-\phi)$, giving the
\emph{minimum detection delay}
\begin{equation}
  D^{*}
  \;=\;
  \frac{z_{\alpha}^{2}}{\mathrm{SNR}_1^{2}}\cdot\frac{1+\phi}{1-\phi}
  \;=\;
  \frac{z_{\alpha}^{2}\, v\, (1+\phi)}
       {N\,\Delta\mathrm{IC}^{2}\,(1-\phi)} .
  \label{eq:dstar}
\end{equation}
On the frozen panel the measured daily slope variance is
$\mathrm{Var}(S_t)\approx 0.0116$ (equivalently $v\approx 0.996$ at
$N\approx 86$), reproducing $D^{*}\approx 722$ trading days; plugging
rounded display inputs ($\phi\approx 0.847$,
$\Delta\mathrm{IC}\approx 0.028$) into Eq.~\eqref{eq:dstar} yields
$\approx 713$, rounding, not a separate model. The qualitative conclusion
$D^{*}\gg H_{\mathrm{det}}$ is robust: even at $\sigma_e^{2}=0.7$ the
delay remains $D^{*}\approx 500$. Eq.~\eqref{eq:dstar} concerns
\emph{signal-decay detectability} and is logically distinct from
Theorem~\ref{thm:detectability}, which bounds \emph{score-aggregate}
regime detection under concept shift; the contribution here is the
synthesis into a readiness ladder and its application to universe
design.\footnote{$\phi$ is the first-order autocorrelation of the daily
cross-sectional RankIC of the \emph{score} series, not of a
factor-neutralised residual-IC series. A rebuilt universe should remeasure
$\phi$ on residual, factor-neutral RankIC before breadth planning;
residualisation typically lowers $\phi$, so the figure used here is a
conservative (pessimistic) input. Eq.~\eqref{eq:dstar} also assumes
factor-neutral scores ($\gamma=0$); under leakage the dichotomy installs
an irreducible floor $\sigma_F^{2}\gamma^{2}$, so the reported $D^{*}$ is
an optimistic lower bound.}

With measured values ($N \approx 86$, $\phi \approx 0.847$,
$\Delta\mathrm{IC} \approx 0.028$, $\rho_e \approx 0.004$,
$H_{\mathrm{det}} = 30$): $D^{*} \approx 722$ trading days, roughly 24
times the budget and 36 times the production hold. Even in the optimistic
limit $\phi \to 0$, $D^{*} \approx 60$ days at $N = 86$: no reduction in
autocorrelation alone brings the delay inside the budget at this breadth.
A complementary joint diagnostic pooling score, $\hat{e}$, $g$, and $G$ at
the $d=10$ mid-hold checkpoint via a group-purged, cross-fitted logistic
separator, defining $\mathrm{SNR}_2 = d'^{2}$ from the out-of-sample
standardised class separation and
$D^{*}_{\mathrm{joint}} = z_{\alpha}^{2}/\mathrm{SNR}_2$ without the AR(1)
inflation, attains $D^{*}_{\mathrm{joint}} \approx 100$ trading days
($\mathrm{SNR}_2 \approx 0.040$): a sevenfold shortening that still
exceeds the budget.

Two inversions of Eq.~\eqref{eq:dstar} make the breadth requirement
concrete. Holding measured $\phi$, $\Delta\mathrm{IC}$, $\rho_e$ fixed,
the minimum breadth bringing the delay within budget is
\begin{equation}
  N_{\min}
  = \frac{z_{\alpha}^{2}\,v\,(1+\phi)}
         {H_{\mathrm{det}}\,\Delta\mathrm{IC}^{2}\,(1-\phi)}
  = N\,\frac{D^{*}}{H_{\mathrm{det}}}
  \approx 2{,}070 \text{ stocks},
  \label{eq:nmin}
\end{equation}
roughly twenty-four times present breadth; equivalently, the smallest IC
shift detectable within budget at $N\approx 86$ is
\begin{equation}
  \Delta\mathrm{IC}_{\min}
  = \sqrt{\frac{z_{\alpha}^{2}\,v\,(1+\phi)}{N\,H_{\mathrm{det}}\,(1-\phi)}}
  = \Delta\mathrm{IC}\,\sqrt{D^{*}/H_{\mathrm{det}}}
  \approx 0.137,
  \label{eq:dicmin}
\end{equation}
about five times the entire measured signal: only near-total collapse or
reversal of skill would be identifiable before the position closes.
Table~\ref{tab:dstar-grid} shows breadth, not autocorrelation, is the
binding constraint at $N = 86$.

\begin{table}[htbp]
  \centering
  \caption{$D^{*}$ at $N = 86$ and $N_{\min}$ for
  $D^{*} \le H_{\mathrm{det}}$ across $\phi$, holding measured
  $\Delta\mathrm{IC}$, $\rho_e$, $z_{\alpha}$, and $H_{\mathrm{det}}$
  fixed (Eqs.~\eqref{eq:dstar}--\eqref{eq:nmin}).}
  \label{tab:dstar-grid}
  \small
  \begin{tabular}{@{}ccc@{}}
    \toprule
    $\phi$ & $D^{*}$ at $N = 86$ (days)
      & $N_{\min}$ for $D^{*}\le H_{\mathrm{det}}$ \\
    \midrule
    $0.847$ (measured) & $722$ & $2{,}070$ \\
    $0.5$ & $179$ & $513$ \\
    $0.4$ & $139$ & $399$ \\
    $0.3$ & $111$ & $317$ \\
    $0.0$ & $60$  & $171$ \\
    \bottomrule
  \end{tabular}
\end{table}

The panel also carries an effective sample of only
$\mathrm{eff}\text{-}N \approx 84$ concurrent independent positions
(average-uniqueness over $P \approx 7{,}400$ overlapping holds;
\citealp{lopezdeprado2018}). The autocorrelation $\phi$ and
$\mathrm{eff}\text{-}N$ both arise from the same overlapping-label
dependence; they gate different quantities (detection delay versus
allocator sample size) and are not double-counted. The same overlap
depresses the count of independent decision episodes available to any
allocator trained on this panel ($E_{\mathrm{eff}} \approx 84$), far
below conventional sample-size requirements for stable off-policy
estimation. Table~\ref{tab:universe-tiers} collects the verdicts.

\begin{table}[htbp]
  \centering
  \caption{Universe-readiness tiers and verdicts on the frozen development
  AI panel. Gate definitions:
  Eqs.~\eqref{eq:dstar}--\eqref{eq:dicmin}; measured inputs:
  Section~\ref{sec:res-readiness}.}
  \label{tab:universe-tiers}
  \small
  \begin{tabular}{@{}c l p{6.4cm} c@{}}
    \toprule
    Tier & Programme goal & Primary gates & AI panel \\
    \midrule
    A & Exploratory signal
      & $|t_{\mathrm{IC}}| > 2$
      & Pass$^{\dagger}$ \\
    B & Walk-forward / UQ
      & $\mathrm{eff}\text{-}N \geq 100$;\ \ $|t_{\mathrm{IC}}| > 3$
      & Fail \\
    C & Live-grade (DSR)
      & Tier~B;\ \ deflated-Sharpe check
      & Fail \\
    D & Alpha-timing
      & $D^{*} \leq H_{\mathrm{det}}$;\ \ $\phi \leq 0.5$;\ \
        $\mathrm{eff}\text{-}N \geq 200$
      & Fail \\
    E & RL / allocator
      & Tier~D;\ \ $\mathrm{eff}\text{-}N \geq 500$;\ \
        $E_{\mathrm{eff}} \geq 500$
      & Fail \\
    \bottomrule
  \end{tabular}
  \par\smallskip
  \begin{minipage}{0.95\linewidth}
    \footnotesize $^{\dagger}$Tier~A certifies signal \emph{existence} on
    the development partition ($|t_{\mathrm{IC}}| > 2$ from the DEV RankIC
    analysis), not profitable alpha-decay timing (ruled out by the $D^{*}$
    results) nor the FINAL-window RankIC threshold of
    Section~\ref{sec:res-base}. $|t_{\mathrm{IC}}| > 3$ follows
    \citet{harvey2016}; $\mathrm{eff}\text{-}N$ follows
    \citet{lopezdeprado2018}.
  \end{minipage}
\end{table}

\section{Cost-Sensitivity Stress Table}
\label{app:cost-sensitivity}

Table~\ref{tab:cost-sensitivity} stresses per-leg transaction costs on the
walk-forward vol-sized production portfolio (109 months,
February~2016--February~2025), holding the monthly turnover series and
return history fixed; only the assumed cost per leg varies.

\begin{table}[!htbp]
  \centering
  \caption{Transaction-cost sensitivity of the production portfolio
    (vol-sized LightGBM, monthly rebalancing, 42.7\% mean turnover).}
  \label{tab:cost-sensitivity}
  \footnotesize
  \setlength{\tabcolsep}{3pt}
  \renewcommand{\arraystretch}{1.05}
  \begin{tabularx}{\linewidth}{@{}%
    >{\raggedright\arraybackslash}X%
    >{\centering\arraybackslash}p{0.09\linewidth}%
    >{\centering\arraybackslash}p{0.09\linewidth}%
    >{\centering\arraybackslash}p{0.10\linewidth}%
    >{\centering\arraybackslash}p{0.10\linewidth}%
    >{\centering\arraybackslash}p{0.08\linewidth}@{}}
    \toprule
    \textbf{Cost assumption} & \textbf{Ann.\ cost} &
    \textbf{Net ret.} & \textbf{Net Sharpe} &
    \textbf{$\Delta$Sharpe} & \textbf{OK?} \\
    \midrule
    0 bps/leg (zero cost)            & 0.00\%  & 88.0\% & 2.765 & ---
      & Yes \\
    5 bps/leg (liquid large-cap)     & 0.51\%  & 87.5\% & 2.749
      & $-$0.016 & Yes \\
    \textbf{10 bps/leg (production)} & \textbf{1.02\%} & \textbf{87.0\%}
      & \textbf{2.734} & \textbf{$-$0.031} & \textbf{Yes} \\
    20 bps/leg (illiquid / small-cap)& 2.05\%  & 85.9\% & 2.703
      & $-$0.062 & Yes \\
    30 bps/leg (high slippage)       & 3.07\%  & 84.9\% & 2.672
      & $-$0.093 & Yes \\
    50 bps/leg (extreme stress)      & 5.12\%  & 82.9\% & 2.610
      & $-$0.155 & Yes \\
    \midrule
    5\% p.a.\ borrow on short leg$^{\dagger}$
      & 5.00\%  & 82.0\% & 2.58 & $-$0.15 & Yes \\
    8\% p.a.\ borrow on short leg$^{\dagger}$
      & 8.00\%  & 79.0\% & 2.51 & $-$0.22 & Yes \\
    \bottomrule
  \end{tabularx}
  \par\smallskip
  {\footnotesize $^{\dagger}$Short-borrow fee stress (illustrative): full
    short notional on dollar-neutral top-10/bottom-10 book; not included in
    the base 10~bps/leg transaction-cost model.}
\end{table}

\noindent\textbf{Caveat.}
This is a turnover-based cost stress test, not an institutional-capacity
analysis: borrow availability, locate failures, crowding, and order-book
impact at scale are not modelled in the per-leg rows
(Section~\ref{sec:limitations}). The borrow-fee rows apply a flat annual
short-borrow charge to the full short leg of the dollar-neutral L/S book
(3--8\% p.a.\ range cited in Section~\ref{sec:limitations}), consistent
with the $\Delta$Sharpe mapping of the per-leg transaction-cost rows
($\approx 0.03$ Sharpe per 1\% annualised drag at $\approx 32\%$ vol).
They reduce net Sharpe by $\approx 0.15$--$0.22$ relative to the
10~bps/leg baseline and are directionally consistent with internal
capacity estimates ($\approx$0.1--0.3 Sharpe drag for hard-to-borrow AI
names). Ten bps per leg remains plausible
for liquid large-caps but optimistic for smaller or hard-to-borrow AI
names, where round-trip costs of 60--160~bps are possible.

\section{Reproducibility}
\label{app:reproducibility}

The rank-displacement DEUP methodology is distributed as the open-source
\texttt{deup} Python package \citep{sanderink2026deup}, which generalises
the pipeline for scikit-learn and gradient-boosted backends, supplies
leakage-correct out-of-fold error construction and purged walk-forward
splitters for finance panels, and documents walk-forward parity on the
rank-displacement protocol used here. Reported numbers were produced from a
private research build frozen at the February~2025 evaluation boundary; the
proprietary equity panel and walk-forward artefacts are not redistributed.
Table~\ref{tab:experiment-registry} lists every main-text falsification
experiment and its verdict (Appendix~\ref{app:experiment-registry}).
A post-freeze monitoring audit trail is retained in the private research
archive; portfolio metrics there are \textbf{not} part of the evidence base
(Section~\ref{sec:limitations}). Monitoring alert logs are summarised for
transparency in Section~\ref{sec:res-monitoring-corr}. Public
evaluation inputs: Fama--French factor returns \citep{french2025},
CIFAR-10-C \citep{hendrycks2019}, and market data accessed under
non-commercial provider terms (Polygon.io, Financial Modelling Prep,
Alpaca Markets). All random processes are seeded
(\texttt{random\_state} $= 42$), including the primary ranker, the error
predictor, and the $5{,}000$-trial permutation tests; the finance pipeline
runs on a versioned, read-only DuckDB feature store, so re-running the
walk-forward loop reproduces bit-identical input matrices.
Table~\ref{tab:software-stack} lists the principal packages.

\begin{table}[ht]
  \centering
  \caption{Software stack. PyTorch is required only for the CIFAR-10-C
    experiment of Section~\ref{sec:cifar_external}.}
  \label{tab:software-stack}
  \small
  \begin{tabular}{@{}llp{0.48\linewidth}@{}}
    \toprule
    \textbf{Package} & \textbf{Version} & \textbf{Purpose} \\
    \midrule
    Python           & $\geq$3.10  & Runtime environment \\
    LightGBM         & $\geq$4.0   & Primary ranker and DEUP error
      predictor $g(x)$ \\
    DuckDB           & $\geq$0.9   & Versioned columnar feature store
      (read-only) \\
    pandas           & $\geq$2.0   & Data wrangling and panel construction \\
    NumPy            & $\geq$1.24  & Numerical array operations \\
    SciPy            & $\geq$1.10  & Spearman rank-IC and permutation
      statistics \\
    scikit-learn     & $\geq$1.3   & Preprocessing and AUROC computation \\
    statsmodels      & $\geq$0.14  & Newey--West HAC variance and ACF
      diagnostics \\
    PyArrow          & $\geq$14.0  & Parquet serialisation for frozen
      snapshots \\
    PyTorch          & $\geq$2.0   & ResNet-18 backbone for CIFAR-10-C \\
    \bottomrule
  \end{tabular}
\end{table}

\section{Signal-Hygiene Checklist}
\label{app:hygiene}

During walk-forward signal generation on the thematic universe, three
data-hygiene defect classes were observed in the research pipeline (not
validated as part of a live track record): a delisted ticker passing the
universe filter after its exit, an acquisition-pending ticker entering a
signal with no tradable market data, and a stale vendor quote persisting
across several consecutive signal generations. They generalise to four
qualitative rules for any thematic long--short ranker whose universe churns:
\begin{enumerate}[leftmargin=*,itemsep=3pt]
  \item \textbf{Static blocklist, refreshed before each signal.} Maintain
  an explicit blocklist of delisted and acquisition-pending tickers and
  apply it before signal generation, independently of the dynamic
  tradability filters, corporate-action state changes propagate to vendor
  universe flags with lag.
  \item \textbf{Reject stale quotes.} Treat any quote older than one
  trading session as missing rather than as a price; a stale quote entering
  a cross-sectional rank silently distorts every neighbouring rank.
  \item \textbf{Reconcile corporate-action feeds first.} Run the
  corporate-action reconciliation (splits, mergers, delistings) strictly
  before signal generation, not as a post-hoc audit.
  \item \textbf{Exclude, never reprice.} A non-investable pick is dropped
  from the book; substituting a proxy price or carrying the position
  forward converts a data defect into a fictitious return.
\end{enumerate}

\section{Meta-Allocator Interface}
\label{app:allocator}

Table~\ref{tab:allocator-interface} specifies the per-date outputs the
framework emits for a downstream meta-allocator, with the section in which
each is defined and validated.

\begin{table}[ht]
  \centering
  \caption{Meta-allocator interface: per-date framework outputs.}
  \label{tab:allocator-interface}
  \small
  \begin{tabularx}{\linewidth}{@{}lllX@{}}
    \toprule
    \textbf{Field} & \textbf{Type} & \textbf{Frequency} &
    \textbf{Defined in} \\
    \midrule
    $\widehat{P}(\text{good day} \mid H(t))$ & float $[0,1]$ & daily &
      Section~\ref{sec:res-gate} (Platt-calibrated) \\
    $\Gamma(t)$ & float $[0,1]$ & daily &
      Eq.~(\ref{eq:gamma-gate}), Section~\ref{sec:method-gate} \\
    $G(t)$ & binary & daily &
      Section~\ref{sec:method-gate} \\
    $\hat{e}(x_{i,t})$ cross-sectional percentile & float $[0,1]$ &
      daily, per position &
      Eq.~(\ref{eq:epistemic-signal}), Section~\ref{sec:method-deup} \\
    $\mathrm{width\_deup}$ / $\mathrm{risk\_score}$ &
      float / float $[0,1]$ & daily, per position &
      Section~\ref{sec:res-conformal} \\
    \bottomrule
  \end{tabularx}
\end{table}

\section{Additional Figures and Tables}
\label{app:extra-floats}

\begin{figure}[ht]
  \centering
  \resizebox{0.95\linewidth}{!}{%
  \begin{tikzpicture}[
      node distance=8mm and 8mm,
      every node/.style={font=\small},
      stage/.style={draw, rounded corners=2pt, align=center,
                    text width=26mm, minimum height=14mm, inner sep=2pt,
                    execute at begin node={\baselineskip=9pt\relax}},
      input/.style={stage, fill=blue!8},
      model/.style={stage, fill=gray!10},
      deploy/.style={stage, fill=green!10},
      eval/.style={stage, fill=orange!10},
      arr/.style={-{Stealth[length=2mm]}, thick},
    ]
    \node[input]                          (s1) {\textbf{1.~Data \&\\Feature Store}\\[1pt]
                                                \scriptsize SIP tape $+$ DuckDB};
    \node[input, right=of s1]             (s2) {\textbf{2.~Universe\\Construction}\\[1pt]
                                                \scriptsize AI-themed,\\[-1pt]
                                                \scriptsize $\leq 100$ stocks};
    \node[input, right=of s2]             (s3) {\textbf{3.~Label\\Construction}\\[1pt]
                                                \scriptsize 20/60/90d\\[-1pt]
                                                \scriptsize excess vs.~QQQ};
    \node[input, right=of s3]             (s4) {\textbf{4.~Feature\\Engineering}\\[1pt]
                                                \scriptsize 50$+$ features,\\[-1pt]
                                                \scriptsize 8 categories};
    \node[model,  below=15mm of s4]       (s5) {\textbf{5.~LightGBM\\Walk-Forward}\\[1pt]
                                                \scriptsize 109 folds,\\[-1pt]
                                                \scriptsize expanding DEV};
    \node[model,  left=of s5]             (s6) {\textbf{6.~DEUP Uncertainty\\Pipeline}\\[1pt]
                                                \scriptsize $g(x),\,a(x),\,\hat{e}(x)$};
    \node[deploy, left=of s6]             (s7) {\textbf{7.~Two-Level\\Deployment}\\[1pt]
                                                \scriptsize $G(t)$ gate $+$\\[-1pt]
                                                \scriptsize $\hat{e}$-cap};
    \node[eval,   left=of s7]             (s8) {\textbf{8.~Evaluation \&\\Robustness}\\[1pt]
                                                \scriptsize Sharpe, PSR/DSR,\\[-1pt]
                                                \scriptsize multi-crisis};
    \draw[arr] (s1) -- (s2);
    \draw[arr] (s2) -- (s3);
    \draw[arr] (s3) -- (s4);
    \draw[arr] (s4) -- (s5);
    \draw[arr] (s6) -- (s7);
    \draw[arr] (s7) -- (s8);
  \end{tikzpicture}
  }
  \caption{Framework of model development. Stages~1--4 (blue) construct
    the point-in-time feature panel; stages~5--6 (gray) fit the
    walk-forward ranker and the DEUP pipeline; stage~7 (green) applies the
    two-level deployment rule; stage~8 (orange) runs the acceptance battery
    of Section~\ref{sec:method-eval}.}
  \label{fig:framework}
\end{figure}
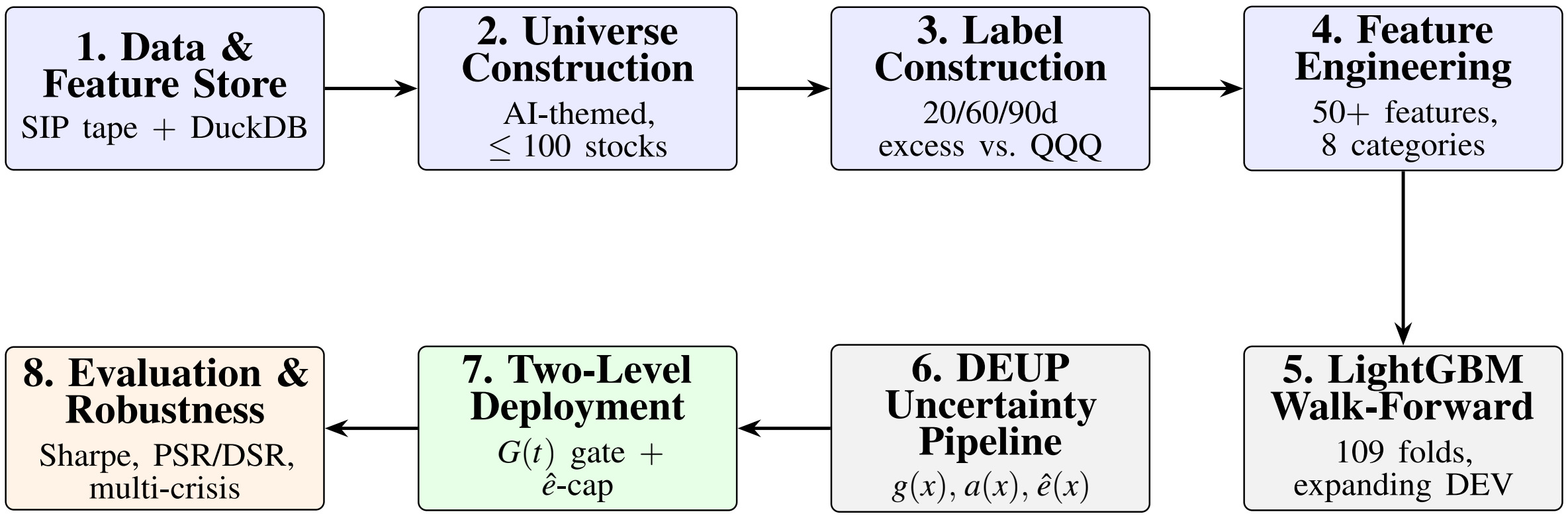

\begin{figure}[ht]
  \centering
  \resizebox{0.88\linewidth}{!}{%
  \begin{tikzpicture}[
      node distance=8mm and 12mm,
      every node/.style={font=\small},
      box/.style={draw, rounded corners=2pt, align=center,
                  minimum height=10mm, minimum width=24mm, inner sep=3pt},
      sigbox/.style={draw, fill=gray!10, rounded corners=2pt, align=center,
                     minimum height=12mm, minimum width=18mm, inner sep=3pt},
      decision/.style={draw, diamond, aspect=2, align=center, inner sep=2pt,
                       minimum width=24mm},
      arr/.style={-{Stealth[length=2mm]}, thick},
    ]
    \node[box] (hreal)  {$H_{\text{real}}(t)$\\[1pt]
                         \scriptsize EWMA matured RankIC};
    \node[box, below=of hreal]  (hdrift) {$H_{\text{drift}}(t)$\\[1pt]
                         \scriptsize feature \& score drift};
    \node[box, below=of hdrift] (hdis)   {$H_{\text{disagree}}(t)$\\[1pt]
                         \scriptsize cross-model spread};
    \node[sigbox, right=15mm of hdrift] (sig) {$h(\cdot)$\\[1pt]
                         \scriptsize weighted\\[-1pt]\scriptsize sigmoid};
    \node[box, right=of sig] (h) {$H(t)\!\in\![0,1]$};
    \node[box, right=of h] (gam) {$\Gamma(t)$};
    \node[decision, right=of gam] (g) {\scriptsize $G(t){=}1$?\\[-1pt]
                         \scriptsize $\Gamma\!\geq\!0.2$};
    \node[box, right=20mm of g, yshift=11mm, fill=green!10] (trade)
        {\textbf{Trade}\\[1pt]
         \scriptsize Vol-sizing $+$ $\hat{e}$-cap\\[-1pt]
         \scriptsize (Section~\ref{sec:method-twolevel})};
    \node[box, right=20mm of g, yshift=-11mm, fill=red!10] (cash)
        {\textbf{Abstain}\\[1pt]
         \scriptsize zero gross\\[-1pt]\scriptsize exposure};
    \draw[arr] (hreal.east)  -- (sig.west);
    \draw[arr] (hdrift.east) -- (sig.west);
    \draw[arr] (hdis.east)   -- (sig.west);
    \draw[arr] (sig)         -- (h);
    \draw[arr] (h)           -- (gam);
    \draw[arr] (gam)         -- (g);
    \draw[arr] (g.north east) -- node[above, sloped, font=\scriptsize,
                                   pos=0.35] {Yes} (trade.west);
    \draw[arr] (g.south east) -- node[below, sloped, font=\scriptsize,
                                   pos=0.35] {No}  (cash.west);
  \end{tikzpicture}
  }
  \caption{Regime-trust gate detail. The three PIT-safe health components
    are combined via a weighted sigmoid into $H(t)$
    (Eq.~\ref{eq:gate}), rescaled to $\Gamma(t)$
    (Eq.~\ref{eq:gamma-gate}), and binarised at the DEV-fitted $0.2$
    threshold. In a multi-strategy framework, $H(t)$ serves as an
    expert-level health signal redirecting notional to active strategies
    rather than as a cash-hold rule.}
  \label{fig:regime-gate}
\end{figure}
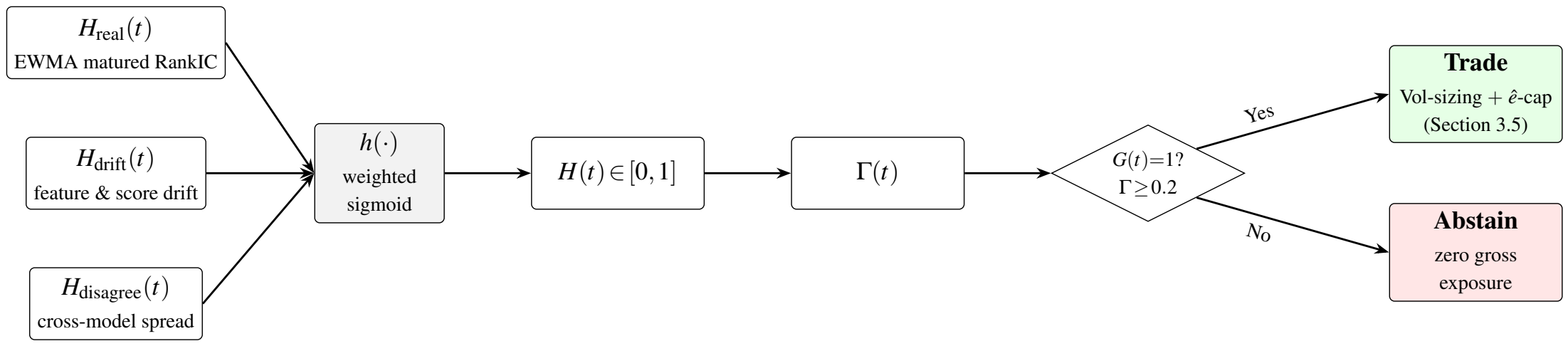

\begin{figure}[!htbp]
  \centering
  \includegraphics[width=\linewidth]{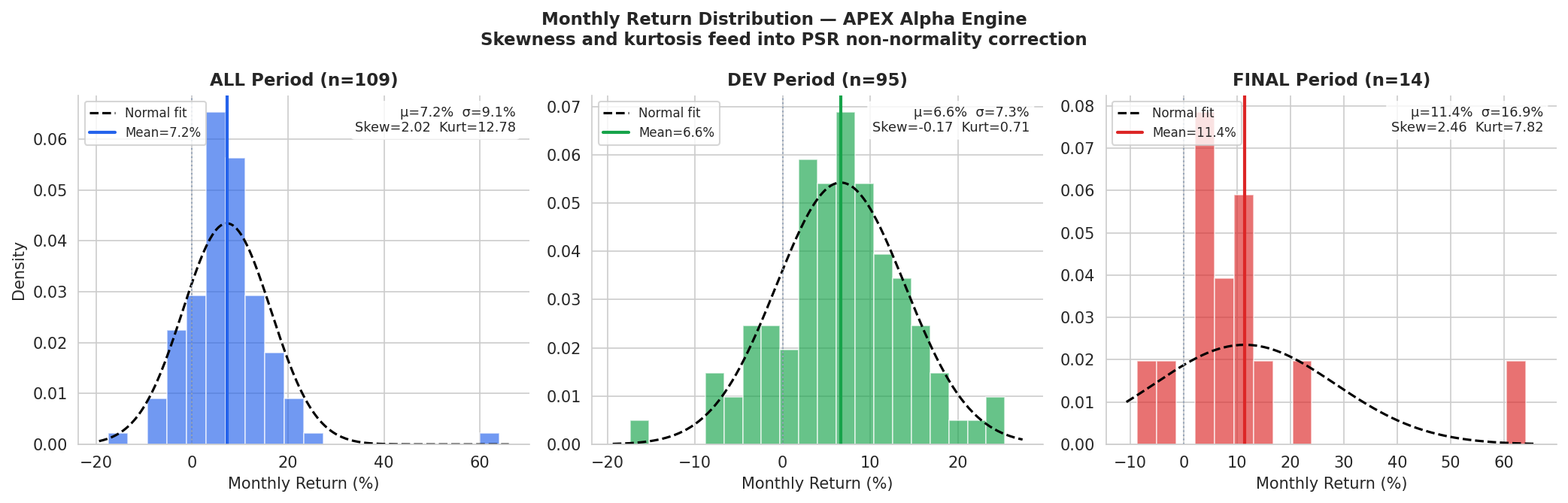}
  \caption{Monthly return distributions of the ungated equal-weight shadow
    portfolio on the vol-sized LightGBM ranker (ALL $n = 109$; DEV $n = 95$;
    FINAL $n = 14$) with fitted normals (dashed). Moments on monthly returns: ALL $\mu = 7.2\%$,
    $\sigma = 9.1\%$, skew $2.02$, excess kurtosis $12.78$; DEV $\mu = 6.6\%$,
    $\sigma = 7.3\%$, skew $-0.17$, excess kurtosis $0.71$; FINAL $\mu = 11.4\%$,
    $\sigma = 16.9\%$, skew $2.46$, excess kurtosis $7.82$ (normal baseline
    $=0$). Eq.~(\ref{eq:psr}) uses raw kurtosis $\widehat{\gamma}_4$, not
    excess kurtosis.}
  \label{fig:return_dist_52}
\end{figure}

\begin{figure}[!htbp]
  \centering
  \includegraphics[width=0.88\linewidth]{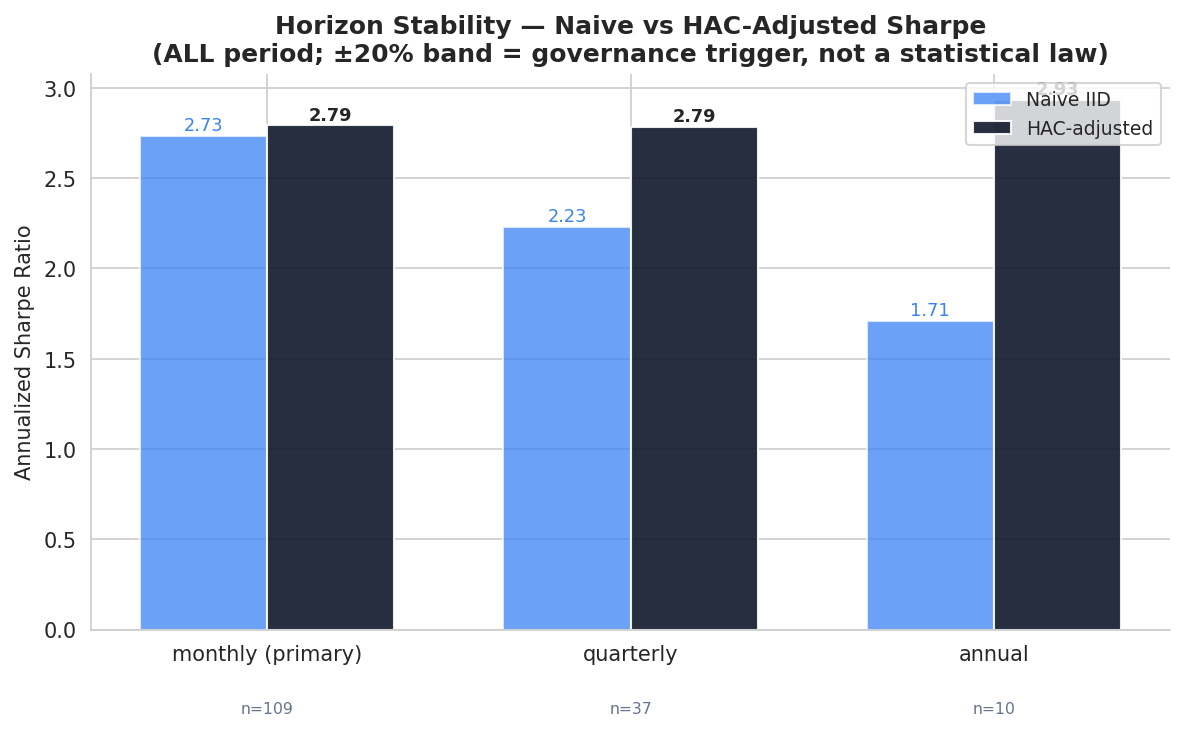}
  \caption{Horizon stability of the annualised Sharpe ratio (ALL period).
    Paired bars: naive IID Sharpe (light) and HAC-adjusted Sharpe (dark) at
    monthly ($n = 109$), quarterly ($n = 37$), and annual ($n = 10$)
    frequencies. The $\pm 20\%$ band is a monitoring threshold, not a
    statistical test.}
  \label{fig:horizon_stab_52}
\end{figure}

\begin{figure}[!htbp]
  \centering
  \includegraphics[width=0.92\linewidth]{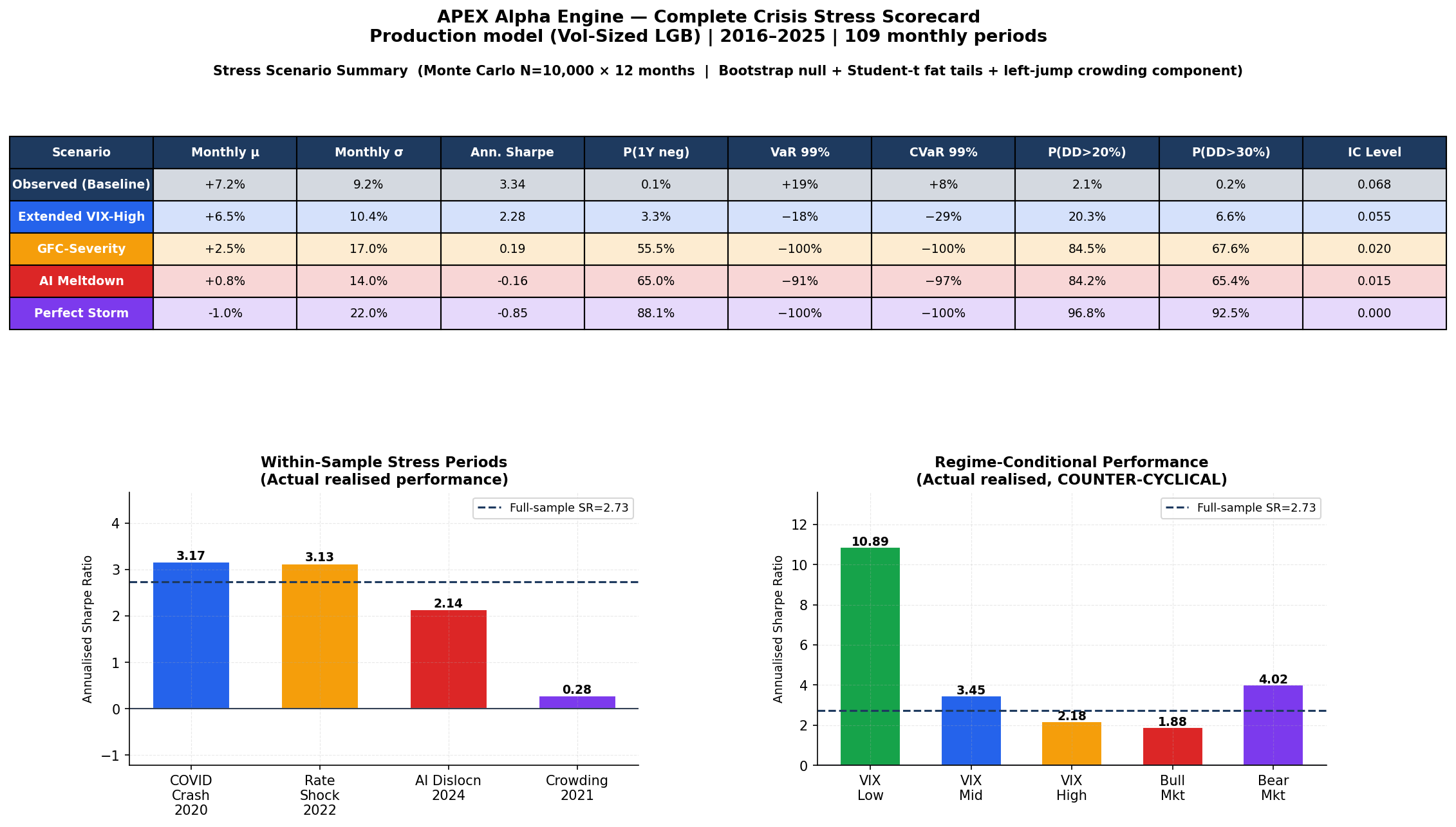}
  \caption{Crisis stress scorecard (vol-sized LightGBM, 2016--2025, 109
    monthly periods). Top: Monte Carlo scenario summary
    ($N = 10{,}000 \times 12$ months; Student-$t$ tails with Bernoulli
    left-jump crowding component). Bottom left: realised annualised Sharpe
    in four historical stress episodes (dashed $= 2.73$, full sample).
    Bottom right: regime-conditional Sharpe ratios (VIX terciles; bull and
    bear regimes).}
  \label{fig:stress_scorecard_52}
\end{figure}

\begin{figure}[!htbp]
  \centering
  \includegraphics[width=0.88\linewidth]{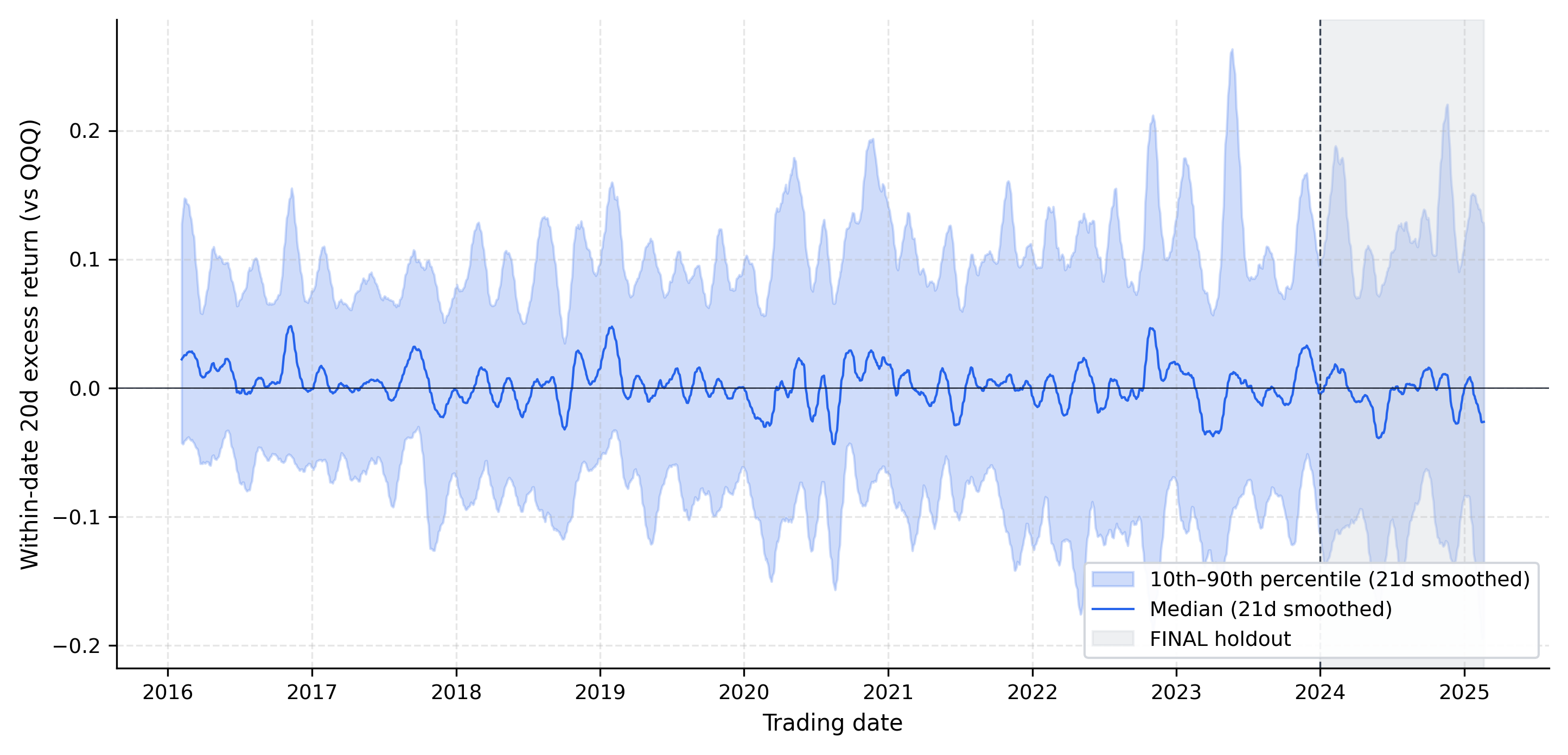}
  \caption{Cross-sectional dispersion of 20-day benchmark-relative excess
    returns: within-date 10th, 50th and 90th percentiles per evaluation
    date, 21-day rolling mean. Shaded: FINAL holdout. Mean 10th--90th
    spread: $0.19$ (DEV) versus $0.24$ (FINAL).}
  \label{fig:ch4-dispersion}
\end{figure}

\begin{table}[ht]
  \centering
  \caption{Feature categories and representative variables. Top-importance
    pair by aggregate LightGBM gain (Category~2): \texttt{adv\_20d} and
    \texttt{adv\_60d} (combined $\approx$20.4\%).}
  \label{tab:feature-categories}
  \small
  \begin{tabular}{@{}lp{0.60\linewidth}@{}}
    \toprule
    \textbf{Category} & \textbf{Representative features} \\
    \midrule
    1.\ Price momentum
      & \texttt{mom\_1m}, \texttt{mom\_3m}, \texttt{mom\_6m},
        \texttt{mom\_12m} \\[3pt]
    2.\ Volatility and liquidity
      & \texttt{vol\_20d}, \texttt{vol\_60d}, \texttt{vol\_of\_vol},
        \texttt{adv\_20d}, \texttt{adv\_60d} \\[3pt]
    3.\ Drawdown and relative strength
      & \texttt{max\_drawdown\_60d}, \texttt{rel\_strength\_1m},
        \texttt{rel\_strength\_3m} \\[3pt]
    4.\ Regime and macro indicators
      & \texttt{vix\_percentile\_252d}, \texttt{market\_return\_21d},
        \texttt{market\_vol\_21d}, \texttt{credit\_ratio},
        \texttt{breadth\_ratio}, \texttt{downside\_rv\_share} \\[3pt]
    5.\ Fundamental quality
      & earnings quality composite, balance-sheet stability,
        profitability score \\[3pt]
    6.\ Earnings events
      & announcement proximity, earnings surprise,
        post-announcement drift (5d) \\[3pt]
    7.\ Missingness masks
      & coverage fraction, staleness indicator,
        new-stock history flag \\[3pt]
    8.\ Feature hygiene diagnostics
      & collinearity flags, importance stability score,
        PIT-scan output (allowlist governance; not direct ranker inputs) \\
    \bottomrule
  \end{tabular}
\end{table}

\begin{table}[ht]
  \centering
  \caption{Feature-importance stability across the walk-forward sample
    (frozen LightGBM specification). Rank correlation compares
    gain-importance orderings from models re-fit on an early window
    (2016--2019) and a late window (2022--2023); top-3 overlap counts
    common features among the three highest-gain inputs.}
  \label{tab:feat-importance-531}
  \small
  \begin{tabular}{lccl}
    \toprule
    Horizon & Rank corr.\ (early vs late) & Top-3 overlap
      & Top-3 features \\
    \midrule
    20d & 0.976 & 3/3 &
      \texttt{adv\_20d}, \texttt{vol\_60d}, \texttt{mom\_12m} \\
    60d & 0.976 & 3/3 &
      \texttt{adv\_20d}, \texttt{mom\_12m}, \texttt{vol\_60d} \\
    90d & 0.952 & 3/3 &
      \texttt{adv\_20d}, \texttt{vol\_60d}, \texttt{mom\_12m} \\
    \bottomrule
  \end{tabular}
\end{table}

\begin{table}[ht]
  \centering
  \caption{Five deepest drawdown episodes of the walk-forward 20-day
    monthly long--short shadow portfolio (LGB baseline score ranking,
    non-overlapping monthly equity curve; full-sample max DD $-21.9\%$ on
    that curve, versus $-18.1\%$ on the headline vol-sized production
    curve). Max~DD is peak-to-trough depth; VIX bucket and market regime
    measured at the trough month. Environment labels are descriptive macro
    annotations, not causal attributions.}
  \label{tab:drawdown-episodes-531}
  \small
  \begin{tabular}{rlllll}
    \toprule
    Max DD & Trough & Duration & VIX & Regime & Environment \\
    \midrule
    $-21.9\%$ & 2021-04 & 9 months & high & bull & Meme-stock mania \\
    $-16.6\%$ & 2024-07 & 8 months & high & bull & AI thematic rally \\
    $-11.5\%$ & 2023-10 & 4 months & high & bear & Rate-hiking cycle \\
    $-8.2\%$  & 2020-09 & 3 months & mid  & bull & COVID recovery \\
    $-8.0\%$  & 2022-01 & 4 months & high & bull & Inflation shock \\
    \bottomrule
  \end{tabular}
\end{table}

\begin{sidewaystable}[p]
  \centering
  \caption{Comparison of uncertainty-quantification methods for
    cross-sectional equity ranking deployment.}
  \label{tab:uq-methods}
  \resizebox{0.90\textheight}{!}{%
  \renewcommand{\arraystretch}{1.18}%
  \begin{tabular}{@{}lccllcll@{}}
    \toprule
    \textbf{Method}
      & \textbf{Epi.}
      & \textbf{Ale.}
      & \textbf{Calibration}
      & \textbf{Cost}
      & \textbf{GBT}
      & \textbf{Finance app.}
      & \textbf{Key limitation} \\
    \midrule
    MC Dropout$^{[1]}$
      & $\checkmark^{*}$
      & $\checkmark^{\dagger}$
      & Moderate; weaker OOD
      & Low--med (multi-pass)
      & $\times$
      & Indirect
      & GBT-incompat.$^{*}$ \\[1pt]
    Deep Ensembles$^{[2]}$
      & $\checkmark$
      & $\checkmark^{\dagger}$
      & Strong; OOD-robust
      & Med--high (multi-model)
      & $\times$
      & Indirect
      & Retraining cost prohibitive walk-forward \\[1pt]
    Bayesian NN (VI)$^{[3]}$
      & $\checkmark$
      & $\checkmark$
      & Conservative; variable
      & High (VI overhead)
      & $\times$
      & Indirect
      & VI approx.\ error; GBT-incompat. \\[1pt]
    Evidential DL (NIG)$^{[4]}$
      & $\checkmark$
      & $\checkmark$
      & Miscal.\ under vol clustering
      & Low (single pass)
      & $\times$
      & Equities, crypto
      & NIG prior unsuited to GARCH dynamics \\[1pt]
    PGBM$^{[8]}$
      & $\times$
      & $\checkmark$ (leaf-weight var.)
      & Moderate (parametric dist.)
      & Low (single GBM)
      & $\checkmark$
      & None
      & Total variance only; no epis.\ decomp. \\[1pt]
    GBT Ensembles$^{[9]}$
      & $\checkmark^{d}$ (proxy)
      & $\checkmark^{d}$ (proxy)
      & Moderate (ensemble variance)
      & Med--high (multi-model; SGLB variant)
      & $\checkmark$
      & None
      & Disagreement proxy $\neq$ Bregman excess-risk \\[1pt]
    UACQR$^{[10]}$
      & $\checkmark^{d}$ (proxy)
      & $\checkmark$
      & Dist.-free; stronger cond.\ cov.
      & Med (bagged quantile regressors)
      & $\checkmark$
      & None
      & Interval-level only; no finance app. \\[1pt]
    \textbf{DEUP}$^{[5]}$
      & \textbf{$\checkmark$ (excess risk)}
      & \textbf{$\checkmark$ (proxy)}
      & \textbf{Adaptive; depends on $\boldsymbol{a(x)}$}
      & \textbf{Low (1 model)}
      & $\checkmark$
      & \textbf{Equities}$^{a}$
      & \textbf{Aleatoric floor non-trivial; no indep.\ ranking
        lit.}$^{a}$ \\[1pt]
    Conformal Pred.$^{[6],f}$
      & Combined
      & Combined
      & Dist.-free; ACI/weighted variants shift-robust
      & Low (cal.\ layer)
      & $\checkmark$
      & Emerging$^{b}$
      & No decomp.; basic CP degrades under shift \\[1pt]
    Cond.\ vol./quant.\ reg.$^{[7]}$
      & $\times$
      & $\checkmark$
      & Strong for cond.\ vol/quantiles
      & Low
      & $\checkmark$
      & Standard (vol, VaR, quantiles)
      & No epistemic component \\
    \bottomrule
  \end{tabular}}\\[4pt]
  \begin{minipage}{0.90\textheight}
    \scriptsize
    \emph{Notes.}~$^{*}$Dropout-as-Bayesian contested
    \citep{lefolgoc2021}.~$^{\dagger}$Epi./ale.\ entangled in predictive
    variance.~$^{a}$Author's preliminary study
    \citep{sanderink2026alpha}; no independent DEUP-on-ranking
    publication.~$^{b}$\citet{schmitt2026}; \citet{kaya2025};
    \citet{fermanian2025}.~$^{d}$Ensemble-variance proxies $\neq$ Bregman
    excess-risk.~$^{f}$EPICSCORE/CREDO
    \citep{cabezas2025,cabezas2026}: epistemic inflation in conformal
    scores, interval-level only.~`Indirect' $=$ benchmark validation
    without a finance case study; `GBT' ($\checkmark$) $=$ applicable
    without architectural change to tree backbones.\\[2pt]
    \emph{Citations.}~[1] \citet{gal2016}; [2] \citet{lakshminarayanan2017};
    [3] \citet{neal1996}; [4] \citet{wong2025}; [5] \citet{lahlou2023};
    [6] \citet{angelopoulos2023} and shift-robust variants
    \citep{barber2023,gibbs2021,gibbs2024}; [7] finance-native
    vol./quantile practice \citep{blasco2024}; [8] \citet{sprangers2021};
    [9] \citet{ustimenko2021}; [10] \citet{rossellini2024}.
  \end{minipage}
\end{sidewaystable}


\begin{thebibliography}{100}
\providecommand{\natexlab}[1]{#1}
\providecommand{\url}[1]{\texttt{#1}}
\expandafter\ifx\csname urlstyle\endcsname\relax
  \providecommand{\doi}[1]{doi: #1}\else
  \providecommand{\doi}{doi: \begingroup \urlstyle{rm}\Url}\fi

\bibitem[Adams and MacKay(2007)]{adams2007}
Ryan~P. Adams and David J.~C. MacKay.
\newblock Bayesian online changepoint detection.
\newblock \emph{arXiv preprint}, 2007.
\newblock arXiv:0710.3742.

\bibitem[Allen et~al.(2025)Allen, Kacperczyk, and Kumar]{allen2025}
Franklin Allen, Marcin~T. Kacperczyk, and Karthik S.~S. Kumar.
\newblock Modern machine learning tools in finance: A critical perspective.
\newblock Technical Report 2025/139, HKU Jockey Club Enterprise Sustainability
  Global Research Institute, 2025.
\newblock SSRN:5439898.

\bibitem[Angelopoulos and Bates(2023)]{angelopoulos2023}
Anastasios~N. Angelopoulos and Stephen Bates.
\newblock Conformal prediction: A gentle introduction.
\newblock \emph{Foundations and Trends in Machine Learning}, 16\penalty0
  (1--2):\penalty0 1--247, 2023.

\bibitem[Arian et~al.(2024)Arian, {Norouzi Mobarekeh}, and Seco]{arian2024}
Hamid~R. Arian, Daniel {Norouzi Mobarekeh}, and Luis~A. Seco.
\newblock Backtest overfitting in the machine learning era: A comparison of
  out-of-sample testing methods in a synthetic controlled environment.
\newblock \emph{Knowledge-Based Systems}, 286:\penalty0 111219, 2024.

\bibitem[Bailey and {L{\'o}pez de Prado}(2012)]{bailey2012}
David~H. Bailey and Marcos {L{\'o}pez de Prado}.
\newblock The {S}harpe ratio efficient frontier.
\newblock \emph{Journal of Risk}, 15\penalty0 (2):\penalty0 3--44, 2012.

\bibitem[Bailey and {L{\'o}pez de Prado}(2014)]{bailey2014}
David~H. Bailey and Marcos {L{\'o}pez de Prado}.
\newblock The deflated {S}harpe ratio: Correcting for selection bias, backtest
  overfitting, and non-normality.
\newblock \emph{Journal of Portfolio Management}, 40\penalty0 (5):\penalty0
  94--107, 2014.

\bibitem[Barber et~al.(2023)Barber, Cand{\`e}s, Ramdas, and
  Tibshirani]{barber2023}
Rina~Foygel Barber, Emmanuel~J. Cand{\`e}s, Aaditya Ramdas, and Ryan~J.
  Tibshirani.
\newblock Conformal prediction beyond exchangeability.
\newblock \emph{The Annals of Statistics}, 51\penalty0 (2):\penalty0 816--845,
  2023.
\newblock \doi{10.1214/23-AOS2276}.

\bibitem[Barroso and Saxena(2021)]{barroso2021}
Pedro Barroso and Konark Saxena.
\newblock Lest we forget: Learn from out-of-sample forecast errors.
\newblock \emph{Review of Financial Studies}, 34\penalty0 (7):\penalty0
  3609--3647, 2021.

\bibitem[{Bickford Smith} et~al.(2025){Bickford Smith}, Kossen, Trollope, {van
  der Wilk}, Foster, and Rainforth]{bickford2025}
Freddie {Bickford Smith}, Jannik Kossen, Eleanor Trollope, Mark {van der Wilk},
  Adam Foster, and Tom Rainforth.
\newblock Rethinking aleatoric and epistemic uncertainty.
\newblock In \emph{Proceedings of the 42nd International Conference on Machine
  Learning (ICML 2025)}, volume 267 of \emph{PMLR}, 2025.
\newblock arXiv:2412.20892.

\bibitem[Blasco et~al.(2024)Blasco, S{\'a}nchez, and Garc{\'i}a]{blasco2024}
Txus Blasco, J.~Salvador S{\'a}nchez, and Vicente Garc{\'i}a.
\newblock A survey on uncertainty quantification in deep learning for financial
  time series prediction.
\newblock \emph{Neurocomputing}, 589:\penalty0 126193, 2024.

\bibitem[Buehler et~al.(2025)Buehler, Horvath, Limmer, and
  Schmidt]{buehler2025}
Hans Buehler, Blanka Horvath, Yannick Limmer, and Thorsten Schmidt.
\newblock Uncertainty-aware strategies: A model-agnostic framework for robust
  financial optimization through subsampling.
\newblock \emph{arXiv preprint}, 2025.
\newblock arXiv:2506.07299.

\bibitem[Cabezas et~al.(2025)Cabezas, Santos, Ramos, and Izbicki]{cabezas2025}
Luben M.~C. Cabezas, Vagner~S. Santos, Thiago~R. Ramos, and Rafael Izbicki.
\newblock Epistemic uncertainty in conformal scores: A unified approach.
\newblock In \emph{Proceedings of the 41st Conference on Uncertainty in
  Artificial Intelligence (UAI 2025)}, volume 244 of \emph{PMLR}, pages
  443--470, 2025.
\newblock arXiv:2502.06995.

\bibitem[Cabezas et~al.(2026)Cabezas, Sloman, Resende, Wu, Caprio, and
  Izbicki]{cabezas2026}
Luben M.~C. Cabezas, Sabina~J. Sloman, Bruno~M. Resende, Fanghui Wu, Michele
  Caprio, and Rafael Izbicki.
\newblock {CREDO}: Epistemic-aware conformalized credal envelopes for
  regression.
\newblock \emph{arXiv preprint}, 2026.
\newblock arXiv:2603.06826.

\bibitem[Cassar and Gerakos(2017)]{cassar2017}
Gavin Cassar and Joseph Gerakos.
\newblock Do risk management practices work? {E}vidence from hedge funds.
\newblock \emph{Review of Accounting Studies}, 22\penalty0 (3):\penalty0
  1084--1121, 2017.
\newblock \doi{10.1007/s11142-017-9403-5}.

\bibitem[Chaudhuri and Lopez-Paz(2023)]{chaudhuri2023}
Kamalika Chaudhuri and David Lopez-Paz.
\newblock Unified uncertainty calibration.
\newblock \emph{arXiv preprint}, 2023.
\newblock arXiv:2310.01202.

\bibitem[Chow(1970)]{chow1970}
C.~K. Chow.
\newblock On optimum recognition error and reject tradeoff.
\newblock \emph{IEEE Transactions on Information Theory}, 16\penalty0
  (1):\penalty0 41--46, 1970.

\bibitem[Cui et~al.(2025)Cui, Feng, Ma, and Su]{cui2025}
Liyuan Cui, Guanhao Feng, Jingyu Ma, and Yongmiao Su.
\newblock Breaks and trends in factor premia.
\newblock \emph{SSRN preprint}, 2025.
\newblock SSRN:5330449.

\bibitem[Deng et~al.(2023)Deng, Suh, Gould, and Zheng]{deng2023}
Weijian Deng, Yumin Suh, Stephen Gould, and Liang Zheng.
\newblock Confidence and dispersity speak: Characterizing prediction matrix for
  unsupervised accuracy estimation.
\newblock In \emph{Proceedings of the 40th International Conference on Machine
  Learning (ICML 2023)}, volume 202 of \emph{PMLR}, pages 7658--7674, 2023.

\bibitem[Depeweg et~al.(2018)Depeweg, Hern{\'a}ndez-Lobato, Doshi-Velez, and
  Udluft]{depeweg2018}
Stefan Depeweg, Jos{\'e}~Miguel Hern{\'a}ndez-Lobato, Finale Doshi-Velez, and
  Steffen Udluft.
\newblock Decomposition of uncertainty in {B}ayesian deep learning for
  efficient and risk-sensitive learning.
\newblock In \emph{Proceedings of the 35th International Conference on Machine
  Learning (ICML 2018)}, pages 1184--1193, 2018.

\bibitem[Eggen et~al.(2025)Eggen, Espe, Grude, Risstad, and
  Sandberg]{eggen2025}
Simen Eggen, Tobias~J. Espe, Kristian Grude, Morten Risstad, and Rune Sandberg.
\newblock Financial time series uncertainty: A review of probabilistic {AI}
  applications.
\newblock \emph{Journal of Economic Surveys}, 40\penalty0 (2):\penalty0
  915--953, 2025.
\newblock \doi{10.1111/joes.70018}.

\bibitem[El-Yaniv and Wiener(2010)]{elyaniv2010}
Ran El-Yaniv and Yair Wiener.
\newblock On the foundations of noise-free selective classification.
\newblock \emph{Journal of Machine Learning Research}, 11:\penalty0 1605--1641,
  2010.

\bibitem[Fama and French(2015)]{fama2015}
Eugene~F. Fama and Kenneth~R. French.
\newblock A five-factor asset pricing model.
\newblock \emph{Journal of Financial Economics}, 116\penalty0 (1):\penalty0
  1--22, 2015.
\newblock \doi{10.1016/j.jfineco.2014.10.010}.

\bibitem[Fermanian et~al.(2025)Fermanian, Humbert, and
  Blanchard]{fermanian2025}
Jean-Baptiste Fermanian, Pierre Humbert, and Gilles Blanchard.
\newblock Transductive conformal inference for full ranking.
\newblock In \emph{Advances in Neural Information Processing Systems 39
  (NeurIPS 2025)}, 2025.

\bibitem[Fort et~al.(2019)Fort, Hu, and Lakshminarayanan]{fort2019}
Stanislav Fort, Huiyi Hu, and Balaji Lakshminarayanan.
\newblock Deep ensembles: A loss landscape perspective.
\newblock \emph{arXiv preprint}, 2019.
\newblock arXiv:1912.02757.

\bibitem[Franc et~al.(2021)Franc, Pr{\r{u}}{\v{s}}a, and
  Vor{\'a}{\v{c}}ek]{franc2021}
Vojtech Franc, Daniel Pr{\r{u}}{\v{s}}a, and V{\'a}clav Vor{\'a}{\v{c}}ek.
\newblock Optimal strategies for reject option classifiers.
\newblock In \emph{Journal of Machine Learning Research}, volume~24, pages
  1--49, 2021.

\bibitem[French(2025)]{french2025}
Kenneth~R. French.
\newblock Kenneth {R}. {F}rench data library.
\newblock
  \url{https://mba.tuck.dartmouth.edu/pages/faculty/ken.french/data_library.html},
  2025.

\bibitem[Gal and Ghahramani(2016)]{gal2016}
Yarin Gal and Zoubin Ghahramani.
\newblock Dropout as a {B}ayesian approximation: Representing model uncertainty
  in deep learning.
\newblock In \emph{Proceedings of the 33rd International Conference on Machine
  Learning (ICML 2016)}, pages 1050--1059, 2016.

\bibitem[Gama et~al.(2014)Gama, {\v{Z}}liobait{\.e}, Bifet, Pechenizkiy, and
  Bouchachia]{gama2014}
Jo\~{a}o Gama, Indr{\.e} {\v{Z}}liobait{\.e}, Albert Bifet, Mykola
  Pechenizkiy, and Abdelhamid Bouchachia.
\newblock A survey on concept drift adaptation.
\newblock \emph{ACM Computing Surveys}, 46(4):\penalty0 44:1--44:37, 2014.
\newblock \doi{10.1145/2523813}.

\bibitem[Garg et~al.(2022)Garg, Balakrishnan, Lipton, Neyshabur, and
  Sedghi]{garg2022}
Saurabh Garg, Sivaraman Balakrishnan, Zachary~C. Lipton, Behnam Neyshabur, and
  Hanie Sedghi.
\newblock Leveraging unlabeled data to predict out-of-distribution performance.
\newblock In \emph{Proceedings of the 10th International Conference on Learning
  Representations (ICLR 2022)}, 2022.
\newblock arXiv:2201.04234.

\bibitem[Garlappi et~al.(2007)Garlappi, Uppal, and Wang]{garlappi2007}
Lorenzo Garlappi, Raman Uppal, and Tan Wang.
\newblock Portfolio selection with parameter and model uncertainty: A
  multi-prior approach.
\newblock \emph{Review of Financial Studies}, 20\penalty0 (1):\penalty0 41--81,
  2007.

\bibitem[Geifman and El-Yaniv(2017)]{geifman2017}
Yonatan Geifman and Ran El-Yaniv.
\newblock Selective classification for deep neural networks.
\newblock In \emph{Advances in Neural Information Processing Systems 30
  (NeurIPS 2017)}, pages 4878--4887, 2017.
\newblock arXiv:1705.08500.

\bibitem[Gibbs and Cand{\`e}s(2021)]{gibbs2021}
Isaac Gibbs and Emmanuel~J. Cand{\`e}s.
\newblock Adaptive conformal inference under distribution shift.
\newblock In \emph{Advances in Neural Information Processing Systems 34
  (NeurIPS 2021)}, pages 1660--1672, 2021.
\newblock arXiv:2106.00170.

\bibitem[Gibbs and Cand{\`e}s(2024)]{gibbs2024}
Isaac Gibbs and Emmanuel~J. Cand{\`e}s.
\newblock Conformal inference for online prediction with arbitrary distribution
  shifts.
\newblock \emph{Journal of Machine Learning Research}, 25\penalty0
  (162):\penalty0 1--36, 2024.

\bibitem[Grinold(1989)]{grinold1989}
Richard~C. Grinold.
\newblock The fundamental law of active management.
\newblock \emph{Journal of Portfolio Management}, 15\penalty0 (3):\penalty0
  30--37, 1989.
\newblock \doi{10.3905/jpm.1989.409211}.

\bibitem[Gu et~al.(2020)Gu, Kelly, and Xiu]{gu2020}
Shihao Gu, Bryan Kelly, and Dacheng Xiu.
\newblock Empirical asset pricing via machine learning.
\newblock \emph{Review of Financial Studies}, 33\penalty0 (5):\penalty0
  2223--2273, 2020.

\bibitem[Guillory et~al.(2021)Guillory, Shankar, Ebrahimi, Darrell, and
  Schmidt]{guillory2021}
Devin Guillory, Vaishaal Shankar, Sayna Ebrahimi, Trevor Darrell, and Ludwig
  Schmidt.
\newblock Predicting with confidence on unseen distributions.
\newblock In \emph{Proceedings of the IEEE/CVF International Conference on
  Computer Vision (ICCV 2021)}, pages 1134--1144, 2021.
\newblock \doi{10.1109/ICCV48922.2021.00117}.

\bibitem[Gustafsson et~al.(2020)Gustafsson, Danelljan, and
  Sch{\"o}n]{gustafsson2020}
Fredrik~K. Gustafsson, Martin Danelljan, and Thomas~B. Sch{\"o}n.
\newblock Evaluating scalable {B}ayesian deep learning methods for robust
  computer vision.
\newblock In \emph{Proceedings of the IEEE/CVF Conference on Computer Vision
  and Pattern Recognition Workshops (CVPRW 2020)}, 2020.

\bibitem[Habibi(2021)]{habibi2021}
Reza Habibi.
\newblock Bayesian online change-point detection in finance.
\newblock \emph{Statistics, Optimization and Information Computing}, 9\penalty0
  (3):\penalty0 580--588, 2021.

\bibitem[Hamilton(1989)]{hamilton1989}
James~D. Hamilton.
\newblock A new approach to the economic analysis of nonstationary time series
  and the business cycle.
\newblock \emph{Econometrica}, 57\penalty0 (2):\penalty0 357--384, 1989.

\bibitem[Harvey et~al.(2016)Harvey, Liu, and Zhu]{harvey2016}
Campbell~R. Harvey, Yan Liu, and Heqing Zhu.
\newblock \ldots and the cross-section of expected returns.
\newblock \emph{Review of Financial Studies}, 29\penalty0 (1):\penalty0 5--68,
  2016.
\newblock \doi{10.1093/rfs/hhv059}.

\bibitem[Harvey et~al.(2018)Harvey, Hoyle, Korgaonkar, Rattray, Sargaison, and
  {van Hemert}]{harvey2018}
Campbell~R. Harvey, Edward Hoyle, Russell Korgaonkar, Sandy Rattray, Matthew
  Sargaison, and Otto {van Hemert}.
\newblock The impact of volatility targeting.
\newblock \emph{Journal of Portfolio Management}, 45\penalty0 (1):\penalty0
  14--33, 2018.
\newblock \doi{10.3905/jpm.2018.45.1.014}.

\bibitem[Hendrycks and Dietterich(2019)]{hendrycks2019}
Dan Hendrycks and Thomas Dietterich.
\newblock Benchmarking neural network robustness to common perturbations and
  corruptions.
\newblock In \emph{Proceedings of the 7th International Conference on Learning
  Representations (ICLR 2019)}, 2019.
\newblock arXiv:1903.12261.

\bibitem[Hendrycks and Gimpel(2017)]{hendrycks2017}
Dan Hendrycks and Kevin Gimpel.
\newblock A baseline for detecting misclassified and out-of-distribution
  examples in neural networks.
\newblock In \emph{Proceedings of the 5th International Conference on Learning
  Representations (ICLR 2017)}, 2017.
\newblock arXiv:1610.02136.

\bibitem[Hinder et~al.(2020)Hinder, Artelt, and Hammer]{hinder2020}
Fabian Hinder, Andr{\'e} Artelt, and Barbara Hammer.
\newblock Towards non-parametric drift detection via dynamic adapting window
  independence drift detection ({DAWIDD}).
\newblock In \emph{Proceedings of the 37th International Conference on Machine
  Learning (ICML)}, volume~119 of \emph{PMLR}, pages 4249--4259, 2020.

\bibitem[Hinder et~al.(2023)Hinder, Vaquet, Brinkrolf, and
  Hammer]{hinder2023}
Fabian Hinder, Valerie Vaquet, Johannes Brinkrolf, and Barbara Hammer.
\newblock On the hardness and necessity of supervised concept drift
  detection.
\newblock In \emph{Proceedings of the 12th International Conference on Pattern
  Recognition Applications and Methods (ICPRAM)}, pages 164--175.
  SCITEPRESS, 2023.
\newblock \doi{10.5220/0011797500003411}.

\bibitem[Huang(2025)]{huang2025}
Jiayi Huang.
\newblock Bayesian change-point detection in multiple simulated and real
  financial time series.
\newblock In \emph{Proceedings of the 2025 International Conference on Economic
  Management and Big Data Application (ICEMBDA 2025)}, pages 904--911, 2025.
\newblock \doi{10.1145/3770177.3770326}.

\bibitem[H{\"u}llermeier and Waegeman(2021)]{hullermeier2021}
Eyke H{\"u}llermeier and Willem Waegeman.
\newblock Aleatoric and epistemic uncertainty in machine learning: An
  introduction to concepts and methods.
\newblock \emph{Machine Learning}, 110\penalty0 (3):\penalty0 457--506, 2021.

\bibitem[Jacquier et~al.(2025)Jacquier, Muhle-Karbe, and
  Mulligan]{jacquier2025}
Antoine Jacquier, Johannes Muhle-Karbe, and Joseph Mulligan.
\newblock In-sample and out-of-sample {S}harpe ratios for linear predictive
  models.
\newblock \emph{arXiv preprint}, 2025.
\newblock arXiv:2501.03938.

\bibitem[Jiang et~al.(2022)Jiang, Nagarajan, Baek, and Kolter]{jiang2022}
Yiding Jiang, Vaishnavh Nagarajan, Christina Baek, and J.~Zico Kolter.
\newblock Assessing generalization of {SGD} via disagreement.
\newblock In \emph{Proceedings of the 10th International Conference on Learning
  Representations (ICLR 2022)}, 2022.
\newblock arXiv:2106.13799.

\bibitem[Kaya and Nguyen(2025)]{kaya2025}
Irem Kaya and Khuong~An Nguyen.
\newblock Conformal prediction for reliable stock selections.
\newblock In \emph{Proceedings of the Fourteenth Symposium on Conformal and
  Probabilistic Prediction with Applications (COPA 2025)}, volume 266 of
  \emph{PMLR}, pages 781--783, 2025.

\bibitem[Kelly(1956)]{kelly1956}
John~L. Kelly, Jr.
\newblock A new interpretation of information rate.
\newblock \emph{Bell System Technical Journal}, 35\penalty0 (4):\penalty0
  917--926, 1956.

\bibitem[Kendall and Gal(2017)]{kendall2017}
Alex Kendall and Yarin Gal.
\newblock What uncertainties do we need in {B}ayesian deep learning for
  computer vision?
\newblock In \emph{Advances in Neural Information Processing Systems 30
  (NeurIPS 2017)}, pages 5574--5584, 2017.

\bibitem[Klibanoff et~al.(2005)Klibanoff, Marinacci, and
  Mukerji]{klibanoff2005}
Peter Klibanoff, Massimo Marinacci, and Sujoy Mukerji.
\newblock A smooth model of decision making under ambiguity.
\newblock \emph{Econometrica}, 73\penalty0 (6):\penalty0 1849--1892, 2005.

\bibitem[Kotelevskii et~al.(2025{\natexlab{a}})Kotelevskii, Goloburda,
  Kondratyev, Fishkov, Guizani, Moulines, and Panov]{kotelevskii2025b}
Nikita Kotelevskii, Maria Goloburda, Vladimir Kondratyev, Alexander Fishkov,
  Mohsen Guizani, {\'E}ric Moulines, and Maxim Panov.
\newblock Multidimensional uncertainty quantification via optimal transport.
\newblock \emph{arXiv preprint}, 2025{\natexlab{a}}.
\newblock arXiv:2509.22380.

\bibitem[Kotelevskii et~al.(2025{\natexlab{b}})Kotelevskii, Kondratyev,
  Tak{\'a}{\v{c}}, Moulines, and Panov]{kotelevskii2025a}
Nikita Kotelevskii, Vladimir Kondratyev, Martin Tak{\'a}{\v{c}}, {\'E}ric
  Moulines, and Maxim Panov.
\newblock From risk to uncertainty: Generating predictive uncertainty measures
  via {B}ayesian estimation.
\newblock In \emph{Proceedings of the 13th International Conference on Learning
  Representations (ICLR 2025)}, 2025{\natexlab{b}}.

\bibitem[Lahlou et~al.(2023)Lahlou, Jain, Nekoei, Butoi, Bertin, Rector-Brooks,
  Korablyov, and Bengio]{lahlou2023}
Salem Lahlou, Moksh Jain, Hadi Nekoei, Victor~I. Butoi, Paul Bertin, Jarrid
  Rector-Brooks, Maksym Korablyov, and Yoshua Bengio.
\newblock {DEUP}: Direct epistemic uncertainty prediction.
\newblock \emph{Transactions on Machine Learning Research}, 2023.

\bibitem[Lakshminarayanan et~al.(2017)Lakshminarayanan, Pritzel, and
  Blundell]{lakshminarayanan2017}
Balaji Lakshminarayanan, Alexander Pritzel, and Charles Blundell.
\newblock Simple and scalable predictive uncertainty estimation using deep
  ensembles.
\newblock In \emph{Advances in Neural Information Processing Systems 30
  (NeurIPS 2017)}, pages 6402--6413, 2017.

\bibitem[{Le Folgoc} et~al.(2021){Le Folgoc}, Baltatzis, Desai, Devaraj, Ellis,
  {Martinez Manzanera}, Nair, Qiu, Schnabel, and Glocker]{lefolgoc2021}
Lo{\"i}c {Le Folgoc}, Vasileios Baltatzis, Sujal Desai, Anand Devaraj, Sam
  Ellis, Octavio~E. {Martinez Manzanera}, Arjun Nair, Huaqi Qiu, Julia
  Schnabel, and Ben Glocker.
\newblock Is {MC} dropout {B}ayesian?
\newblock \emph{arXiv preprint}, 2021.
\newblock arXiv:2110.04286.

\bibitem[Lee(2026)]{lee2026}
Christopher Lee.
\newblock Conformal prediction under distribution shift: A {COVID}-19 natural
  experiment.
\newblock \emph{arXiv preprint}, 2026.
\newblock arXiv:2601.00908.

\bibitem[Leippold et~al.(2022)Leippold, Wang, and Zhou]{leippold2022}
Markus Leippold, Qian Wang, and Wenyu Zhou.
\newblock Machine learning in the {C}hinese stock market.
\newblock \emph{Journal of Financial Economics}, 145\penalty0 (2):\penalty0
  64--82, 2022.

\bibitem[Liao et~al.(2025)Liao, Ma, Neuhierl, and Schilling]{liao2025}
Yuan Liao, Xinjie Ma, Andreas Neuhierl, and Linda Schilling.
\newblock The uncertainty of machine learning predictions in asset pricing.
\newblock \emph{arXiv preprint}, 2025.
\newblock arXiv:2503.00549.

\bibitem[Liu et~al.(2026)Liu, Luo, Wang, and Zhang]{liu2026}
Yuan Liu, Yang Luo, Zhen Wang, and Xin Zhang.
\newblock Uncertainty-adjusted sorting for asset pricing with machine learning.
\newblock \emph{arXiv preprint}, 2026.
\newblock arXiv:2601.00593.

\bibitem[{L{\'o}pez de Prado}(2018)]{lopezdeprado2018}
Marcos {L{\'o}pez de Prado}.
\newblock \emph{Advances in Financial Machine Learning}.
\newblock John Wiley \& Sons, Hoboken, NJ, 2018.

\bibitem[{L{\'o}pez de Prado} et~al.(2025){L{\'o}pez de Prado}, Lipton, and
  Zoonekynd]{lopezdeprado2025}
Marcos {L{\'o}pez de Prado}, Alexander Lipton, and Vincent Zoonekynd.
\newblock How to use the {S}harpe ratio.
\newblock \emph{SSRN preprint}, 2025.
\newblock SSRN:5520741.

\bibitem[Lu et~al.(2023)Lu, Qin, Zhai, Shen, Chen, Wang, Kolouri, Stepputtis,
  Campbell, and Sycara]{lu2023}
Yuzhe Lu, Yilong Qin, Runtian Zhai, Andrew Shen, Ketong Chen, Zhenlin Wang,
  Soheil Kolouri, Simon Stepputtis, Joseph Campbell, and Katia Sycara.
\newblock Characterizing out-of-distribution error via optimal transport.
\newblock In \emph{Advances in Neural Information Processing Systems 36
  (NeurIPS 2023)}, 2023.
\newblock arXiv:2305.15640.

\bibitem[MacKay(1992)]{mackay1992}
David J.~C. MacKay.
\newblock A practical {B}ayesian framework for backpropagation networks.
\newblock \emph{Neural Computation}, 4\penalty0 (3):\penalty0 448--472, 1992.

\bibitem[Manna et~al.(2025)Manna, Sett, Dey, Gu, Schifano, and He]{manna2025}
Arnab Manna, Arka~V. Sett, Dipak~K. Dey, Yuwen Gu, Elizabeth~D. Schifano, and
  Jieying He.
\newblock Conformal prediction inference in regularized insurance models.
\newblock \emph{Applied Stochastic Models in Business and Industry},
  41\penalty0 (5):\penalty0 e70045, 2025.
\newblock \doi{10.1002/asmb.70045}.

\bibitem[Mucs{\'a}nyi et~al.(2024)Mucs{\'a}nyi, Kirchhof, and Oh]{mucsanyi2024}
B{\'a}lint Mucs{\'a}nyi, Michael Kirchhof, and Seong~Joon Oh.
\newblock Benchmarking uncertainty disentanglement: Specialized uncertainties
  for specialized tasks.
\newblock In \emph{Advances in Neural Information Processing Systems 37
  (NeurIPS 2024)}, pages 50972--51038, 2024.
\newblock \doi{10.52202/079017-1614}.

\bibitem[Mulliner et~al.(2026)Mulliner, Harvey, Xia, Fang, and {van
  Hemert}]{mulliner2026}
Alex Mulliner, Campbell~R. Harvey, Chao Xia, Ed~Fang, and Otto {van Hemert}.
\newblock Regimes.
\newblock \emph{Journal of Portfolio Management}, 52\penalty0 (4):\penalty0
  6--25, 2026.
\newblock \doi{10.3905/jpm.2025.1.798}.

\bibitem[Naeini et~al.(2015)Naeini, Cooper, and Hauskrecht]{naeini2015}
Mahdi~Pakdaman Naeini, Gregory~F. Cooper, and Milos Hauskrecht.
\newblock Obtaining well calibrated probabilities using {B}ayesian binning.
\newblock In \emph{Proceedings of the 29th AAAI Conference on Artificial
  Intelligence (AAAI 2015)}, pages 2901--2907, 2015.

\bibitem[Neal(1996)]{neal1996}
Radford~M. Neal.
\newblock \emph{Bayesian Learning for Neural Networks}.
\newblock PhD thesis, University of Toronto, 1996.

\bibitem[Newey and West(1987)]{newey1987}
Whitney~K. Newey and Kenneth~D. West.
\newblock A simple, positive semi-definite, heteroskedasticity and
  autocorrelation consistent covariance matrix.
\newblock \emph{Econometrica}, 55\penalty0 (3):\penalty0 703--708, 1987.
\newblock \doi{10.2307/1913610}.

\bibitem[Nie et~al.(2026)Nie, Mulvey, Poor, Yu, and Huang]{nie2026}
Yizhan Nie, John~M. Mulvey, H.~Vincent Poor, Chenyang Yu, and Haochen Huang.
\newblock Deep generative models meet statistical methods: A generalised
  framework for financial regime identification.
\newblock \emph{Annals of Operations Research}, 2026.
\newblock \doi{10.1007/s10479-026-07041-z}.

\bibitem[Nystrup et~al.(2017)Nystrup, Hansen, Madsen, and
  Lindstr{\"o}m]{nystrup2017}
Peter Nystrup, Bo~William Hansen, Henrik Madsen, and Erik Lindstr{\"o}m.
\newblock Dynamic allocation or diversification: A regime-based approach to
  multiple assets.
\newblock \emph{Journal of Portfolio Management}, 44\penalty0 (2):\penalty0
  62--73, 2017.

\bibitem[Oelschl{\"a}ger and Adam(2023)]{oelschlager2023}
Lennart Oelschl{\"a}ger and Timo Adam.
\newblock Detecting bearish and bullish markets in financial time series using
  hierarchical hidden {M}arkov models.
\newblock \emph{Statistical Modelling}, 23\penalty0 (2):\penalty0 107--126,
  2023.
\newblock \doi{10.1177/1471082X211034048}.

\bibitem[Ovadia et~al.(2019)Ovadia, Fertig, Ren, Nado, Sculley, Nowozin,
  Dillon, Lakshminarayanan, and Snoek]{ovadia2019}
Yaniv Ovadia, Emily Fertig, Jie Ren, Zachary Nado, D.~Sculley, Sebastian
  Nowozin, Joshua Dillon, Balaji Lakshminarayanan, and Jasper Snoek.
\newblock Can you trust your model's uncertainty? {E}valuating predictive
  uncertainty under dataset shift.
\newblock In \emph{Advances in Neural Information Processing Systems 32
  (NeurIPS 2019)}, pages 13991--14002, 2019.
\newblock arXiv:1906.02530.

\bibitem[Plassier et~al.(2025)Plassier, Fishkov, Dheur, Guizani, {Ben Taieb},
  Panov, and Moulines]{plassier2025}
Vincent Plassier, Alexander Fishkov, Victor Dheur, Mohsen Guizani, Souhaib {Ben
  Taieb}, Maxim Panov, and {\'E}ric Moulines.
\newblock Rectifying conformity scores for better conditional coverage.
\newblock In \emph{Proceedings of the 42nd International Conference on Machine
  Learning (ICML 2025)}, volume 267 of \emph{PMLR}, pages 49459--49492, 2025.

\bibitem[Platt(1999)]{platt1999}
John~C. Platt.
\newblock Probabilistic outputs for support vector machines and comparisons to
  regularized likelihood methods.
\newblock In Alexander~J. Smola, Peter Bartlett, Bernhard Sch{\"o}lkopf, and
  Dale Schuurmans, editors, \emph{Advances in Large Margin Classifiers}, pages
  61--74. MIT Press, Cambridge, MA, 1999.

\bibitem[Pun et~al.(2023)Pun, Zheng, and Zhao]{pun2023}
Chi~Seng Pun, Zhongyuan Zheng, and Jingwei Zhao.
\newblock Reinforcement learning for regime-switching portfolio allocation.
\newblock \emph{Quantitative Finance}, 23\penalty0 (7--8):\penalty0 1059--1078,
  2023.

\bibitem[Qian(2011)]{qian2011}
Edward Qian.
\newblock Risk parity and diversification.
\newblock \emph{The Journal of Investing}, 20\penalty0 (1):\penalty0 119--127,
  2011.

\bibitem[Rabanser and Papernot(2025)]{rabanser2025}
Stephan Rabanser and Nicolas Papernot.
\newblock What does it take to build a performant selective classifier?
\newblock In \emph{Advances in Neural Information Processing Systems 38
  (NeurIPS 2025)}, 2025.
\newblock arXiv:2510.20242.

\bibitem[Rossellini et~al.(2024)Rossellini, Barber, and
  Willett]{rossellini2024}
Raphael Rossellini, Rina~Foygel Barber, and Rebecca Willett.
\newblock Integrating uncertainty awareness into conformalized quantile
  regression.
\newblock In \emph{Proceedings of the 27th International Conference on
  Artificial Intelligence and Statistics (AISTATS 2024)}, volume 238 of
  \emph{PMLR}, 2024.

\bibitem[Sanderink(2026{\natexlab{a}})]{sanderink2026alpha}
Ursina Sanderink.
\newblock When alpha breaks: Two-level uncertainty for safe deployment of
  cross-sectional stock rankers.
\newblock \emph{arXiv preprint}, 2026{\natexlab{a}}.
\newblock arXiv:2603.13252.

\bibitem[Sanderink(2026{\natexlab{b}})]{sanderink2026deup}
Ursina Sanderink.
\newblock deup: Direct epistemic uncertainty prediction for scikit-learn
  models.
\newblock \url{https://github.com/ursinasanderink/deup}, 2026{\natexlab{b}}.
\newblock Python package v0.4.2.

\bibitem[Schmitt(2026)]{schmitt2026}
Marcus Schmitt.
\newblock Taming tail risk in financial markets: Conformal risk control for
  nonstationary portfolio {VaR}.
\newblock \emph{arXiv preprint}, 2026.
\newblock arXiv:2602.03903.

\bibitem[Shu et~al.(2024)Shu, Yu, and Mulvey]{shu2024}
Yizhan Shu, Chenyu Yu, and John~M. Mulvey.
\newblock Dynamic asset allocation with asset-specific regime forecasts.
\newblock \emph{Annals of Operations Research}, 346\penalty0 (1):\penalty0
  285--318, 2024.
\newblock \doi{10.1007/s10479-024-06266-0}.

\bibitem[Siahkali et~al.(2026)Siahkali, Verma, and Gupta]{siahkali2026}
Farzan Siahkali, Apoorva Verma, and Vibhor Gupta.
\newblock Coverage guarantees for pseudo-calibrated conformal prediction under
  distribution shift.
\newblock \emph{arXiv preprint}, 2026.
\newblock arXiv:2602.14913.

\bibitem[Singh et~al.(2024)Singh, Sharma, Havaldar, Saket, and
  Raghuveer]{singh2024}
Sagalpreet Singh, Navodita Sharma, Shreyas Havaldar, Rishi Saket, and Aravindan
  Raghuveer.
\newblock Learning from label proportions and covariate-shifted instances.
\newblock \emph{arXiv preprint}, 2024.
\newblock arXiv:2411.12334.

\bibitem[Spears et~al.(2020)Spears, Zohren, and Roberts]{spears2020}
Taylor Spears, Stefan Zohren, and Stephen Roberts.
\newblock Investment sizing with deep learning prediction uncertainties for
  high-frequency {E}urodollar futures trading.
\newblock Technical report, University of Oxford and Oxford-Man Institute of
  Quantitative Finance, 2020.
\newblock SSRN:3664497.

\bibitem[Sprangers et~al.(2021)Sprangers, Schelter, and {de
  Rijke}]{sprangers2021}
Olivier Sprangers, Sebastian Schelter, and Maarten {de Rijke}.
\newblock Probabilistic gradient boosting machines for large-scale
  probabilistic regression.
\newblock In \emph{Proceedings of the 27th ACM SIGKDD Conference on Knowledge
  Discovery and Data Mining (KDD 2021)}, pages 1510--1520, 2021.
\newblock \doi{10.1145/3447548.3467278}.

\bibitem[Stankevi{\v{c}}i{\=u}t{\.e} et~al.(2021)Stankevi{\v{c}}i{\=u}t{\.e},
  Alaa, and {van der Schaar}]{stankeviciute2021}
Kamil{\.e} Stankevi{\v{c}}i{\=u}t{\.e}, Ahmed~M. Alaa, and Mihaela {van der
  Schaar}.
\newblock Conformal time-series forecasting.
\newblock In \emph{Advances in Neural Information Processing Systems 34
  (NeurIPS 2021)}, pages 6216--6228, 2021.

\bibitem[Sudjianto and Zhang(2026)]{sudjianto2026}
Agus Sudjianto and Aijun Zhang.
\newblock {GBDT} geometry for local calibration and model repair.
\newblock Technical report, SSRN Working Paper, 2026.
\newblock SSRN:6815720.

\bibitem[Sukhov(2026)]{sukhov2026}
Sergey Sukhov.
\newblock Bayesian {K}elly criterion with parameter uncertainty: A robust
  framework for position sizing under estimation risk.
\newblock \emph{SSRN preprint}, 2026.
\newblock SSRN:6195358.

\bibitem[Sun and Yu(2025)]{sun2025}
Sophia Sun and Rose Yu.
\newblock Conformal prediction for time-series forecasting with change points.
\newblock In \emph{Advances in Neural Information Processing Systems 38
  (NeurIPS 2025)}, 2025.
\newblock arXiv:2509.02844.

\bibitem[Tsaknaki et~al.(2025)Tsaknaki, Lillo, and Mazzarisi]{tsaknaki2025}
Ioanna-Yvonni Tsaknaki, Fabrizio Lillo, and Piero Mazzarisi.
\newblock Online learning of order flow and market impact with {B}ayesian
  change-point detection methods.
\newblock \emph{Quantitative Finance}, 25\penalty0 (2):\penalty0 307--322,
  2025.
\newblock \doi{10.1080/14697688.2024.2337300}.

\bibitem[Ustimenko et~al.(2021)Ustimenko, Prokhorenkova, and
  Malinin]{ustimenko2021}
Aleksei Ustimenko, Liudmila Prokhorenkova, and Andrey Malinin.
\newblock Uncertainty in gradient boosting via ensembles.
\newblock In \emph{Proceedings of the 9th International Conference on Learning
  Representations (ICLR 2021)}, 2021.

\bibitem[Vovk et~al.(2005)Vovk, Gammerman, and Shafer]{vovk2005}
Vladimir Vovk, Alexander Gammerman, and Glenn Shafer.
\newblock \emph{Algorithmic Learning in a Random World}.
\newblock Springer, New York, 2005.

\bibitem[Wang and Yan(2021)]{wang2021}
Feifei Wang and Xuemin~Sterling Yan.
\newblock Downside risk and the performance of volatility-managed portfolios.
\newblock \emph{Journal of Banking \& Finance}, 131:\penalty0 106198, 2021.

\bibitem[Wang et~al.(2020)Wang, Lin, and Mikhelson]{wanglin2020}
Matthew Wang, Yi-Hong Lin, and Ilya Mikhelson.
\newblock Regime-switching factor investing with hidden {M}arkov models.
\newblock \emph{Journal of Risk and Financial Management}, 13:\penalty0 311,
  2020.
\newblock \doi{10.3390/jrfm13120311}.

\bibitem[White(2000)]{white2000}
Halbert White.
\newblock A reality check for data snooping.
\newblock \emph{Econometrica}, 68\penalty0 (5):\penalty0 1097--1126, 2000.
\newblock \doi{10.1111/1468-0262.00152}.

\bibitem[Wong et~al.(2025)Wong, Chan, and Azizi]{wong2025}
Steven Y.~K. Wong, Jennifer S.~K. Chan, and Lamiae Azizi.
\newblock Quantifying neural network uncertainty under volatility clustering.
\newblock \emph{Neurocomputing}, 614:\penalty0 128816, 2025.
\newblock \doi{10.1016/j.neucom.2024.128816}.

\bibitem[Zadrozny and Elkan(2002)]{zadrozny2002}
Bianca Zadrozny and Charles Elkan.
\newblock Transforming classifier scores into accurate multiclass probability
  estimates.
\newblock In \emph{Proceedings of the 8th ACM SIGKDD International Conference
  on Knowledge Discovery and Data Mining (KDD 2002)}, pages 694--699, 2002.
\newblock \doi{10.1145/775047.775151}.

\bibitem[Zhang and Zhou(2025)]{zhang2025}
Rui Zhang and Peng Zhou.
\newblock Uncertainty quantification based on conformal prediction for
  industrial time series with distribution shift.
\newblock \emph{IEEE Transactions on Industrial Informatics}, 21\penalty0
  (5):\penalty0 3676--3685, 2025.
\newblock \doi{10.1109/TII.2025.3529920}.

\end{thebibliography}
\end{document}